\documentclass[nofootinbib,aps, prd, twocolumn, lengthcheck, superscriptaddress]{revtex4-1}
\usepackage{amsfonts, amssymb, amsmath, graphicx, comment, bm, slashed, caption, subcaption, dcolumn,color}
\newcommand{\beq}{\begin{eqnarray}}
\newcommand{\eeq}{\end{eqnarray}}
\newcommand{\Lc}{{\cal{L}}}
\newcommand{\bd}{\mathbf}

\newcommand{\tr}{\mbox{tr}}

\newcommand{\lqcd}{\Lambda_{ {\rm QCD}} }

\newcommand{\calP}{ \mathcal{P} }

\def\braket#1{\mathinner{\langle{#1}\rangle}}

\usepackage[colorlinks=true,linktocpage=true,linkcolor=blue,citecolor=blue]{hyperref}
\begin{document}
\title{From hadrons to quarks in neutron stars: a review}

\author{Gordon Baym}
\affiliation{Department of Physics, University of Illinois at Urbana-Champaign, 1110 W. Green Street, Urbana, Illinois 61801, USA}
\affiliation{iTHES Research Group, RIKEN, Wako, Saitama 351-0198, Japan}
\affiliation{The Niels Bohr International Academy, The Niels Bohr Institute,  University of Copenhagen, Blegdamsvej 17, DK-2100 Copenhagen \O, Denmark}

\author{Tetsuo Hatsuda}
\affiliation{iTHES Research Group, RIKEN, Wako, Saitama 351-0198, Japan}
\affiliation{iTHEMS Program, RIKEN, Wako, Saitama 351-0198, Japan}
\affiliation{Theoretical Research Division, Nishina Center, RIKEN, Wako 351-0198, Japan}

\author{Toru Kojo}
\affiliation{Key Laboratory of Quark and Lepton Physics (MOE) and Institute of Particle Physics, Central China Normal University, Wuhan 430079, China}
\affiliation{Department of Physics, University of Illinois at Urbana-Champaign, 1110 W. Green Street, Urbana, Illinois 61801, USA}

\author{Philip D. Powell}
\affiliation{Department of Physics, University of Illinois at Urbana-Champaign, 1110 W. Green Street, Urbana, Illinois 61801, USA}
\affiliation{Lawrence Livermore National Laboratory, 7000 East Ave., Livermore, CA 94550}

\author{Yifan Song}
\affiliation{Department of Physics, University of Illinois at Urbana-Champaign, 1110 W. Green Street, Urbana, Illinois 61801, USA}

\author{Tatsuyuki Takatsuka}
\affiliation{Theoretical Research Division, Nishina Center, RIKEN, Wako 351-0198, Japan}
\affiliation{Iwate University, Morioka 020-8550, Japan}

\date{\today}

\begin{abstract}   In recent years our understanding of neutron stars has advanced remarkably, thanks to research converging from many directions.  The importance of understanding neutron star behavior and structure has been underlined by the recent direct detection of gravitational radiation from merging neutron stars.  The clean identification of several heavy neutron stars, of order two solar masses, challenges our current understanding of how dense matter can be sufficiently stiff to support such a mass against gravitational collapse.   Programs underway to determine simultaneously the mass and radius of neutron stars will continue to constrain and inform theories of neutron star interiors.   At the same time, an emerging understanding in quantum chromodynamics (QCD) of how nuclear matter can evolve into deconfined quark matter at high baryon densities is leading to advances in understanding the equation of state of the matter under the extreme conditions in neutron star interiors.  

   We review here the equation of state of matter in neutron stars from the solid crust through the liquid nuclear matter interior to the quark regime at higher densities.   We focus in detail on the question of how quark matter appears in neutron stars, and how it affects the equation of state.   After discussing the crust and liquid nuclear matter in the core we briefly review aspects of microscopic quark physics relevant to neutron stars, and quark models of dense matter based on the Nambu--Jona-Lasinio framework, in which gluonic processes are replaced by effective quark interactions.    
We turn then to describing equations of state useful for interpretation of both electromagnetic and gravitational observations, reviewing the emerging picture of hadron-quark continuity in which hadronic matter turns relatively smoothly, with at most only a weak first order transition, into quark matter with increasing density.    We review construction of {\em unified equations of state} that interpolate between the reasonably well understood nuclear matter regime at low densities and the quark matter regime at higher densities.   The utility of such interpolations is driven by the present inability to calculate the dense matter equation of state in QCD from first principles.   As we review, the parameters of effective quark models -- which have direct relevance to the more general structure of the QCD phase diagram of dense and hot matter -- are constrained by neutron star mass and radii measurements, in particular favoring large repulsive density-density and attractive diquark pairing interactions.  We describe the structure of neutron stars constructed from the unified equations of states with crossover.  Lastly we present the current equations of state -- called ``QHC18" for quark-hadron crossover -- in a parametrized form practical for neutron star modeling.
\end{abstract}


\maketitle

\section{Introduction}

Neutron stars provide a cosmic laboratory in which the phases of cold dense strongly interacting nuclear matter are realized~\cite{haensel,annrev1,Hell:2014xva,Lattimer:2012nd}. Indeed, while heavy ion collision experiments and lattice quantum chromodynamics (QCD) simulations provide insight into the properties of hot and dense QCD, neutron stars are the only known window into the rich structure of cold dense QCD.  Recent astrophysical inferences of neutron star masses, $M$, and radii, $R$, in low mass x-ray binaries \cite{OzelFreire,Ozel2010,Steiner2012,Steiner2012-2,Lattimer2014,Ozel2015,Steiner2015,Steiner2016,fredcole,Alvarez-Castillo:2016oln}, and the wealth of new data, on masses and radii of isolated neutron stars as well, expected from the NICER (the Neutron Star Interior Composition Explorer) experiment \cite{nicer,michi,Miller:2016kae,ozel-nicer,Bogdanov_2008,timing} on the International Space Station will significantly constrain the neutron star equation of state.   Such constraints are crucial for understanding observations of dynamical neutron star phenomena,  from neutron star seismology \cite{kokkotas2001} to binary neutron star inspirals, now detected gravitationally \cite{GW170817} and subsequent mergers detected by multi-messenger electromagnetic signals \cite{GW170817A}. Reliable equations of state, at zero and elevated temperatures, are crucial for predicting the gravitational wave signatures of neutron star--black hole and neutron star--neutron star mergers \cite{Hotokezaka:2011dh,takami,baiotti,sekiguchi,stu1,stu2,bernuzzi} to be detected at gravitational wave observatories present and future,
including LIGO \cite{aLIGO,ligo-1,ligo-2}, Virgo \cite{virgo}, GEO \cite{GEO}, KAGRA \cite{Kagra},  LIGO-India \cite{LIGO-India},  and LISA and other spaced-based observatories \cite{LISA}, as well as via pulsar timing arrays \cite{PTA}. The purpose of this review is to outline our current understanding of the microscopic physics of dense matter in the interior of neutron stars, and from this standpoint to construct families of equations of state useful for interpretation of both electromagnetic and gravitational observations.

    In addition, as the only source of ``data" on cold high density matter in QCD, neutron stars provide a rich testing ground for microscopic theories of dense nuclear matter, providing an approach complementary to probing dense matter in ultrarelativistic heavy ion collision experiments at the Relativistic Heavy Ion Collider (RHIC) in Brookhaven and the Large Hadron Collider (LHC) at CERN.    A major challenge is to understand the facets of microscopic interactions that allow the existence of massive neutron stars. Discoveries in recent years of neutron stars with $M \sim 2$ solar masses ($M_\odot$),  including the binary millisecond pulsar J1614-2230, with mass $1.928\pm0.017 M_\odot$ \cite{fonseca} (the original mass measurement was $1.97\pm  0.04 M_\odot$ \cite{Demorest}), and the pulsar J0348+0432 with mass $2.01 \pm 0.04 M_\odot$ \cite{Antoniadis2013}  present a direct challenge to theoretical models of dense nuclear matter.\footnote{In addition the extreme black widow millisecond pulsars PSR J1957+20 \cite{vankerwijk}, PSR J2215+5135  \cite{Schroeder-Halperin}, and  PSR J1311-3430 \cite{Romani2012,Romani2015} possibly have masses as large as 2.5 $M_ \odot$; however the masses remain uncertain owing to  the need for more complete modeling of the heating of the companion stars by the neutron stars.}    

      The existence of such massive stars  has important implications for dense matter in QCD.  For example, they require a stiff equation of state, i.e., with large pressure for a given energy (or mass) density, and thus rule out a number of softer theoretical models, and at the same time impose severe constraints on the possible phases of dense QCD matter.  In particular, massive neutron stars are difficult (but not impossible) to explain in the context of hadronic models of neutron star matter in which the emergence of strange hadrons around twice nuclear saturation density softens the equation of state and limits the maximum stable star mass.

\begin{figure}
\hspace{-0.5cm}
\includegraphics[width = 0.46\textwidth]{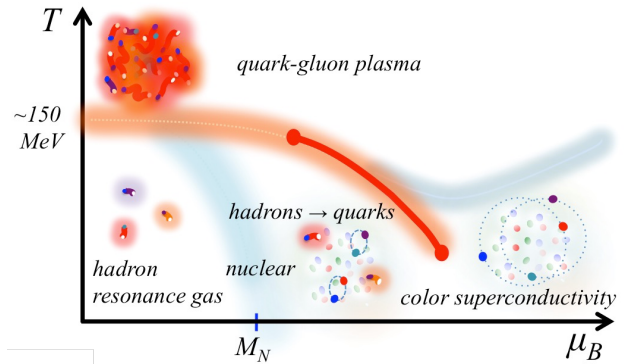}
\caption{
\footnotesize{. 
Schematic phase diagram of dense nuclear matter, in the baryon chemical potential $\mu_B$--temperature $T$ plane.   At zero temperature, nucleons are present only above $\mu_B \sim M_N$, the nucleon mass.  At the low temperatures inside neutron stars,
matter evolves from nuclear matter at low densities to a quark-gluon plasma at high density.  BCS pairing of quarks in the plasma regime leads to
the matter being a color superconductor.    (Low temperature BCS pairing states of nucleons are not shown.)  At higher temperatures, matter becomes a quark-gluon plasma, with a possible line of first order transitions, the solid line, terminating at high temperatures at the proposed Asakawa-Yazaki critical point \cite{asakawa}. In addition, the solid line may terminate in a low temperature critical point \cite{Hatsuda2006}}.   }
 \label{phases}
\end{figure}

\subsection{Phases of dense matter}

  Figure~\ref{phases} summarizes the phases of dense nuclear matter in the baryon chemical potential $\mu_B$ -- temperature $T$ plane \cite{Fukushima2011}.   [The baryon chemical potential, increasing with increasing baryon density, here nucleons, is the derivative of the free energy density with respect to the density of baryons.]  At low temperature and chemical potential the degrees of freedom are hadronic, i.e., neutrons, protons, mesons, etc.; and at high temperature or chemical potential matter is in the form of a quark-gluon plasma (QGP) in which the fundamental degrees of freedom are quarks and gluons.  The nature of the transitions from hadronic to a QGP are sketched in Figs.~\ref{fig:3-window} and \ref{fig:HRG_sQGP_QGP} below.  The temperatures in neutron stars,  characteristically much smaller than 1 MeV (or $10^{10}$ K), are well below the temperature scale in Fig.~\ref{phases}, of order 10-10$^2$ MeV; matter in neutron stars lives essentially along the chemical potential axis in this figure.  The exception is at neutron star births in supernovae where temperatures can be tens of MeV, and in final gravitational mergers where temperatures could reach $\sim 10^2$ MeV.   The main problem on which we focus in this review is the description of such matter, and the resulting models of neutron stars, i.e.,  the profiles of baryons density, etc., as a functions of radius from the center of the star.

\subsection{Neutron star models -- the TOV equation}  
      
  We briefly recall that to construct a model of a neutron star -- once one specifies the equation of state, which gives the pressure, $P$, as a function of the mass density, $\rho = \varepsilon/c^2$, where $\varepsilon$ is the total energy density and $c$ is the speed of light -- one integrates the  general relativistic  Tolman-Oppenheimer-Volkov (TOV) equation of hydrostatic balance \cite{Tolman1939,Oppenheimer1939}:
\beq
  \frac{\partial P(r)}{\partial r} &&= \\
&&   - G_N\frac{\rho(r)+P(r)/c^2}{r\left(r-2G_Nm(r)/c^2\right)}[m(r) + 4\pi r^3 P(r)/c^2],\nonumber
  \label{tov}
\eeq
where $G_N$ is Newton's gravitational constant and
\beq
m(r) = \int^r_0 4 \pi r^{\prime 2}d r^\prime \rho(r^\prime)
\eeq
is the mass inside radius $r$.  Therefore, the mass ($M$) of the neutron star is given by $M=m(R)$ with $R$ the stellar radius.  In practice, to calculate the pressure, $P(\rho)$, one first calculates the energy density, $\varepsilon$, as a function of the  baryon density, $n_B$, and then uses the thermodynamic relation $P= n_B^2\partial (\varepsilon/n_B)/\partial n_B$.   Equivalently, one can also calculate $P$ directly as a function of the baryon chemical potential, $\mu_B$. Generally, we discuss the equation of state in the form $P(\mu_B)$.  As we show in Appendix \ref{sec:scaling}, when the basic scale of the energy entering the equation of state is the proton mass, $m_p$, the typical scale of
the neutron star mass is given by $M\sim m_p/\alpha_{G}^{3/2} = 1.86 M_{\odot}$, and the scale of radii is given by $ (\hbar/m_p c)\alpha_G^{-1/2}$  = 17.2 km, where $\alpha_G = m_p^2 G_N/\hbar c\simeq 0.589 \times 10^{-38}$ is the gravitational fine structure constant. 

\subsection{Microscopic calculations of dense nucleonic matter} 

    Microscopic calculations of the equation of state of neutron-star matter, $\varepsilon(n_B)$, have been based on a variety of inputs.  The approach most firmly founded on experiment in the region of nuclear saturation density, $n_0$ $\simeq$ 0.16 nucleons per fm$^3$, or equivalently a mass density $\simeq 2.7\times 10^{14}$ g cm$^{-3}$, is to use nucleon-nucleon scattering data below 350 MeV and the properties of light nuclei to determine two-body potentials together with a three-nucleon potential~\cite{APR,mpr}.  Quantum Monte Carlo calculations based on such potentials give an excellent account of the binding energies as well as excitation energies of light nuclei \cite{lightreview}.   Quantum Monte Carlo calculations have been applied to neutron star structure, and the radius in particular, in Ref.~\cite{gandolfi}.
    
      Calculations of dense matter, which have primarily been carried out in the limits of pure neutron matter and symmetric nuclear matter, have uncertainties owing both to the interactions used,  especially at densities above $n_0$, and the calculational methods employed \cite{Gandolfi:2013baa,HebelerBogner}.  As discussed below, extensions to neutron star matter in beta equilibrium have been based on interpolation between these two limits, which introduces further uncertainties for matter in neutron stars.  

      Not only are the explicit three-body interactions in nuclear matter not well determined \cite{gandolfi} [see Ref.~\cite{Gandolfi:2013baa} for a detailed analysis of the significant uncertainties in pure neutron matter introduced by three body forces], one must ask at higher density when higher body interactions, e.g., four body, become important.  Most naively one can argue that the relative importance of higher body forces is determined dimensionally, since the ratio of the energy density $E_{n+1}$ from $n+1$ body forces to that from $n$ body forces will be $\sim (4\pi r_0^3/3)n_B $, where $r_0$ is a characteristic length of order the range of the nuclear force,  a hard core radius, $\sim$ 0.5 fm or the range of two pion exchange, $\sim (2m_{\pi})^{-1} $ = 0.7 fm.    
Then the measure of importance of the next order forces becomes the parameter $(4\pi r_0^3 /3)n_B \sim$ (0.1-0.2)$n_B/n_0$.   An alternative way to estimate the importance of higher body forces would be via chiral perturbation theory.   Comparison of four-body interaction energies, based on a subset of possible processes
\cite{kaiser2}, with three-body interaction energies \cite{kaiser1}, suggests prima facie that the relative importance of four body compared with three is measured by the parameter $(g_A/f_\pi)^2(n_B/\Delta) \sim 0.9 n_B/n_0$, times numerical factors of order unity, where  $g_A \simeq$ 1.25 is the pion axial vector-renormalization constant, $f_\pi \simeq$ 93 MeV is the pion decay constant, and $\Delta \simeq$ 293 MeV is the mass difference of the excited nucleonic state $\Delta(1232)$ and the nucleon.  Although a more accurate estimate from chiral perturbation theory remains an open problem, these estimates indicate that at the densities achieved in neutron star interiors, $n_B \gtrsim (3-6) n_0$, a well defined expansion in terms of two-, three-, or more, body forces may not exist.   Furthermore, beyond baryon densities a few times $n_0$ the forces between particles should no longer be describable by static few-body potentials.   At the same time, however, the density remains much too low to treat the matter as weakly interacting quark matter; indeed perturbative QCD (pQCD) begins to become applicable for baryon chemical potentials $\mu_B \gtrsim $ (3-6) GeV, corresponding to baryon densities $n_B \gtrsim $(10-100) $n_0$. 
       
     In addition to the limitations inherent in any few-body potential model, equations of state based on nucleons alone fail to account for the rich variety of hadronic degrees of freedom that enter with increasing density, including $\Delta$'s, strangeness~\cite{Bodmer1971,Witten1984,Farhi1984,Baldo2000,Nishizaki2001,Nishizaki2002,Schulze2011}, and meson condensates \cite{BECnstars} including pionic ~\cite{migdalpi,barshay,sawyer,GBpi,APR,AM} and kaonic~\cite{kaplan,PPT,BB,GS}.  The presence of hyperons in neutron stars is difficult to predict, owing to the uncertainty of the forces between hyperons and nucleons as well as between other hyperons.   While elementary hadronic models indicate the emergence of hyperons at densities $\sim$ (2-4) $n_0$,  mixing of strange degrees of freedom at such low densities softens the equation of state; the presence of hyperons is not obviously compatible with the stiff equation of state required by the existence of $\,\gtrsim 2 M_\odot$ stars \cite{wambach-hyp}. Attempts have been made to avoid the softening of the equation of state  by introducing repulsive  forces between strange hadrons and nucleons and between hyperons as well as 
      $\Lambda$-nucleon-nucleon forces, which shift the emergence of strangeness to higher densities~\cite{Nishizaki2001,Nishizaki2002,Weissenborn2011,Weissenborn2012,lonardoni}.   Lattice gauge theory --     
solving QCD on a space-time lattice using Monte Carlo techniques \cite{Ukawa:2015eka} -- will  in the future be able to provide first principles information on hyperon-nucleon and hyperon-hyperon interaction potentials \cite{halqcd}.

       \begin{figure}[t]
\includegraphics[width = 0.48\textwidth]{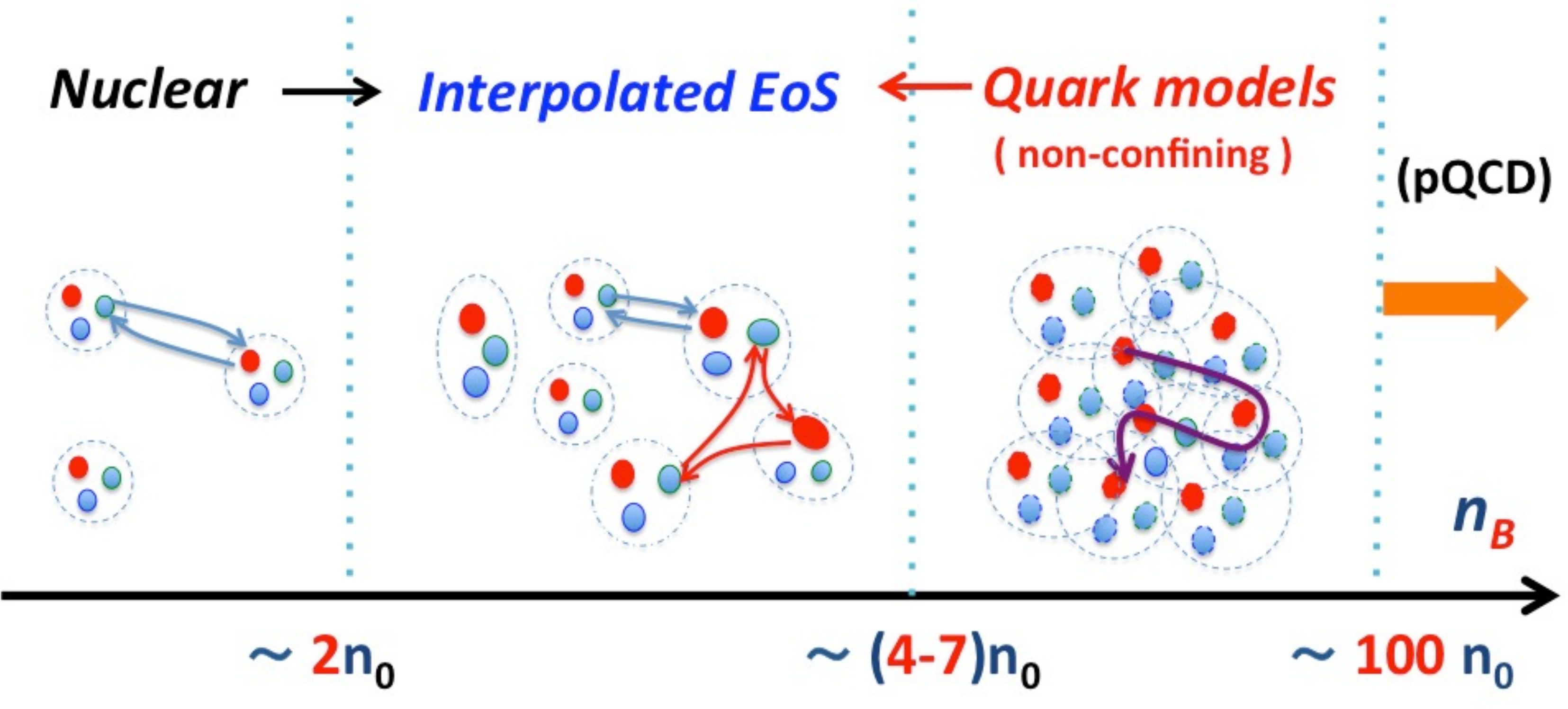}
\caption{\footnotesize{Schematic picture of the transition from nuclear to deconfined quark matter with increasing density.  i) For $n_B \lesssim 2n_0$, the dominant interactions occur via a few ($\sim$1-2) meson or quark exchanges, and description of the matter in terms of interacting nucleons is valid; ii) for $2n_0 \lesssim n_B \lesssim$ (4-7) $n_0$, many-quark exchanges dominate and the system gradually changes from hadronic to quark matter (the range (4-7) $n_0$ is based on geometric percolation theory -- see 
Sec.~\ref{perc}); and  iii) for $n_B \gtrsim$ (4-7) $n_0$, the matter is percolated and quarks no longer belong to specific baryons.  A perturbative QCD description is valid only for  $n_B \gtrsim$ 10-100 $n_0$. 
}  }
\label{fig:3-window}
\end{figure}

       \begin{figure}[t]
\includegraphics[width = 0.48\textwidth]{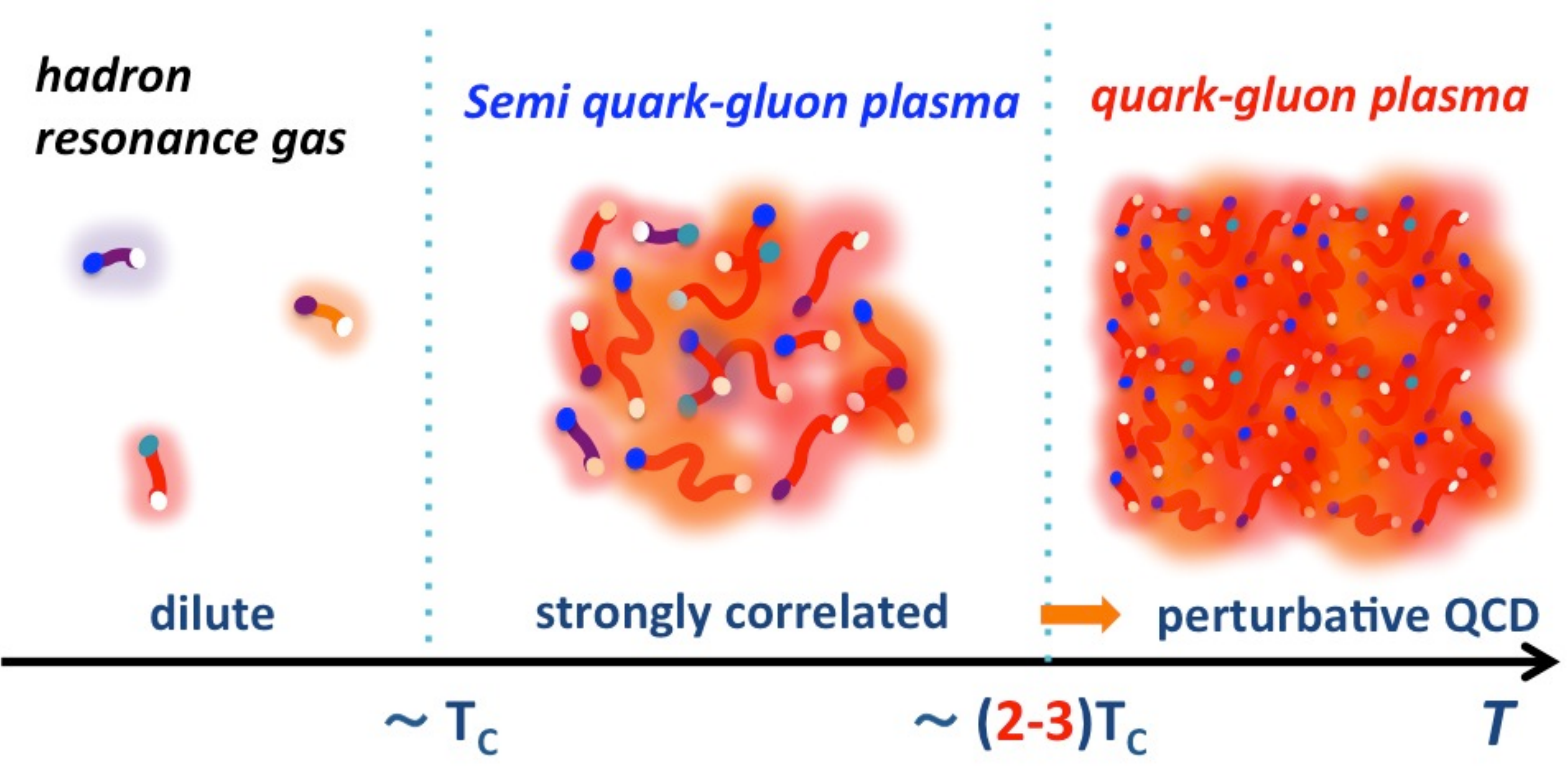}
\caption{\footnotesize{Schematic picture of the crossover transition from the hadronic to quark-gluon plasma phase with increasing temperature.  i) For $T \lesssim T_c$, the system is a dilute gas of hadrons; ii) for $T_c \lesssim T \lesssim$ (2-3)  $T_c$, thermally excited hadrons overlap and begin to form a semi quark-gluon plasma (see text below); and iii) for $T \gtrsim$ (2-3) $T_c$, the matter is percolated and a quasiparticle description of quarks and gluons, including effects of thermal media, becomes valid.}  }  
\label{fig:HRG_sQGP_QGP}
\end{figure}

    Another approach to high density matter has been based on nucleons interacting via elementary meson exchanges  (e.g.,~\cite{dirk,MS}; see~\cite{radius} for a general summary of equations of state).  In such models, multiple meson exchanges and virtual baryonic excitations, which comprise the intermediate states in such theories, again raise the question of whether well-defined ``asymptotic'' laboratory hadrons are the proper degrees of freedom to describe the system at high density.  
    
\subsection{Quark degrees of freedom}    
    
    More realistically, one expects in dense matter a gradual onset of quark degrees of freedom, not accounted for by nucleons interacting via static potentials.  Indeed, as illustrated in Fig.~\ref{fig:3-window}, at a sufficiently high density the matter should percolate, in the sense that their quark constituents are able to propagate throughout the system~\cite{Baym1979,Satz1,Satz2}.     The deconfinement of nuclear matter with increasing density has many similarities to the manner in which atomic gases, when compressed, become gases of itinerant electrons in a background of ions.  Strong nuclear and electromagnetic interactions drive dense matter toward both local color and electrical neutrality.   At low densities, this results in strong correlations between particles, with correlations weakening with increased density.   In the hadronic regime, three quarks bind together to produce a color singlet object.   In the regime between hadronic and quark matter, colored quarks and diquarks appear virtually during quark exchanges between baryons -- essentially the baryon-baryon interactions.   In the quark matter regime a diquark or a pair of quarks can easily find an extra quark nearby to produce local color neutrality, so that the extra quark is weakly correlated with the diquark or pair, as shown in Fig.~\ref{fig:3-window}.

    With increasing baryon density or temperature, the effective degrees of freedom of matter change, possibly accompanied by phase transitions.  The phases of QCD are characterized by a variety of condensates in which a macroscopic number of particles (and antiparticles) are strongly correlated by the strong interaction \cite{BECnstars}. The emergence of condensates reduces the energy of the system, and in addition, condensates break symmetries in QCD, leading to states with lower symmetry than is present in the QCD Hamiltonian.  The condensates, which depend on temperature and baryon density, play an important role in the structure of hadrons, as well as in neutron stars, since the condensation energies are a large fraction of the energy density in a neutron star core.  
    
      Chiral symmetry breaking, caused by the chiral condensation of paired quarks and antiquarks with different chirality (or handedness) -- characterized most simply by a non-vanishing chiral condensate $\langle \bar q q \rangle$, where $q$ is the quark field -- is largely responsible for hadron masses and the existence of the nearly massless Nambu-Goldstone  bosons \cite{Nambu:1961tp}, e.g., the pion as well as the kaon.  As reviewed in \cite{Holt:2014hma}, chiral condensation persists from the vacuum, to nuclear matter \cite{friman}, to high density quark matter.   At high baryon density the condensate $\langle \bar q q \rangle$ is expected to go to zero; however, owing to the formation of further condensates, e.g., by diquark-anti diquark pairs (see below), chiral symmetry is expected to continue to be broken at high densities \cite{dtsonReversal,fukushimaConstruct,Yamamoto,song}.  
          
      In addition, QCD color-magnetic interactions favor the formation of a {\em diquark condensate} of quark pairs in quark matter at low temperatures, 
characterized by a non-vanishing expectation value $\langle q q\rangle$, and         
similar to the Bardeen-Cooper-Schrieffer (BCS) condensate of electron pairs in a superconductor \cite{Alford:2007xm}.  Such a condensate breaks the U(1) symmetry associated with baryon conservation, and leads to low temperature quark matter being a color superconductor.

           Finally, the breaking of scale symmetry in QCD is related to formation of a gluon condensate -- characterized by a non-vanishing expectation value $\langle  F^{\mu\nu}_aF_{\mu\nu}^a\rangle$, where  $F_{\mu\nu}^a$ (with $\mu,\nu$ the space-time and $a$ the color indices) is the gluon field tensor \cite{Shifman:1978bx}.  

      The possible astrophysical role of quarks in stars, and neutron stars in particular, has been discussed ever  since the first proposal of the quark model of hadrons \cite{ivanenko,itoh,chin}.  Because of the difficulty in describing hadrons and quarks within a single framework,  the conventional description of the onset of quark matter has been to regard (hadronic) nuclear and quark matter as distinct phases, to calculate their energy densities using very different models, and then choose the phase with the lower energy density at given baryon density.  In order to guarantee pressure and chemical potential continuity across the transition, one must make a bitangent Maxwell construction, which leads to a first order phase transition from nuclear to quark matter, e.g., \cite{chin,Buballa2005,Benic:2014jia,Alvarez-Castillo:2016wqj}.   Stars fabricated with such a ``hybrid" equation of state --  {\em hybrid stars} -- consist typically of a small quark matter core surrounded by hadronic matter.   Reference \cite{Ranea-Sandoval:2015ldr} presents an illuminating analysis of why the quark cores are typically small.   The conventional picture is based on a thermodynamic comparison of hadronic matter and quark matter at too high a density for the description of hadronic matter to be physical, and at too low a density to apply perturbative QCD to the quark phase.   An alternative description is that with increasing density, neutron star matter undergoes an essentially continuous transformation from the hadronic to quark regimes, a scenario we refer to as {\em hadron-quark continuity}~\cite{Schafer:1998ef,Hatsuda2006,Yamamoto,Bratovic2012,Lourenco2012,Masuda2013-0,Masuda2013,Klahn2013,Masuda:2015kha}.  In this scenario, as the density increases, quark degrees of freedom gradually emerge, with partial restoration of chiral symmetry and the onset of color superconductivity.   
         
\begin{figure}[t]
\includegraphics[width = 0.48\textwidth]{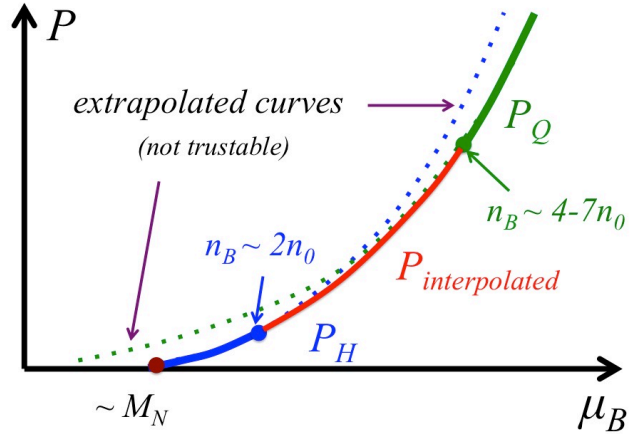}
\caption{\footnotesize{A unified equation of state using a nuclear equation of state for $n_B \lesssim 2n_0$ and a quark matter equation of state for $n_B \gtrsim (4-7)n_0$,  interpolated in the intermediate region.  The dotted curves, the extrapolations of the nuclear and quark matter equations of state, indicate how such extrapolations become unreliable.  }  }
\label{fig:intro_3window}
\end{figure}

   In the picture of QCD exhibiting a continuous evolution from hadronic to quark matter at low temperature neutron star cores are composed of deconfined $u$, $d$, and $s$ quarks.  Strangeness, rather than appearing in matter as hyperons, whose interactions are poorly known, it appears as strange quarks.   Importantly, equations of state exhibiting hadron-quark continuity are consistent with current observational inferences of neutron star radii, as well as neutron star masses $\sim 2 M_\odot$.  The emerging description of dense matter, on which we focus in this review, is that at low densities matter is hadronic, and at high densities is quark matter;  in the intermediate regime, where one does not at present have tools to calculate, one can make physically constrained plausible interpolations between these two limits, arriving at what we shall refer to as a {\em unified equation of state}\footnote{Our usage of this term should not be confused with its prior use  in the literature, e.g.,  \cite{potekhin}, to describe equations of state arising from a consistent physical model in the crust and liquid interior, e.g., in Refs.~\cite{bbp,sly,Togashi:2017mjp}. } does use a   
 to describe matter across the entire range of densities found in the interior of neutron stars \cite{Fraga:2013qra,Kurkela:2014vha,Kojo2014,Kojo:2015fua,qm2015-tk,Whittenbury:2015ziz,Whittenbury:2013wma}.   Figure~\ref{fig:intro_3window} illustrates the construction of a unified equation of state.
  
   The QCD phase transition at finite temperature and low baryon density \cite{Polyakov,Susskind:1979up,Banks:1979fi,Svetitsky}, illustrated in Fig.~\ref{fig:HRG_sQGP_QGP}, is an example of the continuity from hadrons to quarks and gluons.  The smoothness of the evolution of the matter at the transition temperature $T_c$ has been established by lattice Monte Carlo calculations \cite{Aoki:2006we,bazarov-2012,hotQCD,WB}.   At low temperature, space is filled with a dilute gas of hadrons; as the temperature increases this hadronic gas gradually fills the space, until near $T_c$ the hadrons begin to merge, continuously transforming the matter structure and leading to a breakdown of its description purely in terms of hadrons.  Yet, a quasiparticle description based on quarks and gluons does not apply either, for while the hadrons have broken down, the strong correlations remain.   Thus, the system may be described as a strongly correlated quark-gluon plasma \cite{DeTar:1985kx,Hatsuda1985}
or a ``semi-QGP" \cite{Pisarski:2000eq,Dumitru:2010mj,Pisarski:2016ixt} 
in which both hadronic-like and quark-like degrees of freedom can exist.  Indeed perturbative QCD (pQCD) calculations indicate that the quasiparticle picture of a QGP begins to apply only beyond (2-3) $T_c$ \cite{Blaizot:2000fc,Andersen:2011sf}.   This crossover behavior of low density QCD matter has a number of features that are likely to be found in the hadron-quark transition at low temperature. However, it is important to note that while at zero baryon chemical potential physical gluons can be thermally excited as one approaches $T_c$, in cold dense matter gluons appear only virtually, through quantum fluctuations. The properties of gluons at low temperatures, which are much less well known than those at high temperature, is an important subject on which light may be shed through studies of neutron stars \cite{Kojo2014,Kojo:2015fua,qm2015-tk,kenjitoru}.
 
  Calculations of neutron star masses and radii require integration over vastly different density scales, from the crust region to the core.  The outer regions of a neutron star, which are better understood, consist of a solid crust and a nuclear matter liquid just inside the crust, spanning baryon densities up to $\sim$ 2$n_0$.  The neutron star interior, for which densities are $n_B  \sim (2-10) n_0$ is the most difficult to describe from first principles. In this regime quarks and gluons begin to play an important role, but the density is not high enough to apply results from pQCD.  Thus, one is required to introduce some degree of phenomenological modeling.    Understanding the role of confinement and percolation in this regime is a critical and outstanding problem, which renders practical modeling for $n_B \sim$ (2-10)$n_0$ difficult without introducing assumptions on confining forces and percolation.  At higher densities, where matter is a well defined quark-gluon plasma, one faces the difficulty that lattice gauge theory for QCD is at present unable to describe matter at the relevant baryon chemical potentials \cite{Ukawa:2015eka}.  Thus one must introduce models of interacting quark matter in this regime.  Only at much higher densities, beyond those expected in neutron stars, can one apply perturbative QCD for dense matter directly.
  
\subsection{Modelling hadronic and quark matter}  
   
   To illustrate the evolving physics with increasing density, we will generally work with a low density nuclear equation of state at $n_B \lesssim 2n_0$  and a high density quark matter equation of state at $n_B \gtrsim (4-7)n_0$.  In actual calculations we take the specific value $5n_0$. 
 For concreteness, we adopt the Akmal-Pandharipande-Ravenhall (APR) equation of state \cite{APR} for nucleons as a representative nuclear equation of state, while noting that its extremely stiff character at densities well above saturation density provides a rough upper limit for neutron star masses in a purely hadronic description.  
Since cold dense matter in QCD cannot be calculated directly in lattice gauge theory \cite{Ukawa:2015eka},
one must resort to phenomenological models of QCD to describing interacting quarks in neutron stars.     For a quark model at high density as a template, we adopt  the three quark-flavor Nambu-Jona-Lasinio (NJL) model \cite{Vogl:1991qt,Rehberg,Lutz1992,Klevansky:1992qe,Hatsuda1994}, which captures much of the physics needed in neutron stars, and which  has been employed previously in studies of dense QCD matter \cite{Powell,Powell2,Fukushima2011,Abuki,Masuda2013,Masuda2013-0,Takatsuka,Holt:2014hma}.   The fundamental actors in the model are quarks; while the model does not manifestly take gluon degrees of freedom into account, their effects in cold matter are to a large degree modeled by letting the quarks interact via a number of effective interactions reflecting non-perturbative QCD processes involving quarks and gluons at low energy.  We summarize in subsec.~\ref{NJL_int_quarks}  the effective interactions relevant for QCD phenomenology in neutron stars. 
    
   Although we do not review models of the intermediate density nuclear matter lying between the low density hadronic and high density quark regimes, the equation of state in this confinement-dominated regime is, as we shall see, strongly constrained.
At higher densities, where baryons merge with one another, a detailed description of confining forces becomes unnecessary, and the remaining interactions can be inferred from hadronic and nuclear phenomenology, by applying the hadron-quark continuity picture in which the structure and model parameters are adiabatically connected to those in the vacuum.   The requirement that the neutron star equation of state be very stiff, i.e., has a sufficiently large pressure for given energy density -- to allow $2M_\odot$  neutron stars -- tightly constrains the range of effective parameters of these models.   In this way neutron star constraints translate into constraints on model parameters, with possible density dependence, for delineating properties of low energy QCD matter \cite{Whittenbury:2013wma,Klahn2013,Hell:2014xva,Whittenbury:2015ziz,Kojo:2015fua,Masuda:2015kha,Alford:2015gna}.

\vspace{24pt}

\subsection{Determining the mass-radius relation observationally} 
\label{MR} 

     We briefly discuss three current observational developments --  determinations of the neutron star mass-radius relation; binary neutron star mergers, as seen both in gravitational and electromagnetic radiation; and measurements of the thermal properties of neutron stars -- developments which are rapidly leading to a greater understanding of the properties of neutron star interiors.

     Accurate determination of the mass-radius relation of neutron stars strongly constrains the equation of state of neutron star matter.   Beyond simple estimates of radius based on spectroscopic analyses of photons from neutron stars in quiescence (see \cite{timing} and references therein), further information on the relation can be gained from following the evolution of thermonuclear X-ray  bursts on the surfaces of accreting neutron stars in binaries \cite{timing,OzelFreire,fredcole}.    In addition the NICER experiment \cite{nicer,michi,Miller:2016kae,ozel-nicer,Bogdanov_2008,timing} is currently directly measuring the mass and radii of a number of neutron stars.
          
   The photon flux, $F_\infty$ per unit area, from a neutron star at distance $D$, is related to the surface flux, $F_s$, of the star of radius $R$ by $  4\pi D^2 F_\infty = 4\pi R^2F_s/(1+z)^2$,
where $1+z =1/(1-2G_NM/Rc^2)^{1/2}$ is the redshift.  For black-body emission from the surface at an apparent temperature $T_{\rm eff}$, $F_s = \sigma T_{\rm eff}^4$, where $\sigma$ is the Stefan-Boltzmann constant.    In reality,  $T_{\rm eff}$ is not the true surface color temperature,  $T_{\rm color}$,  deduced from thermal fits of the spectra, since it depends on the physics in the atmosphere, e.g., its composition and the surface gravity;  the two temperatures are phenomenologically related by $T_{\rm color} = f_c T_{\rm eff}$ where $f_c \sim 1.3-2$.    The apparent angular size of the neutron star seen by the observer is \cite{fozel},
\beq
 A_{\rm app} \equiv  \frac{F_\infty}{\sigma T_{\rm color,\infty}^4} =  \frac{R_\infty^2}{D^2f_c^4},
 \label{anglesize}
\eeq
in terms of the color temperature seen at the point of observation,   $T_{\rm color,\infty}=  T_{\rm color}/(1+z) $.  The effective radius determining the angular size is $R_\infty = R(1+z)$, which can be understood in terms of light bending by the neutron star \cite{pod2000}.   From observations of $F_{\infty}$, $D$, and $T_{ {\rm color,\infty} }$ one can constrain the relation between $M$ and $R$.  The distance $D$ typically has large uncertainties, $\sim 50\,\%$, while the uncertainty is the level of $5-10\,\%$ for neutron stars in globular clusters. 
 
    There are two important caveats in this procedure.  The first is that neutron stars in strong magnetic fields $\gtrsim 10^{12}$ G are not black-body emitters; atoms at the surface are expected to be highly distorted by the magnetic field, making the surface radiation dependent on its polarization \cite{lai,feryalmag,smitha}.  The second is the assumption that the radiation is from the neutron star surface uniformly,  not from a hot spot on the surface, nor from further out in the atmosphere.
 
   Observations of thermonuclear X-ray bursts on neutron star surfaces (see \cite{OzelFreire,Ozel2015,Steiner2015,Steiner2016,fozel,timing,fredcole}, and references therein) further constrain the mass-radius relation.   In bursts in which the radiation flux exceeds the local Eddington limit (where the radiative force on the atmosphere balances the gravitational force)  the photosphere of the neutron star is lifted;  when it returns to the neutron star surface, the flux at the ``touchdown" point, seen from infinity is,
\beq
F_{\rm Edd,\infty} = \frac{G_NMc \,}{(1+z)\kappa_e D^2},
\eeq
where $\kappa_e$ is the electron-scattering opacity in the atmosphere.   
Combining measurements of the touchdown flux, which is independent of $T_{{\rm eff}}$,  with information from the quiescent flux $F_{\infty}$ leads to useful constraints on $M$ vs. $R$.

 \begin{figure}[t]
\vspace{-0.cm}
\includegraphics[width = 0.5\textwidth]{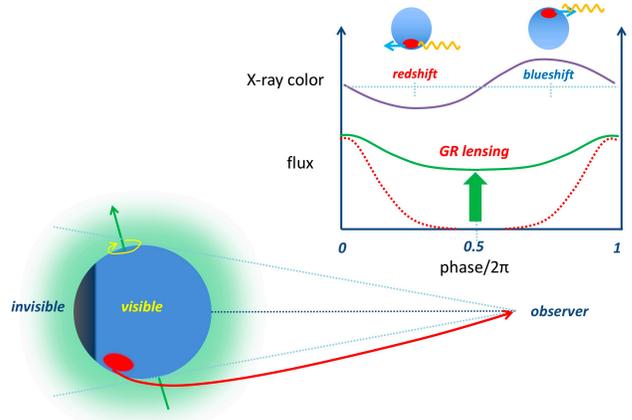}
\caption{
\footnotesize{
Gravitational lensing, which is stronger for neutron stars with larger compactness $M/R$, permits one to see part of the back side of a neutron star.  The inset shows the X-ray flux from a single (red) hot spot on a rotating neutron star, the green line, as a function of the phase of the rotation.  In the absence of gravitational lensing, the flux from a point hot spot, shown as the dotted red line, would vanish for part of a rotational cycle; lensing shrinks the invisible region, allowing the spot to be seen for a larger fraction of the phase, and reduces the contrast between the flux from the brightest and darkest regions.  
Combining general relativistic effects, the velocity deduced from Doppler shifts of the spot color, 
and the rotational frequency allows one to constrain the neutron star mass and radius. 
}}
\label{fig:GR_lensing}
\end{figure}

     The ongoing NICER experiment \cite{nicer} aims to determine the masses and radii of several nearby rotating neutron stars by accurately monitoring their X-ray pulse profiles in time.  Neutron stars with strong magnetic fields have hot spots at the magnetic polar caps, heated by charged particles moving along the magnetic flux lines.   The pulse profiles are periodically modulated by brightness changes associated with these temperature non-uniformities on the neutron star surface.    The key is that gravitational bending of light by the strong gravitational field of neutron stars, dependent on $M/R$, allows one to partially ``see" the back side of the neutron stars.  Figure~\ref{fig:GR_lensing} schematically illustrates the basic mechanism for a single hot spot.   With stronger bending, more of the star is visible, and thus the measurable ratio of the amplitudes between the brightest and darkest points in the profile decreases.   The bending for slowly non-rotating stars, described by the Schwarzschild metric, depends only on $M/R$, and thus to get information on 
$M$ and $R$ separately requires looking at sufficiently rapidly rotating stars, above $\sim$300 Hz in practice, in which corrections from the Kerr metric of rotating neutron stars come into play.  The pulse profiles reflect both general relativistic and Doppler effects;  comparison with waveform modeling, including as a function of color, allows one to extract $M$ and $R$, as well as other parameters such as the quadrupole moment and moment of inertia, which play a role in determining the metric of a strongly gravitating rotating star \cite{ozel-nicer,Bogdanov_2008}.  Eventually NICER should determine neutron star radii and masses to an accuracy $\sim$ 5-10\%.

\subsection{Binary neutron star mergers} 
\label{GW} 

The recent multi-messenger detection of the merger of two neutron stars GW170817 \cite{GW170817,GW170817A} opens a new path to determining neutron star properties both by comparing the measurable gravitational signals with those computed by fully general relativistic simulations with given QCD equations of state \cite{Hotokezaka:2011dh,takami,baiotti,sekiguchi,stu1,stu2}, as well as utilizing the electromagnetic signals accompanying the merger \cite{shapiro17A,m17,mm17,stu17B,rezzolla17,shibata17,janka17,radice17}.
A binary neutron star system at large distance is described by Newtonian mechanics, but as the stars approach each other in the late inspiral phase the stars deform each other, prior to merging in tens of seconds.   The tidal deformability of the stars, discussed below, is very sensitive to their compactness.   When the orbital frequency reaches $\sim 500$ Hz (the current limit of LIGO detectability), the detectable gravitational signals begin to distinguish between different equations of state.   The gravitational waveform just before the merger, especially its frequency, is again strongly dependent on the compactness of the stars; smaller stars can approach more closely before crushing into each other, thus reaching higher orbital frequency.

    Neutron star--neutron star mergers explore the possible existence of very heavy neutron stars and light black holes. 
After the merger, the coalesced star can either collapse into a black hole, or, if the star is not too massive ($\lesssim 3\, M_\odot$) and spinning rapidly, it could remain as a metastable hypermassive star, with high differential rotation and temperature $\sim 10-100$ MeV.  The final object produced by the merger GW170817 has a mass of order 2.7 $M_\odot$, which exceeds the cold non-rotating neutron star mass bound proposed in \cite{lawrence2015,rezzolla17}.   The observed jets favor the object being a black hole \cite{shapiro17A,m17,mm17,stu17B},
since jet formation takes place through conversion of rotational energy into magnetic \cite{stu1,stu2}; others argue for it to be a long-lived massive neutron star \cite{shibata17}.   The event has also been used to place constraints on the radii of the initial neutron stars \cite{janka17}, as well as to suggest a lower bound on the tidal deformability of the stars \cite{radice17}.    

    As two neutron stars approach each other in a merger each tidally deforms the other.  For two stars, A and B of mass $M_{\rm A}$ and $M_{\rm B}$, whose centers are separated by a large distance $\vec R_0$
     (with $|\vec R_0| \gg G_NM_{\rm A}/c^2, G_NM_{\rm B}/c^2$), 
 the quadrupolar tidal gravitational field from star B felt at position $\vec r$ in star A (measured from the center of A) is 
\beq
   \Phi_{\rm tidal}(\vec r\,)  = \frac{G_N M_{\rm B}}{2R_0^3}\left(r^2 - 3(\vec r\cdot \hat R_0\,)^2\right) = \frac12{\cal E}_{ij}r^ir^j, \nonumber\\
   \label{tide1}
\eeq
where ${\cal E}_{ij} =  \partial^2 \Phi_{\rm tidal}/\partial r_i \partial r_j$ is the quadrupolar tidal field tensor.
Such a tidal potential distorts star A, producing a quadrupole moment,
\beq
    Q_{ij} = \int d^3r \ \rho(r)\left(r_ir_j - \frac13 \delta_{ij}r^2\right),
 \eeq 
related to the tidal field in linear order by
\beq
 Q_{ij} = - \lambda_{\rm A} {\cal E}_{ij}, 
    \label{tide2}
\eeq
 thus defining (relativistically) the tidal deformability $\lambda_{\rm A}$  of star~A.  Explicitly
\beq
    Q_{zz} = -2Q_{xx}=-2Q_{yy}\equiv -\lambda_{\rm A}{\cal E}_{zz} =2\lambda_{\rm A} \frac{G_N M_{\rm B}}{R_0^3}; \nonumber\\
   \label{tide3}
\eeq
dimensionally $\lambda_{\rm A}$ is of order\footnote{A simple, qualitative and instructive estimate of $\lambda$ is that for a non-relativistic self-gravitating spherical star of uniform density with mass $M_{\rm A}$ and unperturbed radius $R$.   A prolate deformation of the surface of the star by $\delta R = P_2(\cos\theta)\epsilon R$, produces a quadrupole moment, $Q_{zz} = (2/5)M_{\rm A}R^2 \epsilon$, and increases the gravitational energy of the sphere by $\Delta E_{\rm def} = (3/25)G_NM_{\rm A}^2\epsilon^2/R$.   In addition the energy of the star in the external tidal field (\ref{tide1}) is $\Delta E_{\rm tidal} = - (3/5)G_N M_{\rm A} M_{\rm B} R^2\epsilon/R_0^3$.   Minimization of $\Delta E = \Delta E_{\rm def} + \Delta E_{\rm tidal}$ with respect to $\epsilon$ yields
$\epsilon = (5/2)(M_{\rm B}/M_{\rm A})(R/R_0)^3$, so that $Q_{zz} = M_{\rm B} R^5/R_0^3$, and thus $\lambda_{\rm A}  =  R^5/2G_N$,  $k_2=3/4$ \cite{poisson}, and $\Lambda_A = 16(R/R_s)^5$.  Cf. \cite{kip} for a discussion of tidal deformation in terms of energetics.   Fully relativistic calculations for stars with realistic equations of state yield
much smaller $\Lambda$ than predicted by this schematic calculation, since more mass lives at smaller radii than in a uniform density sphere.}  $R^5/G_N$.    The tidal deformability can be written in terms of the dimensionless Love number, $k_2$, as
\footnote{The
conventional geophysical definition of  $k_2$ is a factor of 2 larger \cite{monk} than that here.}
\beq
    \lambda =  \frac{2}{3} k_2 \frac{R^5}{G_N}.
\eeq

    Gravitational radiation waveforms in neutron star mergers depend on the dimensionless tidal deformability, $\Lambda$, defined by
\beq
      \Lambda = 32\frac{\lambda G_N}{R_s^5} =  \frac{2}{3} k_2 \left( \frac{ R c^2}{M G_N} \right)^5 \,.     
      \label{eq:dimlessLambda}
\eeq
where $R_s = 2MG_N/c^2$ is the Schwarzschild radius of the star; see  Sec.~\ref{sec:constraints}.  The strong dependence of $\Lambda \sim (R/M)^5$ indicates that (for fixed $k_2$) the dimensionless tidal deformability is a good measure of the neutron star compactness, $M/R$.   For an introduction to calculations of $\lambda$ for general relativistic stars see \cite{Hinderer2008}, and for the effects of tidal deformations on gravitational merger waveforms see \cite{Hinderer2008a,Hinderer:2009ca,read2013} and references therein. 

\subsection{Neutrino cooling and transport} 

     Understanding transport and neutrino cooling properties of cold dense matter is essential to determining the phase structure and low energy degrees of freedom in neutron stars \cite{tsuruta,yakovlev,Potekhin:2015qsa}.   The interiors of stars older than a few hundred years are nearly isothermal;  however the surface temperatures depend on the thermal transport from the interior through the crust to the surface (see  \cite{yakovlev}).    When external heating of a star, e.g., by accretion from a companion, is not large, the long-time evolution of neutron star temperatures is dictated by the neutrino luminosity for the first $\sim 10^5 - 10^6$ years, after which photon luminosity from the surface dominates.  

    Neutrino cooling by hadronic matter depends strongly on the fraction of nucleons that are protons.  If the proton fraction is less than some 10-15\%, the proton and electron Fermi momenta are too small for the direct process of neutrino emission, $n \to p+ e^- + \bar{\nu_e}$,  $p \to  n+ e^+ + \nu_e$, to conserve both energy and momentum.  Neutrino emission takes place rather with transfer of momentum to a bystander nucleon, via the  modified URCA process, $n+n \rightarrow n+p+ e^- + \bar{\nu}_e$, and $p+n \rightarrow n+n+ e^+ + {\nu}_e$ --  
leading to the standard, or minimal cooling scenario  \cite{Pethick92,yls,Page:2013hxa}.    
For larger proton fraction, and hence Fermi momentum, the direct URCA processes, $n \rightarrow p+e^- + \bar{\nu}_e$, $p+e^- \rightarrow n+{\nu}_e$, are kinematically allowed in cooling, leading to a rapid or enhanced cooling scenario.  Owing to the extra nucleons in the modified URCA process the cooling rate is suppressed by a factor $\sim (T/T_f)^2$ compared with the direct URCA process, typically some five to six orders of magnitude, where $T_f$ is the Fermi temperature of the nucleons.   

    BCS pairing of nucleons has important physical effects on the cooling of stars.   By suppressing the density of states, the rates of neutrino emissions are reduced by a factor $\sim \exp(-2\Delta/T)$, with $\Delta$ the pairing gap.    With pairing, processes involving pair formation or breaking, such as $n+n \rightarrow [nn] + \nu + \bar{\nu}$, also lead to neutrino emission, where the initial $n$ are excited quasiparticles, and $[nn]$ indicates a condensate pair; these can become important in the regime where pairing suppresses the modified URCA process.    On the other hand,  the heat capacity of paired matter is similarly suppressed.  The net effect of pairing on the cooling of neutron stars depends in detail on models of the equation of state of matter and pairing, as well as on their masses \cite{yakovlev,yls}.
    
   ``Exotic" matter, e.g., pion and kaon condensates, hyperon mixing as well as quark matter, can lead to enhanced cooling via direct URCA processes.  Owing to the theoretical uncertainty of the possible phases inside neutron stars and the uncertainties of the age and surface temperatures of observed neutron stars, it is not yet possible at the moment to conclude whether fast cooling by direct URCA processes as well as by exotic components is taking place.  
In any case, the cooling of neutron stars needs to be investigated using equations of state compatible 
with observed $\sim 2 M_\odot$ neutron stars.
   
   Accretion of matter onto neutron stars leads to thermonuclear bursts on the surface \cite{timing,fozel,fredcole}; observation of the cooling of accretion-heated neutron stars, from a few days to a few tens of years after outbursts, is a innovative way to constrain neutrino luminosity and probe the outer regimes of neutron stars \cite{Wijnands:2017jsc}.   The basic idea is that, following the injection of heat, the neutron star interior acts as a calorimeter.  Using this approach, Cumming et al. \cite{cumming} have deduced a lower bound to the integrated heat capacity of neutron stars from observations of accretion outbursts, from which they argue against matter below 2$n_0$ having a very low specific heat (e.g., as with a color-flavor locked quark core); this result is consistent with the present unified equations of state discussed below. 

\subsection{Outline}

  In this review we first describe the more familiar properties of neutron stars, the crust in Sec.~\ref{sec:crust}, and the liquid nuclear matter in the outer core in Sec.~\ref{sec:outer}.  Although these regions have been well studied for many years, open questions remain, as we discuss.   We then turn to describing quark matter at high density in Sec.~\ref{sec:quark_EOS}, describing effective models for the quark and gluon sectors.  In the following section \ref{sec:hybrid}  we discuss general aspects of a unified equation of state capable of connecting the low density hadronic and high density quark matter regimes, and continue in Sec.~\ref{sec:3-window} with a more detailed analysis of two separate possible unifications, one with a conventional first order hadron-quark phase transition, and another realizing hadron-quark continuity.    We then outline the relationship between the structure of the QCD phase diagram and the neutron star equation of state, discussing hadron-quark continuity.   We review explicit constructions of  unified equations of state valid at all densities for a number of quark model parameter sets, in terms of interpolating between a low density nuclear equation of state at $n_B \lesssim 2n_0$  and a high density quark matter equation of state at $n_B \gtrsim 5n_0$ (Fig.~\ref{fig:intro_3window}).  In this construction we take the crossover between the two regimes to occur around $n_B \sim (2-3) n_0$, below which the strangeness density is small and various hadronic equations of state exhibit very similar properties.  In Sec.~\ref{sec:constraints} we turn to the astrophysical consequences of the unified equations of state, and indicate connections between effective quark model parameters and neutron star mass and radius observations.   We compare several model equations of state reflecting hadron-quark continuity to existing constraints obtained from astrophysical data and find consistency with current inferences~\cite{Ozel2015,Steiner2015}.  Finally, in Sec.~\ref{sec:massive_stars} we summarize the impact of parameter variations on the maximum stable neutron star mass and discuss how observational data of massive neutron stars can provide additional constraints on microscopic model parameters.  In addition, we mention open problems, including the need to develop finite temperature equations of state for modeling gravitational waveforms in neutron star--neutron star and neutron star-- black hole mergers.    In Appendix \ref{sec:scaling} we review scaling properties of the TOV equation, and in Appendix \ref{masses} we review effects of a repulsive quark vector interaction on quark masses, chiral symmetry breaking, and pairing gaps.   Lastly in Appendix~\ref{sec:para_EoS}, we present the equations of state with hadron-quark crossover in a parametrized form -- called QHC18 -- useful for modeling neutron stars.  We generally adopt units $\hbar=1$, and in discussing QCD will also generally take $c=1$.

\section{The crust}\label{sec:crust}

  The crust plays an important role in observable phenomena in neutron stars, even though its mass and thickness are relatively small.   Such phenomena include thermal conduction, which establishes the temperature drop between the core and surface; superfluid neutron dynamics at densities above neutron drip, which have been invoked to understand pulsar glitches and quasi-periodic oscillations; and more generally the elastic properties of the crust, which play a role in oscillations of the star; see, e.g., \cite{timing,RLS,rmodes,watts,qpo} and references therein.   Although as we indicate below, the mass and especially the thickness of the crust are insensitive to the details of the equation of state, the equation of state itself, as well as dynamical phenomena in the crust, remain an important problem.    The crust can be divided into three regions:  a sequence of nuclei below the surface which become increasingly neutron rich with depth into the star; the neutron drip regime, in which the continuum neutron states in the matter are occupied; and at the highest densities in the crust, a sequence of ``pasta" nuclei, including rods and sheets, which account for essentially half the mass of the crust.   Representative calculations of the equation of state in the crust, particularly below the pasta region, include Refs.~\cite{bps,sly,HZD}, while details of the pasta phases can be found in Refs.~\cite{dgr,hsy,ohy,wk,wis,Watanabe:2004tr,Maruyama:2012bi,Iida:2013fra,Horowitz:2004yf,Maruyama:2005vb,urban}.  At a baryon density of order one half nuclear matter density, $n_0$, the matter undergoes a first order phase transition from a solid crust to liquid nuclear matter.

       \begin{figure}[t]
\includegraphics[width = 0.48\textwidth]{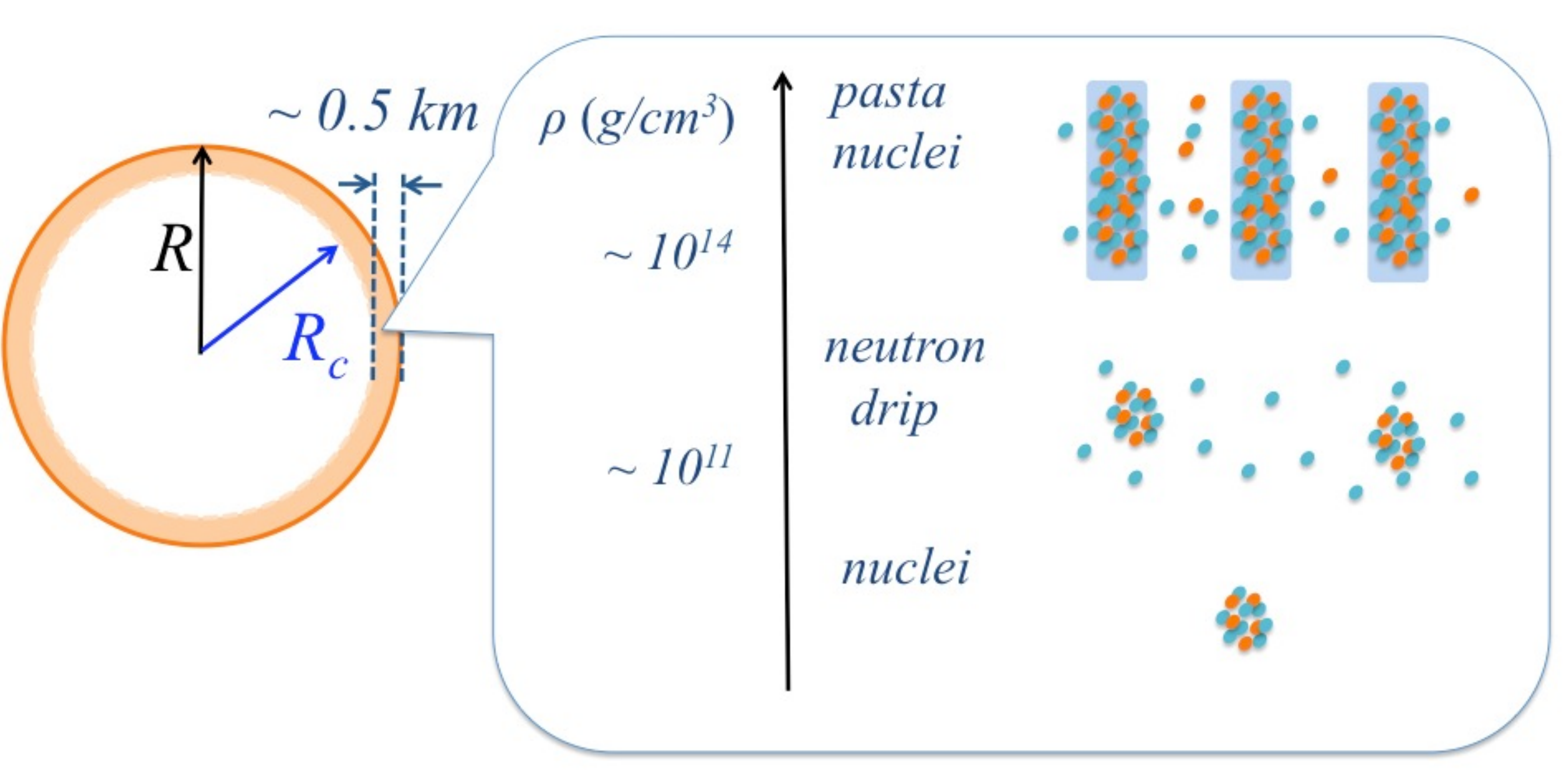}
\caption{\footnotesize{The crust of a neutron star, between the stellar radius $R$ and the radius of the liquid core $R_c$. The mass density and properties of matter change vastly in the crust region. As the density increases, the nuclei in the crust undergo neutron drip, and then become unstable and forming various ``pasta'' phases.   At higher density ($n_B\gtrsim 0.4n_0$) the matter becomes liquid nuclear matter.}  }  
\label{fig:crust_pasta}
\end{figure}

Up to densities of order 10$^{11}$ g/cm$^{3}$ one can use the properties of laboratory nuclei to deduce the sequence of increasingly neutron rich nuclei with depth \cite{pichon}.  Beyond this point, however,  the details of the nuclei that are present, as well as the neutron drip point, at mass density $\sim 4\times 10^{11}$ g/cm$^3$  where continuum nuclear states are first occupied, are sensitive to the shell structure of very neutron-rich nuclei.  The spin-orbit force, which is critical in determining the usual closed shell structure at neutron or proton numbers 20, 28, 50, 82, \dots,  is expected to decrease in more neutron-rich nuclei \cite{pudliner};  the details are still very much in flux, both experimentally and theoretically \cite{sakurai}.  

  To determine the nature of the nuclei beyond neutron drip one must generalize the description of isolated nuclei to allow for the neutron gas outside the nuclei.   Common methods are microscopic calculations based on Skyrme interactions \cite{douchin,pearson},  on generalized liquid drop models \cite{bbp,bennett}, and on chiral effective field theory \cite{HebelerBogner}.    Even though the gross properties of matter in the crust are much better understood than they are at higher densities, there are still a number of uncertainties, typically at the $~ 10 \%$ level, in the structure of the nuclei owing to uncertainties in the effective interactions used.  The superfluid properties of the nuclei immersed in the neutron fluid \cite{pearson,avogadro,avogadro1,chamel}
as well as the superfluid properties of the neutron fluid itself \cite{apw,gps} are also uncertain.
  
   Determining the position of the inner edge of the crust is simple in principle:  given the pressure, $P$, as a function of the baryon chemical potential, $\mu_B$, for the phase with nuclei and for the uniform liquid phase, the ground state for matter with a given $\mu_B$ is the phase with the higher pressure.  The phase transition occurs when $P$ and $\mu_B$ of the two phases are equal and, since $n_B=\partial P/\partial \mu_B$, the density discontinuity across the transition is given by the difference of the slopes of the $P$ versus $\mu_B$ curves for the two phases.  This calculation, demanding detailed calculations of the pasta phases, is complex in reality.   One simple approach to estimating the position of the boundary is to compare the nucleonic pressure in the liquid interior with the pressure of a pure neutron gas (in which $\mu_B$ is the neutron chemical potential $\mu_n$) at the same  $\mu_B$, and locating the baryon chemical potential where the two curves cross.   In this comparison one holds the electron density fixed, and the electron contribution to the pressures plays no role. Again the slopes of $P$ vs. $\mu_B$ at this point give the densities across the transition.   This procedure is simply equivalent to neglecting the surface and Coulomb energies of the nuclei in the phase with nuclei,  so that its thermodynamics is determined only by the dripped neutron fluid.    Detailed calculations are given in Ref.~\cite{lorenz}.
  
   A second way to estimate the position of the transition is again to start in the uniform phase and decrease the density until matter becomes unstable to formation of small amplitude modulations of the proton and neutron densities.  In the absence the Coulomb interactions, the electron density remains uniform, and the conditions for stability to long-wavelength density fluctuations are first that the partial neutron and proton compressibilites are positive,
 \beq
\varepsilon_{nn}>0\,\,\,{\rm and}\,\,\,\varepsilon_{pp}>0,
\label{instab2}
\eeq 
where $\varepsilon(n_n,n_p)$ is the energy per unit volume of uniform nuclear matter and $\varepsilon_{ij}=\partial^2\varepsilon/\partial n_i\partial n_j=\partial \mu_i/\partial n_j=\partial \mu_j/\partial n_i$; these conditions are generally satisfied for energy functions commonly employed.   The second condition is that
\beq
\varepsilon_{nn}\varepsilon_{pp}>\varepsilon_{np}^2.
\label{instab1}
\eeq
When the two terms in Eq.~(\ref{instab1}) become equal, matter becomes unstable to long-wavelength sinusoidal proton and neutron density waves.  In nuclear matter at densities $\lesssim n_0$, $\varepsilon_{np}$ is negative as a consequence of the strong attractive s-wave interaction between neutrons and protons.  Thus the neutron density is large where the proton density is large.   The instability argument can be extended to finite wavelengths by including the Coulomb interaction and surface energy \cite{bbp}; the onset of instability occurs at a lower density when these effects are included.   As found in Ref.~\cite{kai}, the density in the liquid at which the instability occurs is in the range (0.075-0.088) fm$^{-3}\simeq$  (0.47-0.55)$\,n_0$, similar to that found in Ref.~\cite{lorenz}; the estimate based on comparing pure neutron matter with the matter in the liquid interior occurs at a somewhat higher density \cite{lorenz} (but still below $n_0$).

  The thickness of the crust as well as its mass can be deduced from the TOV equation (\ref{tov})
without having a detailed knowledge of the equation of state in the crust.   Hartle \cite{hartle1978} gave initial estimates of the mass and thickness of the crust, and we give improved versions of his arguments here~\cite{zdunik,sotani}.  We focus first on the thickness of the crust. For matter at zero temperature containing a single component, or in full chemical equilibrium, one has  $dP=n_B d\mu_B$ and $\rho c^2+P=\mu_B n_B$; the TOV equation can then be rewritten as
\beq
  \frac{\partial \ln \mu_B(r)}{\partial r} =- \frac{G_N}{c^2}\frac{m(r) + 4\pi r^3 P(r)/c^2}{r(r-2Gm(r)/c^2)}.
  \label{tovmu}
\eeq
In evaluating the right side in the crust, we can to a good approximation replace $m(r)$ by the total mass, $M_c$, of the liquid core of the star, and neglect the pressure term $4\pi P(r) r^3/c^2$  compared with $M_c$.  
 
When $\mu_B(r)$ is continuous, as in fully catalyzed matter, we can readily integrate Eq.~(\ref{tovmu}) from  the outer edge of the core, at radius $R_c$, to the surface of the star, at radius $R$, to find
\beq
       \frac{\, \mu_B (R_c) \,}{ \mu_B (R) } = \left(\frac{1-R_{sc}/R}{1-R_{sc}/R_c} \right)^{1/2},
\eeq
where $\mu_B(R)$ is the chemical potential at the surface of the star and $R_{sc} = 2 M_c G_N/c^2$ is the Schwarzschild radius of the core.     Solving for $R$ we find,
\beq
     \frac{R-R_c}{R_c} = \frac{\zeta}{[R_{sc}/(R_c-R_{sc})] -\zeta},
     \label{Rs}
\eeq
where
\beq
   \zeta \equiv  \left(\frac{\,\mu_B (R_c) \,}{\mu_B (R) }\right)^2 -1.
\eeq
Both $\mu_B (R_{c})$ and $\mu_B(R)$ 
are very close to the nucleon mass.  For example in \cite{bbp}, $\mu_B (R_{c}) - m_n \simeq$ 15 MeV,
while at the stellar surface, $\mu_B(R) - m_n \simeq -$8 MeV,  
the binding energy of nucleons in the outermost nuclei ($^{57}$Fe, ideally).  
Thus $\zeta \simeq 2(\mu_{i} - \mu_B(R) )/m_nc^2$, 
and for the numbers just cited, $\zeta \simeq$ 0.05.
On the other hand, $R_{sc}/R_c \sim 0.3 M/M_\odot$, so that to a good approximation the $\zeta$ can be neglected in the denominator of Eq.~(\ref{Rs}), and we find
\beq
        \frac{\, R-R_c \,}{R_c}  \simeq 2\, \frac{\, \mu_B (R_c) - \mu_B(R) \,}{m_nc^2}\frac{\, R_c-R_{sc} \,}{R_{sc}};
        \label{thick}
\eeq
this equation gives the thickness of the crust in terms of the radius and Schwarzschild radius of the core, and the baryon chemical potential difference between the inner and outer radius of the crust.   Remarkably, the thickness of the crust is insensitive to the  details of the equation of state in the crust.   For the numbers taken above, we find a crust thickness of order 0.5 km.

  The calculation above assumes that the baryon chemical potential is continuous in the crust, as it must be for fully catalyzed matter. In chemical equilibrium the neutron chemical potential and pressure are continuous across a phase transition between different species of nuclei, although the baryon and electron densities need not be continuous.    We comment briefly on the situation when matter in the outer regions of the crust is not in full chemical equilibrium, as 
in accreting stars, where, e.g., the neutron star can have a layer of hydrogen above the fully catalyzed matter.   Across  such a transition between elements the pressure is continuous; however, in general, the baryon chemical potential can have a discontinuity, $\Delta \mu$ in going from larger to smaller radii, since the baryons are bound in nuclei.  For a single discontinuity, $\mu_{c} - \mu_B(R)$  in Eq.~(\ref{thick}) is replaced by $\mu_{i} - \mu_B(R) - \Delta\mu$.   
In other words only the continuous changes in the chemical potential from the edge of the core to the edge of the star determine the thickness of the crust.  

  If the matter is fully catalyzed out to a pressure $P_\chi$, beyond which the star has a layer of atoms of a given species ($A,Z$), then the thickness of this added layer is simply 
\beq
   \Delta R \simeq \frac{2}{m_n}\left[\mu_{AZ}(P_\chi)  - \mu_{AZ}(0)\right]   \left(\frac{R_c}{R_{sc}} -1\right);
\eeq
the chemical potential difference here is given by the equation of state $P(\mu)$ in the added layer.

  To estimate the mass of the crust, we again neglect the pressure terms on the right of the TOV equation, replace $m(r)$ by $M_c$, and integrate the resulting equation from $R_c$ to $R$.   Since the pressure vanishes at the stellar radius, we find
\beq
     P_c &=& R_{sc}\int_{R_c}^R  r^2 dr  \frac{\rho(r)c^2}{r^3(r-R_{sc})}.
 \eeq
The integral is dominated by the contributions from close to the inner radius of the crust, and thus we may
to a good approximation replace the factor $r^3(r-R_{sc})$ in the denominator of the integrand by $R_c^3(R_c-R_{sc})$. 
The crust mass, 
\beq
     M_{crust} = \int_{R_s}^R 4\pi r^2 \rho(r) dr,
\eeq 
is then approximately
\beq
  M_{crust} \simeq 4\pi R_c^3 \frac{P_c}{c^2} \left(\frac{R_c}{R_{sc}}  -1\right),
\label{mcrust}
\eeq  
again not dependent on the details of the equation of state in the crust.
For a characteristic value of $P_c \simeq$ 0.5 MeV/fm$^{3}$, we find 
$M_{crust} \sim 10^{-2} M_\odot$.   

The baryon density at the transition, in the range (0.47-0.55) $n_0$ \cite{kai}, is uncertain.  The related uncertainty in $P_c$ at the transition leads, as one sees from Eq.~(\ref{mcrust}), to an uncertainty in the mass of the crust at a level $\sim$ 8\%.   Although the lattice appears to be stable \cite{kobyakov,kobyakov2}, an additional subtlety in the physics of the crust is the question of whether protons in the crust can also drip out of the nuclei \cite{lorenz}.

 \section{Liquid nuclear matter in the outer core}\label{sec:outer}

     The historic approach to describing nuclear matter in the outer region of the core is first to calculate the energy densities of both symmetric nuclear matter and neutron matter and  then interpolate between them to describe matter at the finite proton fraction expected for matter in beta equilibrium between the protons, electrons and neutrons, viz., 
\beq
\mu_n = \mu_p +\mu_e,
\label{betaeq} 
\eeq
where the rest masses are included in the chemical potentials $\mu_i$.   As described briefly in the introduction, the basic calculational method is to first determine nucleon-nucleon potentials from scattering data supplemented by three-body interactions, and to then employ large scale computational methods to solve the many-body Schr\"odinger equation in the presence of these potentials; a representative such calculation is that of Akmal, Pandharipande, and Ravenhall (APR) \cite{APR}.     It is common to employ an energy density of symmetric matter consistent with the empirical nuclear binding energies and compressibilities, thus reducing theoretical uncertainties. 
A further approach is via chiral effective field theory \cite{chiral}, which provides a systematic development of two and higher body interactions \cite{kai,tews,drischler}.   

    Various microscopic calculations of (pure) neutron matter, reviewed in Ref.~\cite{ggc2015}, are in relatively good agreement up to nuclear matter density $n_0$,  where the dominant two-body interaction between nucleons is well characterized by nucleon--nucleon scattering data,    Zero range three body interactions in neutron matter give no contributions to the energy density, as a consequence of the Pauli principle, which prevents three neutrons being at the same point, because two of them must be in the same spin state.   However, realistic three-body interactions have a finite range, and thus at densities of interest in neutron stars, they 
 give comparable contributions to the 
 energy densities of neutron matter and symmetric nuclear matter, and  greatly stiffen the neutron matter equation of state.
  While the limit of validity of nuclear matter calculations based on interacting nucleons is uncertain, it is not unreasonable to use them up to a density $\sim 2 n_0$.

   To interpolate between pure neutron matter and symmetric nuclear matter, one can to first approximation take the energy density of uniform nuclear matter to be a quadratic function of the neutron excess, 
 $\delta=1-2x$, where $x= n_p/n_B$ is the proton fraction; then, 
\beq
  E(n_B,x)  \simeq E(n_B,0) -4x(1-x)n_B S(n_B),
  \label{quadinterp}
\eeq
where $S(n_B) = \left(E(n_B,0) - E(n_B,1/2)\right)/n_B$ is the  density-dependent symmetry energy.    This interpolation leads typically to a value of $S(n_0)$ at nuclear saturation density close to the empirical symmetry energy for nuclei with roughly equal numbers of neutrons and protons, $\approx 32$ MeV  \cite{kai}.   An alternative approach is to take the kinetic energy to be that of the free gas at given $n_B$ and $x$, with possible effective mass corrections \cite{APR}, and apply a quadratic interpolation only between the interaction-energy density of pure neutron matter and symmetric nuclear matter \cite{APR,kai}. To determine the proton fraction in the liquid interior, one then imposes the condition of beta equilibrium (\ref{betaeq}).   Typically the proton fraction just within the liquid interior varies from about $\sim 3\%$ at the crust-core boundary to $\sim 5\%$ at $n_0$. Initial direct Monte Carlo calculations of asymmetric nuclear matter are given in Ref.~\cite{glcs}.

 It should be noted that BCS pairing of nucleons in the outer regions of the core does not produce significant corrections to the equation of state there, since energy gaps $\Delta$ are $\sim$ MeV, so that condensation energies, of order $\Delta^2/E_F$, are small compared with the Fermi energy, $E_F$.   Such pairing is not taken into account in the APR equation of state.  Uncertainties in the equation of state arise from uncertainties in the three-body interactions with increasing density \cite{kai},  and also from the onset of neutral pion condensation, which APR finds to occur at $\sim$ 0.2 fm$^{-3}$ in matter in beta equilibrium, and at $\sim$ 0.3 fm$^{-3}$ in matter in symmetric nuclear matter.   The transition to the pion condensed state found in APR is first order.\footnote{This result is consistent with a general argument of Dyugaev \cite{dyugaev,migdal-rpts} 
that as condensation is neared the pion-pion scattering amplitude in the medium changes from repulsive to attractive owing to exchange of soft pion modes.}  
However,  APR did not consider spatially non-uniform neutral condensates (e.g.,  $ \sim \cos kz$ for small amplitude condensation), nor did they examine the order parameter of the condensed state; as a consequence the energy of the ``pion condensed phase" reported in APR is an upper bound to the energy of the pion condensed state.  The transition in APR is to a uniform phase characterized by a large tensor correlation length and enhancement of spin-isospin correlations  \cite{AM}.

\section{Quark matter at high density} \label{sec:quark_EOS}

    As an introduction to 
more detailed applications in Sec.~\ref{generalstructure}, we begin in this section by discussing the elementary physics of quark matter, the expected form of neutron star matter at densities well beyond $n_0$.   

\subsection{Noninteracting quark matter} \label{nonintqm}

 The simplest model of quark matter takes into account the bare quark kinetic energy density, $\varepsilon_K$, and the bag constant, $B$, which is the energy density difference between the non-perturbative vacuum in QCD and the perturbative vacuum.   The zero of energy is commonly taken to be that of the non-perturbative vacuum, compared to which the perturbative vacuum has a positive energy, $B$.  For illustration, let us consider 
 a gas of massless quarks with $N_f$ flavors.
 At low temperature the quarks form a degenerate Fermi sea, with quark density
\beq
   n_q = 2N_cN_f \int_0^{p_F} \frac{d^3p}{(2\pi)^3} =  N_cN_f\frac{p_F^3}{3\pi^2}, 
\eeq
where $N_c = 3$ is the number of quark colors, the 2 is for spin, and $p_F$ is the quark Fermi momentum.   The bare quark kinetic energy density is then
\beq
     \varepsilon_{K}  &=& 2N_cN_f\int_0^{p_F} \frac{d^3p}{(2\pi)^3} |p| = \frac{3N_cN_f}{4\pi^2} p_F^4,
\eeq
Then the total energy density is $ \varepsilon =   \varepsilon_{K}  + B$.
Since the quark chemical potential is $\mu_q = \partial \varepsilon/\partial n_q = p_F$, the baryon chemical potential $\mu_B=3\mu_q$, the pressure and the energy density read
\beq
P(p_F) = a p_F^4 - B \,,~~~~~\varepsilon(p_F) = 3 a p_F^4 + B \,,
\label{freeqeos}
\eeq
where $a =N_cN_f/4\pi^2$.  These expressions give the equation of state which is 
 valid for non-interacting massless quarks at high density.
{\em The bag constant},  $B$, is the difference in the energy densities of the perturbative vacuum -- devoid of all particles and condensates -- and the non-perturbative vacuum, which is the true ground state of QCD, including chiral and gluon condensates.

 With this equation of state, the maximum mass of a hypothetical {\em quark star} made of free quarks,
 for $N_c=N_f=3$,  scales with $B$  \cite{Witten1984,Book} (see also Appendix A) as
\beq
M_{ {\rm max} } \simeq 1.78\, \left(\frac{155\, {\rm MeV} }{B^{1/4} }\right)^2 M_\odot  ,
\label{idealgasM}
\eeq
while the corresponding radius scales as
\beq
R \simeq 9.5 \left(\frac{155\, {\rm MeV} }{B^{1/4} }\right)^2 \, {\rm km}.
\label{idealgasR}
\eeq
%

\subsection{The bag constant \label{bagconst}}

  Various estimates of $B$ have been made, for a variety of physics models and choices of which energies to include in $B$ (and which to include in the quarks). As a result of these variations, comparison of the precise values obtained in different models is not generally meaningful.
For example, in the early MIT bag model, $B \simeq$ (145-155)  MeV$^4 \simeq $  (60-80) MeV/fm$^3$.   
The bag constant calculated in the NJL model (see Eq.~(\ref{bnjl}) below) is $\simeq$ (218 MeV)$^4$ = 296 MeV/fm$^3$ \cite{Buballa2005}. On the other hand, Novikov et al.~\cite{novikov}, in a  QCD based calculation, find $B \simeq$ (250-300 MeV)$^4 \simeq$  (500-1000) MeV/fm$^3$, 
an order of magnitude larger than the MIT value.
We note here that it may not be possible to calculate the properties of the gluons, at least in the density range relevant for neutron stars, by applying perturbative QCD, i.e., the gluons might remain non-perturbative.   The properties of such matter have been discussed in the context of quarkyonic matter \cite{McLerran:2007qj}, discussed in Sec.~\ref{perc}.

\subsection{Condensates in the QCD vacuum and quark matter} 
\label{condensates}

\begin{figure}
\hspace{-0.5cm}
\includegraphics[width = 0.48\textwidth]{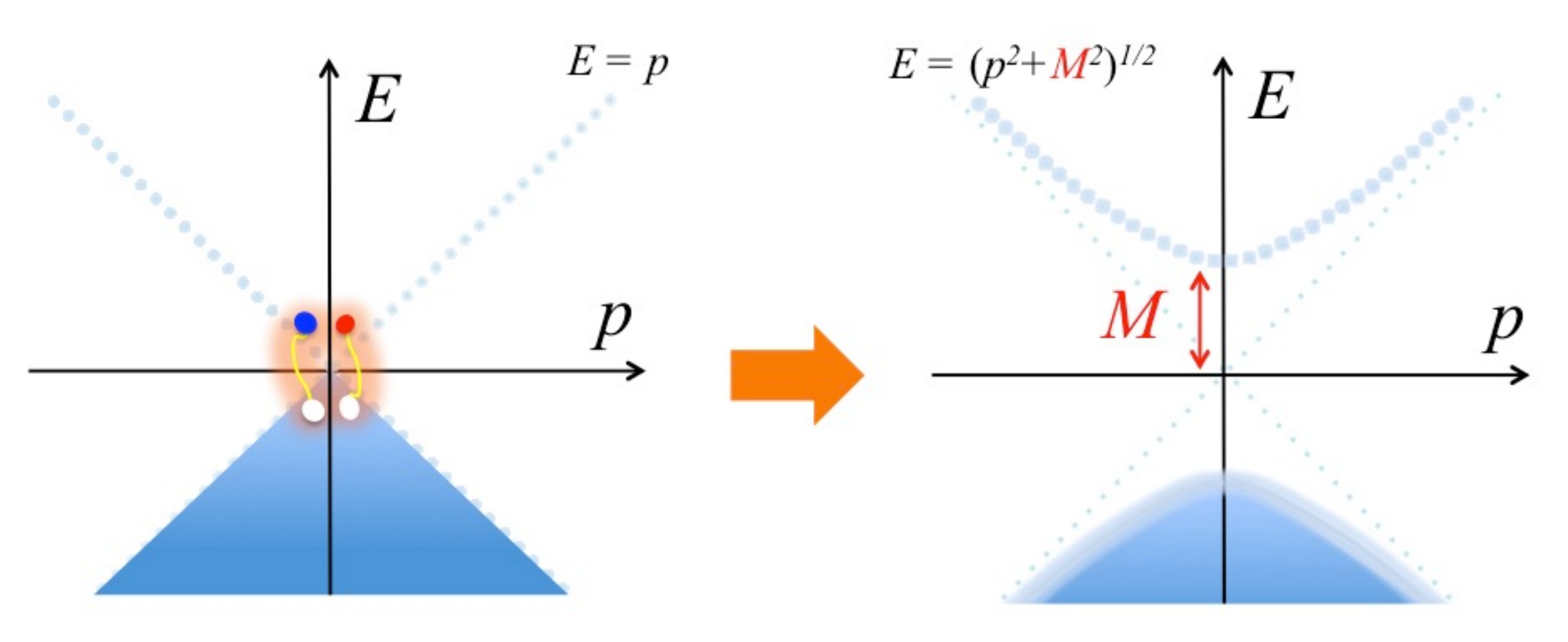}
\caption{
\footnotesize{
Chiral symmetry breaking via quark-antiquark pairing.  Condensation of pairs opens a gap $M$ in the quark dispersion relation, changing the structure of the Dirac sea. The energy density of the symmetry-broken Dirac sea is smaller than that of the symmetric sea by $B \sim \lqcd^4$ (see text). 
}} 
\label{fig:ChSB}
\end{figure}

\begin{figure}
\hspace{-0.5cm}
\includegraphics[width = 0.26\textwidth]{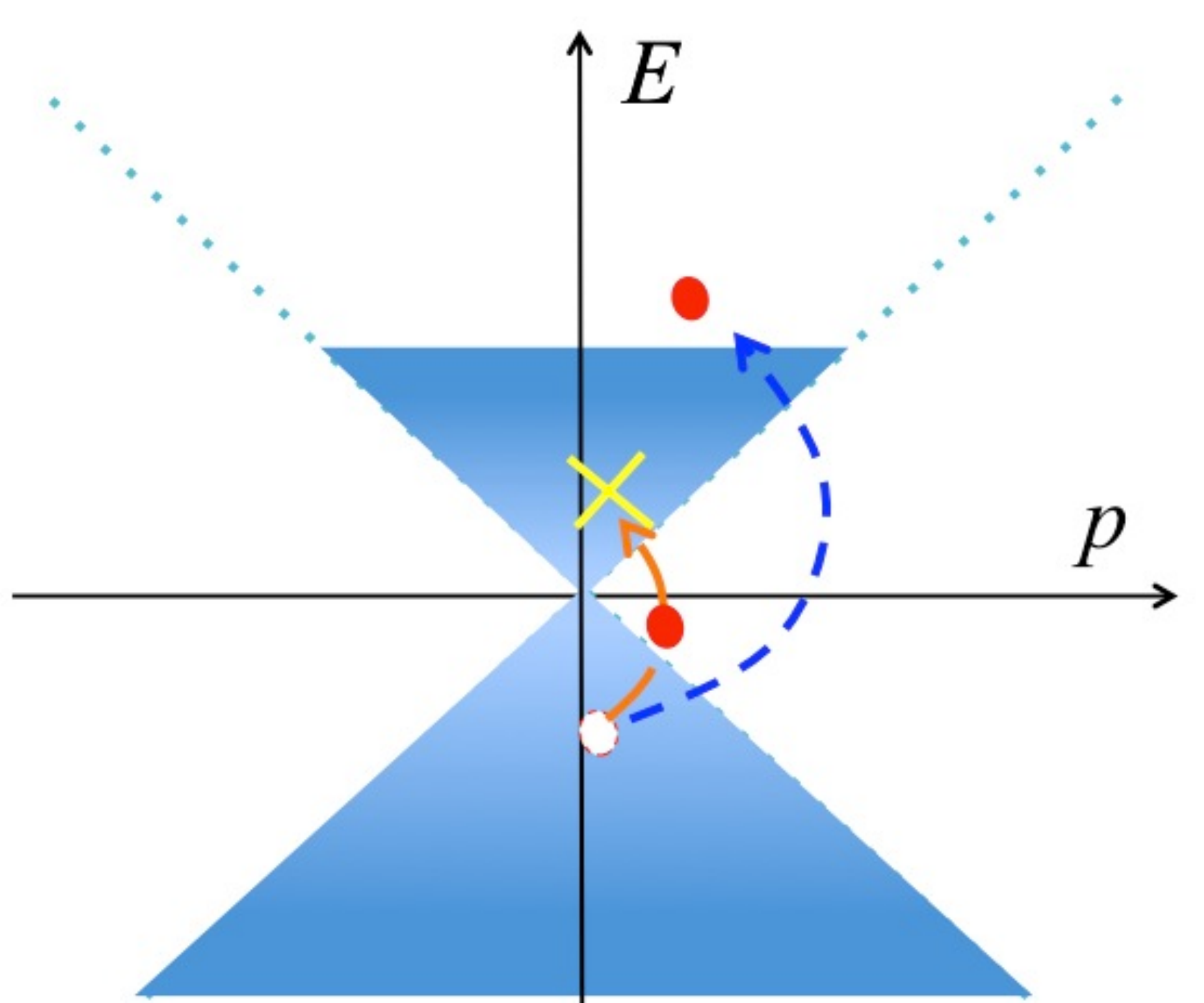}
\caption{
\footnotesize{
Quark-antiquark pairing at high baryon density. In creating a hole in the Dirac sea, the quark in the Dirac sea must, by the Pauli principle, be outside the Fermi sea. 
}} \label{fig:forbidden_ChSB}
\end{figure}

\begin{figure}
\hspace{-0.5cm}
\includegraphics[width = 0.26\textwidth]{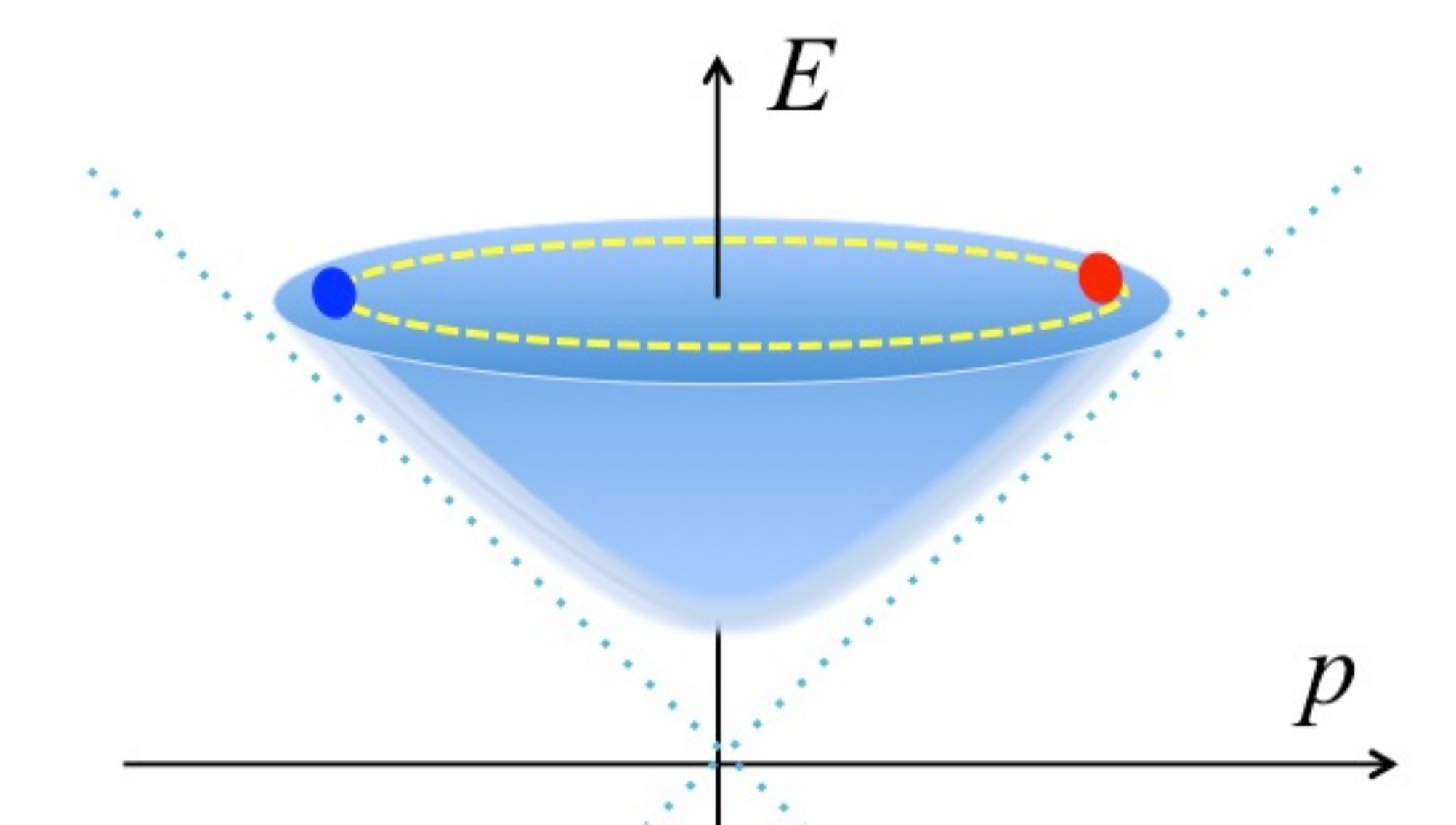}
 \caption{
\footnotesize{
Quark-quark pairing, which leads to color superconductivity at high baryon density. The condensation of pairs opens a gap $\Delta$ near the quark Fermi surface. 
}} \label{fig:Diquark_pair}
\end{figure}

  The strong interactions of quarks and gluons in QCD cause a variety of condensation phenomena.  Fundamental is the formation of a chiral condensate, made of quark-antiquark pairs with different chirality, e.g., a left-handed (i.e., spin antiparallel to momentum) quark and right-handed antiquark (Fig.~\ref{fig:ChSB}).    The approximate chiral symmetry of QCD, as a consequence of which left and right handed quarks comprise essentially independent sectors, is broken by the chiral condensate because it couples quarks of different chiralities.
    Since quarks and antiquarks are bound in the condensed ground state, the creation of a quark excitation requires the breaking of a pair, and the associated energy cost; as a consequence, a quark acquires an effective, or constituent, quark mass, $M$.  This mass, dependent on the strength of the QCD interactions, has a characteristic magnitude, $\lqcd \sim 200\,{\rm MeV}$, the dynamical scale of QCD, much larger than the  bare, or current, quark mass, $m_q$, entering the QCD Hamiltonian.   The emergence of the chiral condensate changes the structure of the Dirac sea, giving rise to a non-perturbative QCD vacuum. The difference in energy density between the (perturbative) chirally symmetric and (non-perturbative) symmetry-broken Dirac seas, with different effective masses $M_{\rm eff}$, at zero temperature and baryon chemical potential formally defines the bag constant: 
\beq
B \equiv \varepsilon(M_{ {\rm eff} } = m_q) - \varepsilon(M_{ {\rm eff} } =M).
\label{quark_bag_const}
\eeq

   At high baryon density, quark-antiquark pairing is no longer energetically favored in the presence of the quark Fermi sea,  Fig.~\ref{fig:forbidden_ChSB}, since to create an antiquark (a hole in the Dirac sea), the quark originally occupying the Dirac sea must now occupy a previously unoccupied high energy state outside the Fermi sea.   One could imagine at first sight that  chiral symmetry would then be restored.  However, in the presence of a quark Fermi sea, pairings employing the degrees of freedom near the Fermi surface become possible, and these can continue to break chiral symmetry \cite{ARW,pisarskirischke}.   Of particular importance is {\em diquark} pairing, in which two quarks (or two quark-holes) near the Fermi surface are paired as electrons are  Cooper paired in an ordinary superconductor, Fig.~\ref{fig:Diquark_pair}.  The diquark pairs, macroscopic in number, form a diquark condensate, and as a condensate of paired electrons gives rise to electromagnetic superconductivity, a condensate of diquarks, which has color charges, gives rise to color superconductivity.  A quark excitation requires the breakup of a condensed pair, costing energy $2\Delta$, where $\Delta$ is the pairing gap.  Condensation reduces the energy density of the system by the energy gain for a single pair, $\Delta$, times the phase space available for such pairs, $\sim 4\pi p_F^2 \Delta$, where $p_F$ is the quark Fermi momentum. This energy reduction, $\sim \Delta^2 p_F^2$ plays a very important role in quark matter equations of state, as we discuss below.  The modeling of these effects will be discussed in subsec.~\ref{NJL_int_quarks}.
     
   The diquark pairing interaction is most attractive in the color-antisymmetric, flavor-antisymmetric, and spin-singlet channel.  Flavor asymmetry implies that quarks in a pair must have different flavors.   At densities relevant to neutron stars, this complicates the pairing, because the $(u,d,s)$-quarks have different quark masses and electric charges, introducing a considerable imbalance in the size of their Fermi seas.  Many possibilities for the preferred pairing structures have been discussed (reviewed in \cite{Alford:2007xm}), but here we discuss only the simplest candidates; two flavor, or 2SC pairing, in which $u$- and $d$-quarks pair at chemical potentials not large enough for significant strangeness to appear, and the color-flavor-locked, or CFL phase, in which the chemical potential is sufficiently high that the $u$, $d$, and $s$ all quarks participate in the pairing.    The detailed structure of these condensates is given in Eq.~(\ref{diquark_field}), below.

\subsection{Nambu--Jona-Lasinio model for interacting quarks}
\label{NJL_int_quarks}

   Owing to the inability of lattice gauge theory calculations to describe cold matter at finite baryon density, due the {\em fermion sign problem} \cite{Ukawa:2015eka}, one needs to adopt phenomenological models of interacting quarks in order to describe dense quark matter in neutron star cores \cite{Powell,Powell2,Fukushima2011,Abuki,Masuda2013,Masuda2013-0,Takatsuka}.   We discuss here the frequently employed Nambu--Jona-Lasinio (NJL) model  \cite{Vogl:1991qt,Klevansky:1992qe,Hatsuda1994}, which replaces the full QCD interactions with effective quark-quark interactions, while at the same time suppressing explicit gluonic degrees of freedom.  In this subsection we write down the NJL model and discuss the physics of the 
effective  interactions.\footnote{Readers not well acquainted with the effective theories of QCD described here can skip the details of this subsection and continue on to Sec.~\ref{sec:hybrid}.}
In describing the NJL model we adopt standard Dirac notation with the Minkowski metric $g_{\mu\nu}$ = diag(1,-1,-1,-1), and $\gamma^0$  and $\gamma_5$ both Hermitian.

   The Lagrangian of the three-flavor NJL model is
\beq
\Lc = \overline{q} ( \gamma^\mu p_\mu  - \hat{m}_q +\mu_q \gamma^0) q + \Lc^{(4)} + \Lc^{(6)}   ,   \label{eq:NJL_Lagrangian}
\eeq
where $q$ is the quark field operator with color, flavor, and Dirac indices, $\bar q = q^\dagger \gamma^0$, $\hat{m}_q$ the quark current mass matrix, $\mu_q$ the (flavor dependent) quark chemical potential and $\Lc^{(4)} = \Lc^{(4)}_\sigma + \Lc^{(4)}_d + \Lc^{(4)}_V$ and $\Lc^{(6)} = \Lc^{(6)}_\sigma + \Lc^{(6)}_{\sigma d}$ are four and six-quark interaction terms, chosen to reflect the symmetries of QCD.   

   The first of the four-quark interactions, a contact interaction with coupling constant $G>0$,
\beq
\Lc^{(4)}_\sigma & = & G \sum^8_{j=0} \left[ (\overline{q} \tau_j q)^2 + (\overline{q} i \gamma_5 \tau_j q)^2 \right] = 8 G \mbox{tr} (\phi^\dagger \phi)   ,  
 \label{eq:L4_sigma}
\eeq
describes spontaneous chiral symmetry breaking, where $\tau_j$ ($j= 0,\ldots,8$) are the generators of the flavor-U(3) symmetries, and in Eq.~(\ref{eq:L4_sigma}), 
\beq
\phi_{ij} = (\overline{q}_R)^j_a (q_L)^i_a
\eeq
 is the chiral operator with flavor indices $i,j$ (with summation over the color index $a$); the right and left quark chirality components are defined by  $q_{R,L} = \frac12(1\pm \gamma_5)q$.
 
   The second of the four-quark terms describes the scattering of a pair of quarks in the s-wave, spin-singlet, flavor- and color-antitriplet channel; this interaction leads to BCS pairing of quarks:
 \beq
\Lc^{(4)}_d & = & H \sum_{A,A^\prime = 2,5,7} \big[ \left(\overline{q} i \gamma_5 \tau_A \lambda_{A^\prime} C \overline{q}^T \right) \left(q^T C i \gamma_5 \tau_A \lambda_{A^\prime} q \right)   \nonumber   \\
	       && \hspace{23mm} + \left(\overline{q} \tau_A \lambda_{A^\prime} C \overline{q}^T \right) \left(q^T C \tau_A \lambda_{A^\prime} q \right) \big]   ,   \nonumber   \\
	    & = & 2 H \tr (d^\dagger_L d_L + d^\dagger_R d_R),
 \label{eq:L4_d}
\eeq
with $H>0$.  Here $\tau_A$ and $\lambda_{A^\prime}$ ($A,A'=2,5,7$) are the antisymmetric generators of U(3) flavor and SU(3) color, respectively, and 
\beq
(d_{L,R})_{ai} = \epsilon_{abc} \epsilon_{ijk} (q_{L,R})^j_b {\cal C} (q_{L,R})^k_c 
\label{diquark_field}
\eeq
are diquark operators of left- and right-handed chirality,  with ${\cal C}=i\gamma^0\gamma^2$ the charge conjugation operator. The diquark pairing interaction leads as well as an attractive correlation between two quarks inside confined hadrons and, in constituent quark models, plays a role in
 the observed mass splittings of hadrons \cite{DeRujula:1975qlm,Anselmino:1992vg,Selem:2006nd,Jaffe:2004ph}. 
This interaction, in weak coupling, arises from single gluon exchange;  however at the densities of interest in neutron stars,  the non-linearities of QCD prevent direct calculation of this interaction, and so one must treat it phenomenologically.

In addition
\beq
\Lc^{(4)}_V & = & -g_V (\overline{q} \gamma^\mu q)^2   ,   \label{vec}
\eeq
with $g_V > 0$, is the Lagrangian for the phenomenological vector interaction, as in Eq.~(\ref{eq:energy}), which produces universal 
repulsion between quarks \cite{Kunihiro:1991qu}. 

    The six-quark interactions represent the effects of the instanton-induced QCD axial anomaly, which breaks the $U(1)_A$ axial symmetry of the QCD Lagrangian. 
    The resulting Kobayashi-Maskawa-'t Hooft (KMT) interaction leads to an effective coupling between the chiral and diquark condensates of the form~\cite{Kobayashi,'tHooft:1986nc}:
\beq
\Lc^{(6)}_\sigma & = & -8 K ( \det \phi + \mbox{h.c.})   ,   \label{eq:L6_sigma}   \\
\Lc^{(6)}_{\sigma d} & = & K^\prime ( \tr [(d^\dagger_R d_L) \phi] + \mbox{h.c.})   ,   \label{eq:L6_sigma_d}
\eeq
where $K$ and $K^\prime$ are positive constants. Provided that $K^\prime \simeq K$ (which one expects on the basis of the Fierz transformation connecting the corresponding interaction vertices) the six-quark interactions encourage the coexistence of the chiral and diquark condensates.  

\subsection{Mean field equation of state}

Having reviewed the structure of the NJL model, we now discuss its application to constructing the  
equation of state for dense quark matter.   For simplicity, we restrict our considerations to mean-field theory.    The  quark density is
\beq
 n_q =\sum_{i=u,d,s} \braket{q^\dagger_i q_i},
\eeq
with an implicit sum over color and spin.  Similarly the flavor-dependent chiral condensate is 
\beq
\sigma_i & = & \braket{\overline{q}_i q_i},
\eeq
with $i=u,d,s$; with our conventions $\sigma$ is generally negative.
The diquark mean fields are 
\beq
d_j &= \braket{q^T C \gamma_5 R_j q} ,
\label{diquark_MF}
\eeq
with matrices
\beq
\left( R_1, R_2, R_3 \right) \equiv \left( \tau_7 \lambda_7, \tau_5 \lambda_5, \tau_2 \lambda_2 \right) \,.
\eeq
The three diquark condensates correspond to $(ds, su, ud)$ quark pairings, respectively.  In the 2SC phase, only $ud$-pairs condense, while in the CFL phase all $(ds,su,ud)$-pairs are present.

The inverse of the mean field single particle propagators can be read off from the mean field Lagrangian, 
\beq
S^{-1} (k) = \begin{pmatrix}
		~ \gamma^\nu k_\nu - \hat{M} + \hat{\mu} \gamma^0 ~&~ \gamma_5 \Delta_k R_k ~\\
		~ - \gamma_5 \Delta^\ast_k R_k ~&~ \gamma^\nu k_\nu - \hat{M} - \hat{\mu} \gamma^0 ~
	     \end{pmatrix}, \nonumber\\
	       \label{eq:Sinv}
\eeq
where the effective mass matrix is diagonal, with elements
\beq
M_i = m_i - 4 G \sigma_i + K |\epsilon_{ijk}| \sigma_j \sigma_k + \frac{K^\prime}{4} \hspace{.5mm} |d_{i}|^2   ,
\label{Mq}
\eeq
while the three diquark pairing amplitudes,
\beq
\Delta_k= -2 d_k \left(H - \frac{K^\prime}{4} \hspace{.5mm} \sigma_i \right)   ,
\eeq
and the effective chemical potential matrix,
\beq
\hat{\mu} = \mu_q - 2 g_V n_q + \mu_8 \lambda_8 + \mu_Q Q \,,   \label{eq:mu_eff}
\eeq
are color and flavor dependent. 

   The inverse propagator  (\ref{eq:Sinv}) is a $72\times 72$ matrix at each momentum, whose eigenvalues, $\epsilon_j$ can be calculated by numerical inversion \cite{Powell}. The eigenvalues are four-fold degenerate (2 for spin times 2 from the Nambu-Gor'kov pairing structure).  The single particle contribution to the thermodynamic potential density,
\beq   
   \Omega(\mu_q,T) = \varepsilon - Ts - \mu_q n_q
   \label{Omega},
\eeq 
which is the negative of the pressure, $P$,  
is then
\beq
\Omega_{ {\rm single} } & = & - 2 \sum^{18}_{j=1} \int^\Lambda \!\!\frac{d^3 \bd{k}}{(2\pi)^3} 
\left[ T \ln \left( 1 + e^{- |\epsilon_j | /T }  \right)+ \frac{ \Delta \epsilon_j}{2} \right] \,,\nonumber \\
\eeq
where $\Delta \epsilon_j = \epsilon_j -\epsilon_j ^{\rm free}$, with $\epsilon_j ^{\rm free}$ the eigenvalues in the non-interacting quark system;
here $\Lambda$ is an ultraviolet cutoff. The dependence on $\mu_q$ is hidden in the eigenvalues $\epsilon_j$. As in mean field treatments, ``condensate" terms, here
\beq
\Omega_{ {\rm cond} } & = & \sum^3_{i=1} 
\left[ 2G\sigma^2_i + \left(H - \frac{K^\prime}{2} \, \sigma_i \right) |d_i|^2 \right],   \nonumber \\
&& \hspace{10mm} - 4 K \sigma_1 \sigma_2 \sigma_3 - g_V n_q^2
\eeq
appear in $\Omega$ (as a consequence of avoiding double counting).

   While the quark matter thermodynamic potential is formally $\Omega^{ {\rm bare} }_{q} = \Omega_{ {\rm single} } + \Omega_{ {\rm cond} }$, one must further choose the ``zero" of $\Omega$ so that it vanishes in the vacuum, $\mu_q = T = 0$.  This choice is important because the absolute energy, not just energy differences, directly enters in the general relativistic TOV equation (\ref{tov}) for neutron star structure.    The correctly normalized quark-matter thermodynamic potential is thus
\beq
\Omega_q (\mu_q,T) \equiv \Omega^{{\rm bare}}_{q } (\mu_q,T) -\Omega^{{\rm bare}}_{q } (\mu_q=T=0).
\eeq

   In addition one must include effects of leptons -- electrons and muons -- that may be present in the matter;  the $\tau$ lepton's large mass  prevents it from playing a role in dense matter.  The Lagrangian for the leptons  is 
\beq
 \Lc_l  = \sum_{l=e,\mu} \overline{\psi}_l ( \gamma^\nu p_\nu - m_l) \psi_l ,
\eeq    
with $\psi_l$ the lepton field operator and $m_l$ the lepton mass. The lepton contribution to the thermodynamic potential is
\beq
\hspace{-0.3cm}
\Omega_l = -2 T \sum_{l=e,\mu} \sum_{\lambda=\pm}  \int \frac{d^3 \bd{k}}{(2\pi)^3} \,\ln \left(1 + e^{-(E_{l} + \lambda \mu_Q)/T} \right), \nonumber\\
\eeq
with $E_{l} = \sqrt{\bd{k}^2 + m^2_l}$.

\subsection{Electric and color neutrality constraints}

    The matter comprising neutron stars must be both electrically and color neutral, since any long range charge or color imbalance would be prohibitively expensive energetically.  To achieve electrical neutrality in quark matter one must allow for the possibility of electrons and muons being present.   
        
In modeling the equation of state, neutrality is simplest to achieve by introducing the charge chemical potential, $\mu_Q$,  and color chemical potentials, and tuning these chemical potentials to keep the color and charge densities zero. The charge chemical potential couples to the charge density in the Lagrangian through a term
\beq
 \Lc_Q  = \mu_Q \left[ q^\dagger Q q - \sum_{l=e,\mu} \psi_l^\dagger \psi_l \right]  , 
 \eeq
where $Q = \mbox{diag} (2/3,-1/3,-1/3)$ is the quark charge operator in flavor space.

  Dependent on the particular diquark pairing scheme,  the diquark pairing interaction, (\ref{eq:L4_d}), can lead to a violation of color neutrality.  For example, pairing of only red and green quarks (as in the 2SC phase) leads to a decreased energy per particle of these colors, and thus an increase of red and green quarks compared with unpaired blue quarks, leaving the system with a net anti-blue color density.   In realistic QCD this color imbalance is exactly cancelled by the appearance of a non-zero coherent gluon field \cite{Gerhold,Dietrich}.  However, this mechanism is outside the scope of the NJL model, and thus one must restore color neutrality by hand \cite{Gerhold,Dietrich,Iida,Steiner,Buballa2},  most generally by introducing eight independent color chemical potentials. For the diquark pairing structures discussed in Eq.~(\ref{diquark_MF}) all color densities except $n_3=\langle q^\dag \lambda_3 q\rangle$ and $n_8=\langle q^\dag \lambda_8 q\rangle$ automatically vanish. Thus, including the term~\cite{Powell2,Iida,Steiner,Buballa2}
\beq
\Lc_{3,8} =  \mu_3 q^\dagger \lambda_3 q  + \mu_8 q^\dagger \lambda_8 q   ,
\eeq
is sufficient ensure color neutrality.

\subsection{Minimizing the thermodynamic potential}

   The total thermodynamic potential is $\Omega = \Omega_q + \Omega_l$.  The thermodynamic state of the system is determined by minimizing the free energy with respect to the six condensates $\{\sigma_i$,\,$d_k$\} and the quark density $n_q$, under the conditions of electrical and color charge neutrality expressed by the conditions
\beq
n_j = - \frac{\, \partial \Omega \,}{\, \partial \mu_{j}}  = 0 ,  \label{eq:constraint_1}
\eeq
with $j = Q, 3, 8$.  In addition the condensates are determined by the six ``gap equations,"
\beq
0 = - \frac{\, \partial \Omega \,}{\, \partial \sigma_i \,} = -\frac{\, \partial \Omega \,}{\, \partial d_i },
\eeq
where all derivatives are taken at fixed quark chemical potential.  Finally the quark density is 
\beq
  n_q = - \frac{\partial \Omega}{\partial \mu_q} .
\eeq
In order to construct the equation of state, one can solve these 10 equations self-consistently, using the method outlined in~\cite{Abuki:2005ms}, to construct first the energy density $\varepsilon(\mu_q,T)$ and the pressure $P(\mu_q,T)$, then finally the desired $P(\varepsilon)$.

\subsection{Parameter sets \label{params}}

  The NJL model for dense quark matter 
  contains two distinct sets of parameters: $\{\Lambda, m_u, m_d, m_s, G, K\}$ and $\{g_V, H, K'\}$. The first set is fixed by matching to QCD vacuum phenomenology.   To be specific, we consider primarily the set by Hatsuda and Kunihiro (HK) \cite{Hatsuda1994} (Table~\ref{tab:couplings}), which gives the vacuum effective masses for the light quarks, $M_{u,d} \simeq 336$ MeV, and the strange quark, $M_s\simeq 528$ MeV.  The NJL model with this set of parameters yields (e.g. Ref.~\cite{Buballa2005}) the ``bag constant" [cf. Eq.~(\ref{quark_bag_const})], 
 \beq
B_{ {\rm NJL} } \equiv
\left[\, \varepsilon(M_{ {\rm eff} } = m_q) - \varepsilon(M_{ {\rm eff} } =M) \,\right]_{T=\mu_q=0} \nonumber\\
 \simeq (218\, {\rm MeV})^4 = 296\, {\rm MeV/fm^{3} } \,.
\label{bnjl}
\eeq 

  The second set of parameters is not well fixed by QCD vacuum phenomenology, but it is natural to expect their values to be characterized by the QCD momentum scale $\lqcd$, in the absence of anomalous mechanisms to suppress these couplings. In addition, the $K^\prime$ terms couple the diquark condensate to the chiral condensate, and thus the value of $K^\prime$ is strongly correlated with the values of $g_V$ and $H$; most, if not all, of the effects of its variation can be absorbed into variations of $g_V$ and $H$.  In the present analysis we assume as a first orientation that the values of these coefficients are of the same order of magnitude as the first set of parameters.
\begin{center}
\begin{table}[t]
\caption{\footnotesize{Three common parameter sets for the three-flavor NJL model: the average up and down bare quark mass $m_{u,d}$, strange bare quark mass $m_s$, coupling constants $G$ and $K$, and three-momentum cutoff $\Lambda$~\cite{Hatsuda1994,Rehberg,Lutz1992}.}}
\begin{tabular}{|c|c|c|c|c|c|}
\hline 
   & $\Lambda$ (MeV) & \ $m_{u,d}$ (MeV) & \ $m_s$ (MeV) & \ $G \Lambda^2$ \ & \ $K \Lambda^5$ \ \\ \hline \hline
HK  & 631.4 & 5.5 & 135.7 & 1.835 & 9.29 \\ \hline
RHK & 602.3 & 5.5 & 140.7 & 1.835 & 12.36 \\ \hline
LKW & 750.0 & 3.6 & 87.0 & 1.820 & 8.90 \\ \hline
\end{tabular}
\label{tab:couplings}
\end{table}
\end{center}

Attempts have been made to estimate the magnitude as well as the medium dependence of $g_V$  in the NJL model (with and without coupling to a Polyakov loop\footnote{The Polyakov loop~\cite{Polyakov,Svetitsky,lo}, which is essentially an order parameter for confinement, plays an important role in determining the structure of the QCD phase diagram at finite temperature~\cite{Powell,Powell2}.  The effective potential for this order parameter fitted to the lattice QCD data at zero density~\cite{Fukushima2004,Rossner2007}, vanishes in the zero temperature limit. 
}) by using lattice QCD inputs or other phenomenological 
 considerations, see e.g.   \cite{Kunihiro:1991qu,Bratovic:2012qs,steinheimer,kenjitoru}.
In the following we describe only the situation with $g_V$ constant in the regime where quarks are relevant for neutron stars,  $n_B \gtrsim 5n_0$, since dependence of $g_V$ on the baryon density arises primarily from the modifications of the gluons by the matter, which 
 is argued to be small \cite{toru-screen}.
 
  Detailed results for the equation of state in the quark phase are given in Sec.~\ref{sec:3-window} and in Appendices~\ref{masses} and \ref{sec:para_EoS}.

\section{Constructing the neutron star equation of state: general considerations \label{sec:hybrid}  }

 Having laid out the basic physics in both the lower density hadronic regime and the higher density quark regime, we turn now to an examination of the general characteristics of the equation of state, paying particular attention to its stiffness and the corresponding implications for neutron star structure.   The most convenient thermodynamic potential for studying the equation of state is the pressure, $P$, as a function of the baryon chemical potential, $\mu_B$.  The pressure must be a continuous function of $\mu_B$, and since the baryon density is given by $n_B= \partial P/\partial \mu_B$, the pressure is monotonically increasing; furthermore the curvature, $\partial^2 P/\partial\mu_B^2 = \partial n_B/\partial \mu_B$, must be positive, else the system is unstable against density fluctuations.  The thermodynamically preferred phase maximizes the pressure at given $\mu_B$, and thus in the presence of competing phases the one with the higher pressure is favored.  In addition, a kink, or discontinuous change of slope, in $P(\mu_B)$ indicates a first order transition (illustrated in Fig.~\ref{fig:1storder} below), while a sudden change in curvature (with continuous slope) indicates a second order transition.

 \subsection{The stiffness of the quark matter equation of state \label{stiffequation of state}}

  We ask now the effects of the bag constant, pairing, and the vector repulsion on the stiffness of the quark matter equation of state.  To see the physics we write the total energy density in the schematic form
\beq
\varepsilon = A n_B^{4/3} +B - Cn_B^{2/3} + D n_B^2,
 \label{eq:energy}
\eeq
where the first term is the kinetic energy for massless quarks (we ignore corrections to the kinetic energy from finite quark masses), $B$ is the bag constant, $C \propto \Delta^2$  \cite{Alford:2004pf} measures the energy contribution of pairing, 
and $D \propto g_V$ measures the strength of the density-density repulsion, $g_V n_q^2$  \cite{Kunihiro:1991qu}.
We assume here that $B$ and $\Delta$ are density independent for simplicity (although at high density, $\Delta \sim \mu_q g^{-5}\exp(-3\pi^2/g\sqrt2 )$ \cite{damson}, where $g$ is the scale dependent QCD coupling constant); then
differentiating (\ref{eq:energy}) yields the baryon chemical potential,
\beq    
    \mu_B=\frac{\partial \varepsilon}{\partial n_B} =&& \frac43 A  n_B^{1/3} - \frac23 C n_B^{-1/3} + 2D n_B, \nonumber\\
\eeq  
and the pressure,
\beq
P =  n_B^2\frac{\partial (\varepsilon/n_B)}{\partial n_B} =\frac13 A n_B^{4/3} +\frac13 C n_B^{2/3}+ D n_B^2  - B. \nonumber \\
\label{pn}
\eeq

We note that in general a term in the energy density of the form $\alpha n_B^\gamma$, leads to a term in the pressure $(\gamma-1)\alpha n_B^\gamma$.  To see how  the pairing interaction and the repulsive interaction
 affect the stiffness of the equation of state, $P(\varepsilon)$, that is, the magnitude of the pressure for a given energy density,  we must evaluate their pressure at a fixed energy density.  Thus we ask how the pressure varies as $\alpha$ is varied at fixed energy density.   In order to keep $\varepsilon$ fixed with varying  $\alpha$ one must vary the density $n_q$;  thus 
\beq
    \frac{\partial P}{\partial \alpha}\Big|_\varepsilon &=&
     \frac{\partial P}{\partial \alpha}\Big|_{n_B} - c_s^2 \frac{\partial \varepsilon}{\partial \alpha}\Big|_{n_B} 
          = (\gamma -1 - c_s^2)  n_B^\gamma,
     \label{palpha}
\eeq
where $c_s^2 = \partial P/\partial \varepsilon|_\alpha $ is the square of the thermodynamic sound speed. 

Since for the repulsive interaction $D n_B^2$, one has $\gamma = 2$ and $\alpha=D$, 
we conclude that increasing the strength of this interaction will always lead to higher pressure and thus a stiffer equation of state, as long as  the causality condition,  $c_s^2 \le 1$ is satisfied. 
We also observe that increasing a density-independent pairing gap, $\Delta$ (corresponding to $\gamma=2/3$,
with $\alpha = - C$) increases the pressure: $\partial P/\partial C|_\varepsilon =  \left(\frac13 + c_s^2\right)n_B^{2/3}$, and stiffens the equation of state.

  We see from this exercise that the effect of interactions on the stiffness of the equation of state depends not only on the sign of the interaction, but also on the power of the density it involves.  Generally, as indicated in Eq.~(\ref{palpha}), increasing the strength of a repulsive interaction with $\gamma > 1 + c_s^2$ stiffens the equation of state, as does increasing the strength of an attractive interaction with $\gamma < 1+ c_s^2$.   A larger value of the bag constant, for which $\gamma = 0$, softens the equation of state.

\begin{figure}
\includegraphics[width = 0.4\textwidth]{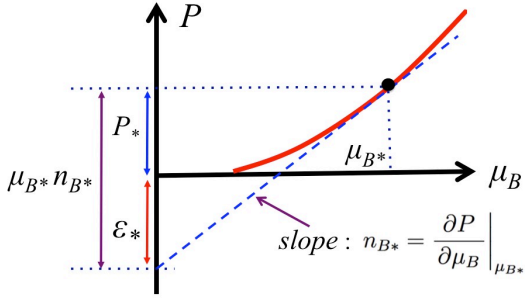}
\vspace{-0.5cm}
\caption{\footnotesize{ Graphical analysis of the equation of state, $P (\mu_B)$.  The slope of the tangent line a given point $(\mu_{B*}, P_*)$, is the baryon density, $n_{B*}=\partial P/\partial \mu_B |_{\mu_{B*}}$, and its intercept on the $P$ axis is the negative of the energy density $\varepsilon_*$.  }}
\vspace{-0.0cm}
\label{fig:stiffeg0}
\end{figure}

\subsection{Graphic determination of the stiffness of the equation of state \label{graphical}}
   
   Figure~\ref{fig:stiffeg0} demonstrates how to determine graphically the relative stiffness of an equation of state from its $P(\mu_B)$ curve \cite{Kojo:2015fua}. The slope of the curve at a given point $(\mu_{B*}, P_*)$ is the baryon density, $n_B$, for the specified  $\mu_B$.  Thus, from the zero temperature thermodynamic identity, $P = \mu_B n_B -\varepsilon$, we find that the tangent curve intercepts the $P$ axis at the point $ -\varepsilon_\ast$,  the negative of the energy density at chemical potential $\mu_{B*}$.  A stiff equation of state is characterized by a large pressure for given energy density $\varepsilon$ (or mass density $\rho = \varepsilon/c^2$), or equivalently by a small energy density for given pressure. 

\begin{figure}
\includegraphics[width = 0.45\textwidth]{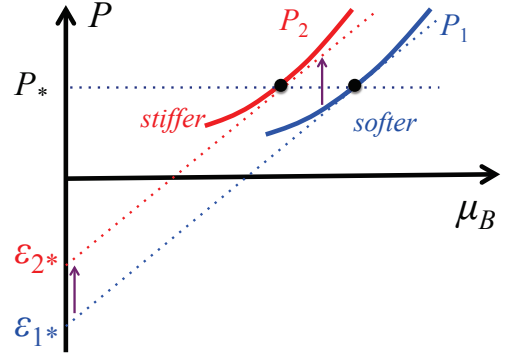}
\vspace{-0.0cm}
\caption{\footnotesize{Comparison of two equations of state in which the pressure curves have the same shape, but $P_2$ is shifted toward lower chemical potential, relative to $P_1$.   The equation of state $P_2$ is stiffer than $P_1$ because $\varepsilon_{1^\ast} < \varepsilon_{2^\ast}$.  A reduction of the bag constant, $B$, would lead precisely to such a shift from $P_1$ to $P_2$.}}
\vspace{-0.0cm}
\label{fig:stiffeg1}
\end{figure}

\begin{figure}
\includegraphics[width = 0.45\textwidth]{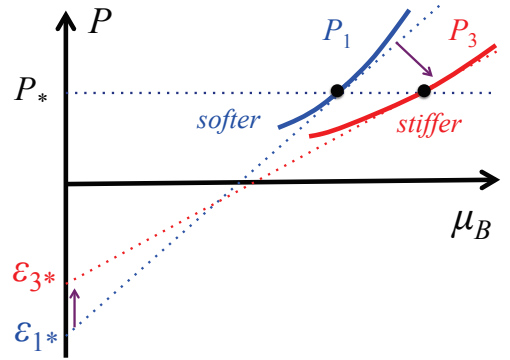}
\vspace{-0.0cm}
\caption{\footnotesize{Increasing the vector repulsion decreases the pressure at fixed $\mu_B$ and decreases the slope at fixed $P$.  The result is that $P_3$, with larger vector repulsion, is a stiffer equation of state than $P_1$.
}}
\vspace{-0.0cm}
\label{fig:stiffeg2B}
\end{figure}

\begin{figure}
\includegraphics[width = 0.45\textwidth]{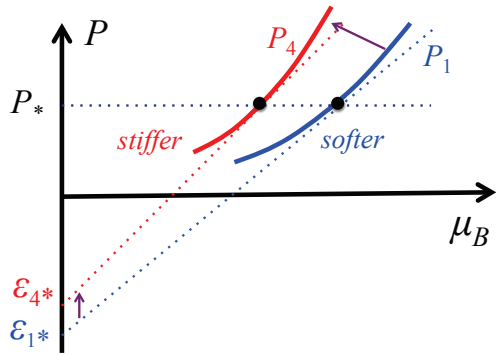} 
\vspace{-0.5cm}
\caption{\footnotesize{Increasing the pairing increases the pressure at fixed $\mu_B$ and  increases the slope at fixed $P$.  The net result is that $P_4$, with larger pairing, is stiffer than $P_1$.
}}
\vspace{-0.0cm}
\label{fig:stiffeg3}
\end{figure}

    The smaller the slope of $P$ at given $\mu_B$, the stiffer the equation of state.  Similarly, the smaller is $\mu_B$ for a given $P$ and slope, as illustrated by the two curves $P_1$ and $P_2$ in Fig.~\ref{fig:stiffeg1}, the stiffer is the equation of state.  To see the effects of the various terms in  Eq.~(\ref{eq:energy}) graphically, we write the terms generically, as before, as $\alpha n_B^\gamma$ and note the thermodynamic identity (cf. Eq.~(\ref{palpha})),
    \beq
    \frac{\partial P}{\partial \alpha}\Big|_{\mu_B}&=&
     \frac{\partial P}{\partial \alpha}\Big|_{n_B} - n_B \frac{\partial\mu_B}{\partial \alpha}\Big|_{n_B}    = -n_B^{\gamma},
     \label{palphamu}
\eeq
where we regard $n_B$ as a function of $\mu_B$.
Thus increasing the bag constant ($\gamma$=0) decreases the pressure at fixed $\mu_B$, and softens the equation of state, as illustrated in Fig.~\ref{fig:stiffeg1}.  Similarly, increasing the strength of the repulsive vector interaction ($ \gamma=2$) leads to a lower slope and a stiffer equation of state, as shown by the shift from $P_1$ to $P_3$ in Fig.~\ref{fig:stiffeg2B}.    On the other hand increasing the pairing strength increases the pressure at fixed $\mu_B$, and at the same time it increases the slope of $P$ vs. $\mu_B$, with the net effect of stiffening the equation of state, as illustrated by the shift from $P_1$ to $P_4$ in Fig.~\ref{fig:stiffeg3}.


\subsection{Hybrid equations of state}
\label{generalstructure}

\begin{figure}
\includegraphics[width = 0.4\textwidth]{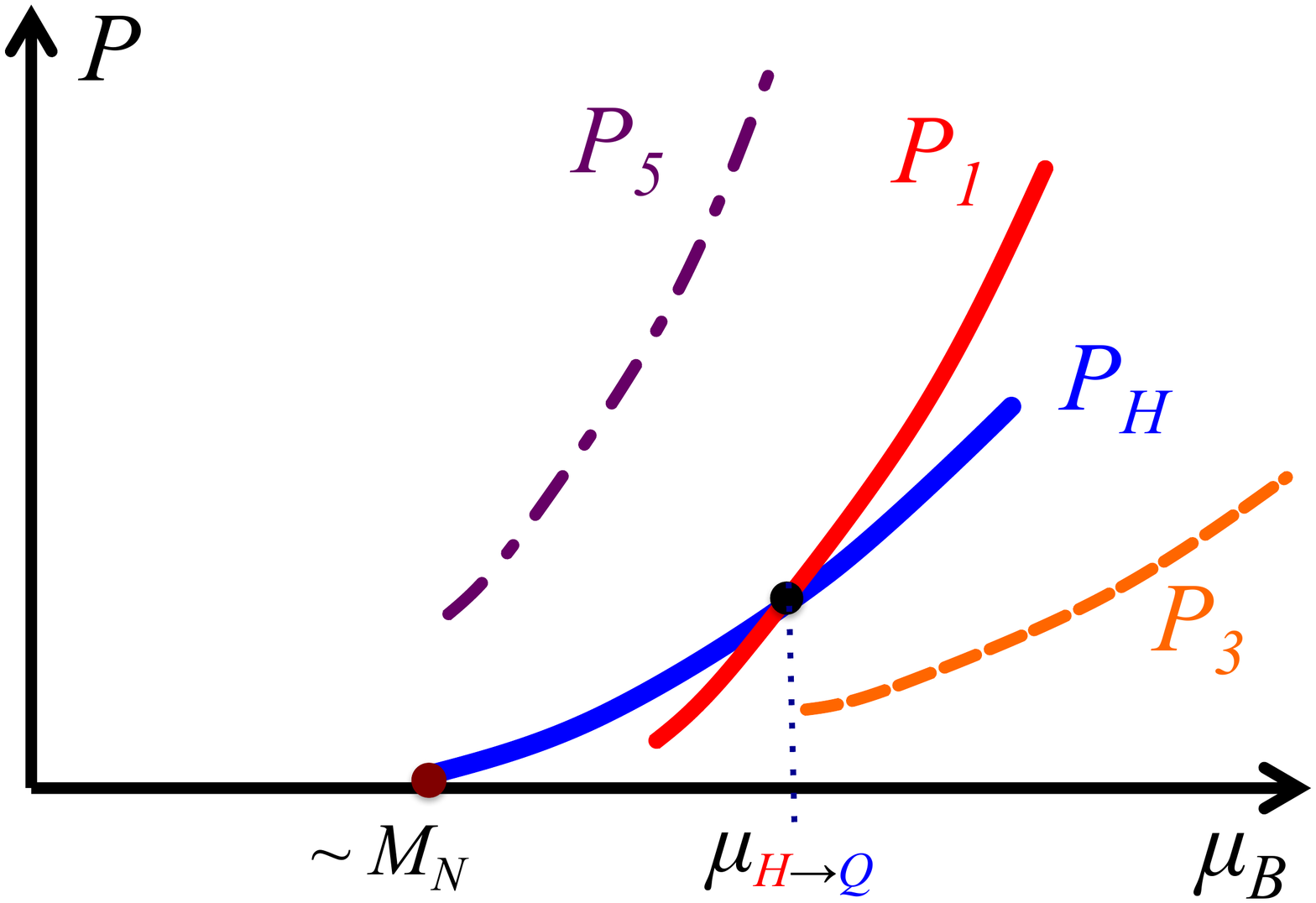}
\vspace{-0.5cm}
\caption{\footnotesize{Conventional construction of a hybrid equation of state from independent hadronic ($P_H$) and quark ($P_{1}$) equations of state.  In such a construction a first order phase transition occurs at $\mu_{H\rightarrow Q}$, and stiff quark equations of state,  e.g.,  $P_{5}$ (cf. $P_2$ in Fig.~\ref{fig:stiffeg1} and $P_4$ in \ref{fig:stiffeg3}) and $P_{3}$ (cf. Fig. \ref{fig:stiffeg2B}), which are incompatible with such a construction are excluded. }}
\vspace{-0.0cm}
\label{fig:hybrideg}
\end{figure}

    Hybrid equations of state assume that matter can be in one of two distinct phases, hadronic or quark.   The favorable phase, hadronic at low densities and quark at high densities, has the higher pressure at fixed chemical potential, with a first order transition between the two phases.    Figure~\ref{fig:hybrideg} shows the construction of a hybrid equation of state in $P$ vs.~$\mu_B$.    To ensure the existence of hadronic matter at low density one demands that the quark pressure, $P_Q$, intersects the hadronic pressure, $P_H$, from below, while to ensure the existence of quark matter at high density, one requires that $P_Q$ be greater than $P_H$ above the chemical potential where the two curves intersect.

   In Fig.~\ref{fig:hybrideg} we also show two quark equations of state, $P_{3}$ and $P_{5}$, which are stiffer than $P_{1}$ ($P_5$ corresponds to either $P_2$ in Fig.~\ref{fig:stiffeg1} or $P_4$ in Fig.~\ref{fig:stiffeg3}, while $P_3$
corresponds to the curve in Fig.~\ref{fig:stiffeg2B}).  The conventional hybrid construction rejects $P_{5}$ because it does not allow hadronic matter at low density.  (In the strange matter hypothesis, where three-flavor quark matter is assumed more stable than nuclear matter at low density \cite{Witten1984,Farhi1984,Alcock,Alpar} the equation of state would be of the form $P_{2}$.)  The relatively stiff quark equation of state $P_{3}$ would also be rejected in constructing a hybrid equation of state because it does not intersect the hadronic pressure $P_H$, and therefore would not in this construction be considered an acceptable model of quark matter at high density.  Thus, large classes of stiff quark matter equations of state -- precisely those consistent with stable massive neutron stars -- must be rejected in a conventional hybrid equation of state construction.  Quark equations of state consistent with such a construction are generally soft, so that within the conventional description, one concludes that massive neutron stars can at most have a small quark matter core (e.g., see Fig.~18 of Ref.~\cite{APR});  Ref.~\cite{AHP} summarizes qualitatively possible conditions on the quark matter equation of state that would support neutron stars of two solar masses. 

   In the conventional construction of a hybrid equation of state one looks for an intersection of the 
pressures of the hadronic and quark matter equations of state as functions of the baryon chemical potential (or equivalently the
energy per baryon, $\varepsilon/n_B$,  as functions of $1/n_B$, the volume per baryon), and makes a Maxwell construction to equate pressures and baryon chemical potentials between the two phases.   
An implicit assumption in this procedure is that both equation of states are reliable in the vicinity of the intersection.
 However, the typical intersection, corresponding to $n_B \sim (2-5) n_0$, 
 is exactly in the region where the hadronic equation of state becomes uncertain due to
 many-body forces and hyperon forces, and the quark equation of state becomes uncertain due to the effects of confinement.
The unified approach, to which shortly we turn in subsec.~\ref{3windowpicture}, allows one to relax 
 the requirement on the intersection
and obtain a description of dense matter which permits certain classes of  stiff equations of state with 
 quark matter.

  The fundamental problem with conventional hybrid equation of state constructions is that they assume that both hadronic and quark matter equations of state are reliable near their intersection.  When the intersection occurs at small chemical potential, one implicitly assumes that the quark pressure at low density is reliable; but no viable quark model calculations are available in the low density regime, due to the difficulty of modeling critically important confining effects.  Similarly, when the intersection occurs at large chemical potential, one accepts the hadronic equation of state at high density, where many-body forces are rapidly enhanced, rendering the equation of state seriously uncertain, even if one continues to assume that hadronic degrees of freedom correctly describe the matter.  One cannot reliably compare hadronic and quark matter pressures across the entire density domain.

  These considerations suggest that the conventional hybrid construction places overly stringent requirements on the form of quark matter equation of state by accepting the predictions of hadronic models above their regimes of validity.  The unified approach, to which shortly we turn in subsec.~\ref{3windowpicture}, allows one to relax these requirements and obtain a description of dense matter which permits certain classes of stiff quark equations of state.

\subsection{Thermodynamics of finite temperature QCD at zero baryon density \label{finiteT} }

       To further motivate the unified construction of the equation of state of cold dense matter, and the emerging role of hadron-quark continuity in the phase diagram, we briefly consider implications of the lattice results \cite{hotQCD,WB,Cheng2010,Borsanyi2010} for the structure of the equation of state at zero chemical potential, $P(T) = Ts-\varepsilon$, with $s$ the entropy density, above and below the crossover at zero chemical potential, Fig.~\ref{fig:HRG_sQGP_QGP}.     These calculations indicate that matter composed of light quarks with finite masses undergoes a rapid continuous crossover with increasing temperature from a hadronic to a quark-gluon phase at the pseudocritical or ``deconfinement" temperature $T_c\sim$(150-155) MeV, with smooth restoration of approximate chiral symmetry. 
(Were the 
$u,d,s$-quarks all massless, then at very high temperatures chiral symmetry would be completely restored in a first order phase transition,  distinguishing the symmetry broken phase and the symmetry restored phase (see e.g., \cite{Fukushima2011}). 

\begin{figure}
\begin{minipage}{0.95\hsize}
\includegraphics[width = 0.9\textwidth]{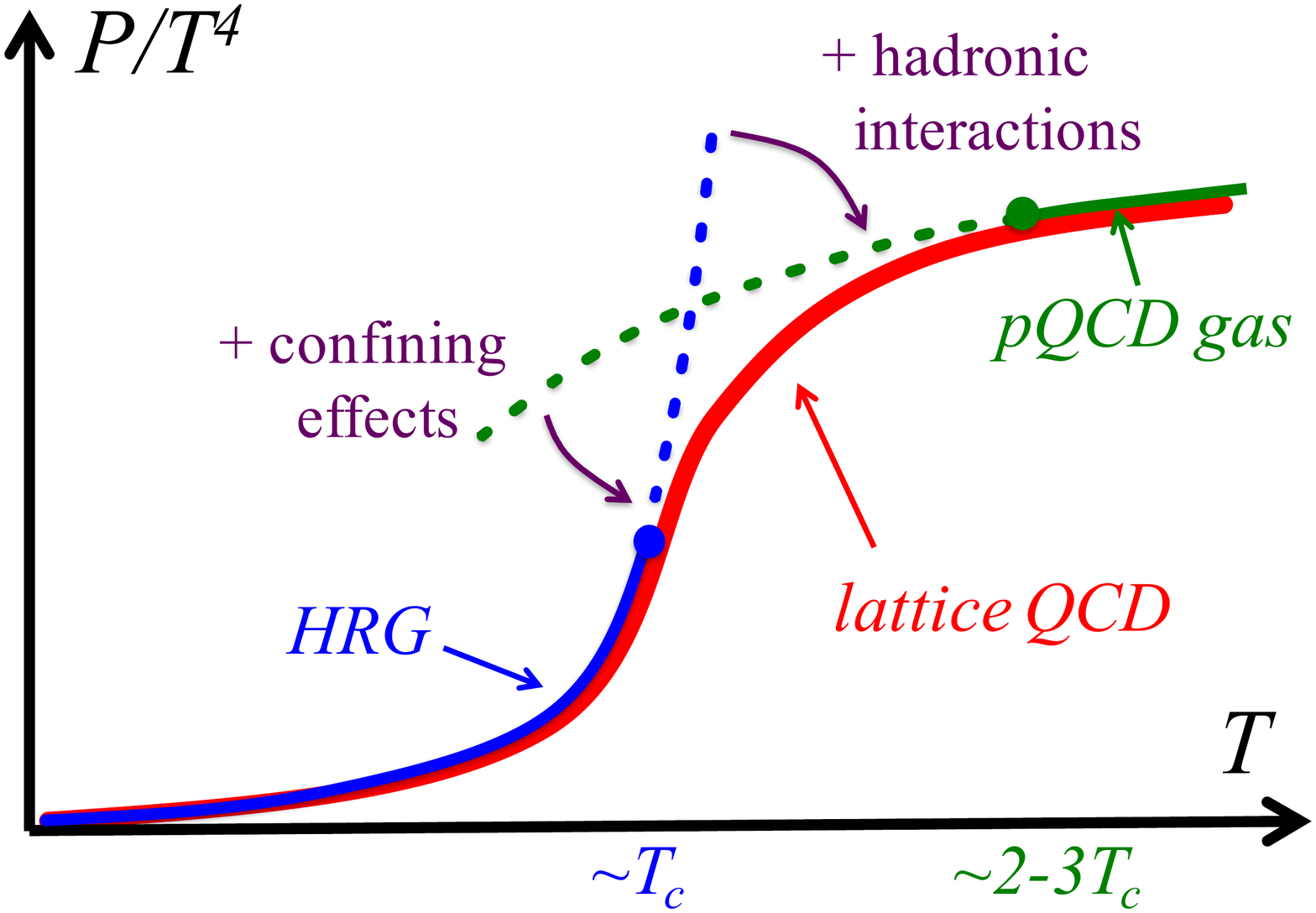}
\end{minipage}
\begin{minipage}{0.95\hsize}
\includegraphics[width = 0.9\textwidth]{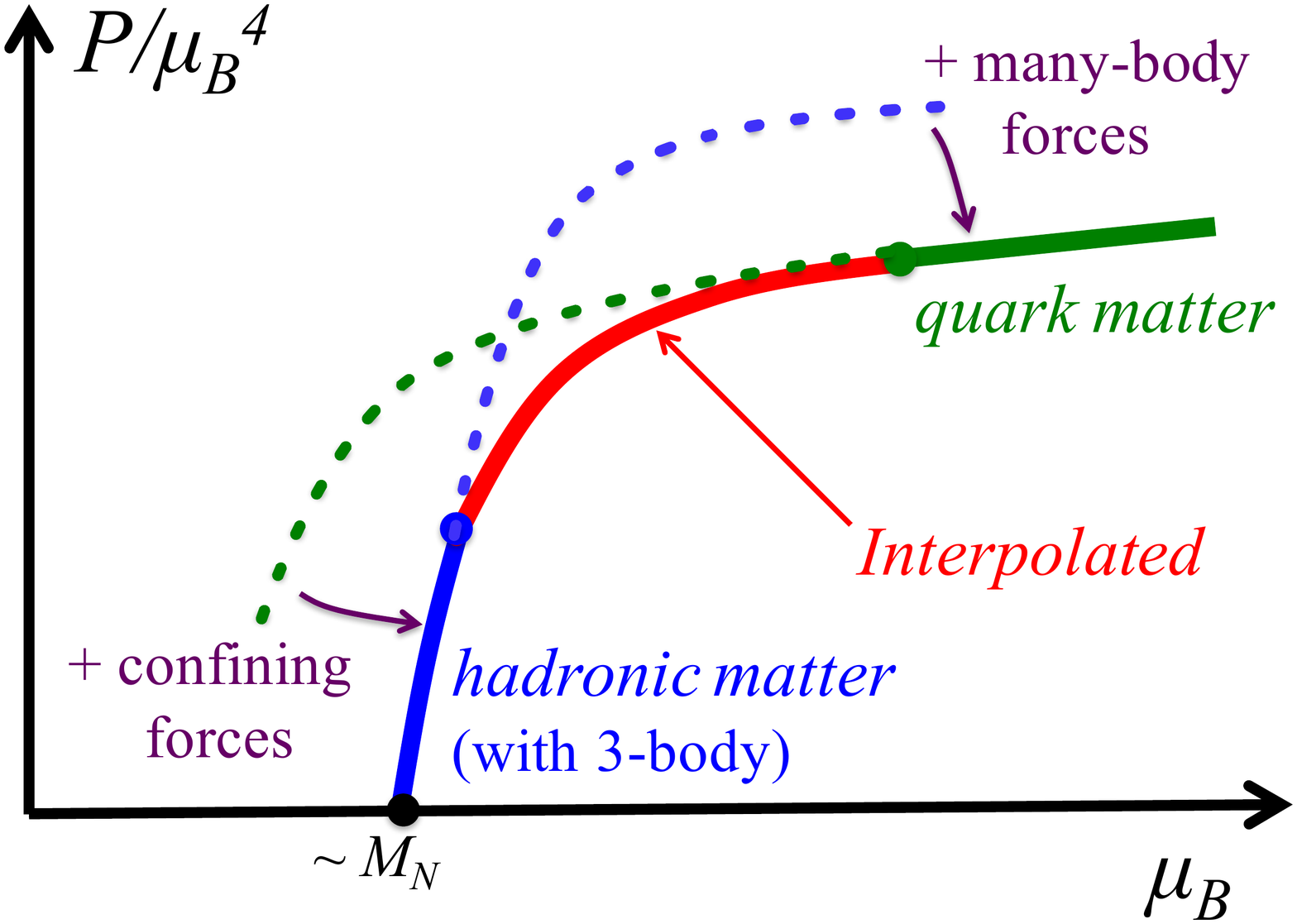}
\end{minipage}
\caption{
\footnotesize{
Schematic representations of the pressure of dense matter at finite temperature and zero baryon chemical potential (upper panel) and finite baryon density and zero temperature (lower panel).  The bold lines represent model predictions within the domain of applicability of each picture, while dotted lines show extrapolations beyond the domains of validity.  Upper panel: pressure of the hadron resonance gas (HRG), the perturbative quasiparticle gas of quarks and gluons (pQCD gas), and lattice QCD calculations.  Lower panel: pressure of hadronic matter with two- and three-body forces, and of a deconfined quark gas.  The interpolated pressure is constructed using the hadronic gas pressure below $\mu_{BL}$ and the quark gas pressure above $\mu_{BU}$.  }}\label{fig:finiteTmu}
\end{figure}

  At zero baryon density and low temperatures, matter is well described by the non-interacting hadron resonance gas (HRG) model \cite{hotQCD,WB,HRG}.   However, as the temperature approaches $T_c$  the hadron resonance gas model strongly overestimates the pressure compared with lattice calculations.   Physically, this discrepancy arises from the large overlap of thermally excited hadrons, whose interactions can no longer be neglected.   On the other hand, at temperatures $\gtrsim$ (2 -- 3) $T_c$, 
the matter can be described reasonably well by a weakly interacting quasiparticle picture of quarks and gluons, a perturbative or {\it pQCD} gas.\footnote{The lattice results do not appear to reproduce the non-interacting Stefan-Boltzmann limit, which one sees only at temperatures beyond those considered in the calculations.    One can understand this difference by considering terms in pQCD to order $g^5$, where $g$ is the QCD coupling constant.}
However, with decreasing temperature the gas pressure calculated by taking interaction effects into account perturbatively is well above that calculated on the lattice, since the lack of confining effects allows an artificially enhanced population of quarks and gluons without triggering the kinetic energy cost of their confinement into hadrons or glueballs.   The behavior of the extrapolated pressures are illustrated in the upper panel of Fig.~\ref{fig:finiteTmu}.

  While we understand both the  low and high temperature limits qualitatively, the intermediate temperature regime, $T_c \lesssim T \lesssim (2-3) T_c$, has a qualitative trend considerably different from the hadron resonance gas or the perturbative QCD gas models~\cite{Asakawa1995}. Matter in this region is a strongly correlated quark-gluon 
plasma~\cite{DeTar:1985kx,Hatsuda1985,Pisarski:2000eq,Dumitru:2010mj,Pisarski:2016ixt}.  The message from understanding finite temperature matter at $n_B=0$ 
is that were one simply to adopt the resonance gas picture at low temperature and the perturbative QCD picture at high temperature and then apply a Maxwell construction to find the phase transition between the two phases,
the transition would necessarily be first order and the resulting hybrid equation of state near the critical point
would depend highly on model artifacts (compare the intersection of the dotted lines in the top panel of Fig.~\ref{fig:finiteTmu} with the smooth lattice results).  However, if one instead restricts the use of each model equation of state to its domain of applicability, and interpolates between the two pictures one can obtain a physically sensible pressure.  By analogy, in finite density matter, as depicted in the bottom panel of Fig.~\ref{fig:finiteTmu}, 
a smooth interpolation between hadronic and quark matter allows a wider range of
equations of state, while avoiding the artifacts introduced by applying models beyond their regimes of validity.  One should keep in mind, however, that such construction depends on the choice
of the interpolating functions, which brings its own source of ambiguity.

\subsection{Hadron-quark continuity and percolation \label{perc} } 
      
   The existence of the crossover from the hadronic to the quark phase at zero baryon chemical potential raises a very instructive question \cite{qm2015-gb}, as it seems to imply that since quarks are free in the plasma phase above the crossover at low $\mu_B$ and finite $T$, by continuity free quarks would have a probability, albeit small, to be present in matter below the crossover, e.g., there could be free quarks running around in air.  This situation is reminiscent of the very tiny possibility of finding free electrons in air, as a result of thermal ionization.  But since there cannot be free quarks in confined low density matter, the correct conclusion is that even above the crossover, there are no free quarks (except in the very high $T$ asymptotically free regime); rather the matter must consist of complicated clusters of gluons and quarks both above and below the crossover, as illustrated in Fig.~\ref{fig:3-window} for matter at finite $\mu_B$ and zero temperature, and Fig.~\ref{fig:HRG_sQGP_QGP} for matter at $\mu_B=0$ and finite temperature.
   
   The crossover, and deconfinement more generally, can be characterized as a {\em percolation} transition -- in which the region in which the quarks can roam freely changes from finite in extent to the system size -- as first proposed in \cite{Baym1979} in terms of classical percolation theory for dense matter at zero temperature  and further amplified by Satz and coworkers \cite{Satz1,Satz2} in terms of quark mobilities; also \cite{lottini}.   
As seen in Fig.~\ref{fig:3-window}, percolation at finite $\mu_B$ and zero temperature proceeds as baryons exchange many quarks and lose their identities. Eventually baryons overlap and quark matter is formed.   The regions of space in which quarks can move around are color singlets, from nuclear to quark matter domain.  Similarly at finite temperature (for $\mu_B$  small compared with the nucleon mass)
the clusters, Fig.~\ref{fig:HRG_sQGP_QGP}, are isolated as single thermal pions which become more and more connected as $T$ increases,  through the gluon and quark exchanges responsible for the interactions of the pions, until the clusters fill enough of space that a single quark can propagate from one end to the other.  As at finite density, the regions of space in which quarks can move around are always net color singlets.  At the percolation transition the sizes of the color singlet regions change from always being finite in the hadronic regime to being the size of the entire system, e.g., the collision volume in a heavy-ion collision.\footnote{While it is easiest to visualize the transition as classical percolation in which regions of space available to the quarks overlap, a more precise picture of the transition is in terms of the probability of quarks being able to traverse the system, as in Fig.~ \ref{fig:3-window}.  One must take into account the quantum nature of the hadrons in discussing their overlap.   For example even though the electron wavefunction in a simple hydrogen atom has an exponentially small tail extending to infinity, one would not claim that any two hydrogen atoms in a gas at whatever distance are always overlapping.  Similarly, in hadrons, the pion and other $\bar q q$ structures (as well as the wee partons) extend well beyond the hadron core.   Understanding the quantum percolation transition in dense matter and its relation to Anderson localization of particles in a disordered system remains an open question \cite{soukoulis}.}

   A critical aspect of hadron-quark continuity is the possibility that the quark and hadron phases have the same symmetry structure.  While  phases with different (exact) symmetries are separated by a first or second order phase transition, the symmetries of the superfluid baryon phase in hadronic matter can, as Sch{\"a}fer and Wilczek \cite{Schafer:1998ef} elegantly discussed, be smoothly connected to those in the CFL phase in quark matter; thus there need not be a sharp phase transition separating the hadron and quark matter phases.\footnote{Hadron-quark continuity in states of finite angular momentum between hadronic matter and superfluid quark matter in the CFL phase is more subtle.  While superfluid hadronic matter and quark matter each carry angular momentum in quantized vortices, owing to quarks having baryon number 1/3, a triply quantized vortex in the hadronic regime would carry the same angular momentum per baryon as a singly quantized $U(1)_B$ vortex in the quark regime \cite{iida3}.  Thus in a rotating neutron star in which the nuclear superfluid evolves with depth into a CLF quark phase, one would at first expect that at some point a surface of boojums (where three low density hadronic vortices merge into one high density CFL vortex) between the low density hadronic and high density quark matter regions.  
However, in the CFL phase, a single $U(1)_B$ vortex is unstable against transforming into three color flux tubes \cite{balachandran,
taeko,alford-vort}, suggesting that at the boojum three low density vortices transform into three color flux tubes \cite{boojum}. In fact, however, each low density vortex can transform directly to a single such non-Abelian vortex without a boojum, consistent with hadron-quark continuity \cite{fluxtube}. }

   In contrast, the conventional picture of dense nuclear matter is that there exists a first order (chiral) phase transition beginining at zero temperature and terminating at a high temperature the critical point \cite{asakawa,Stephanov:2004wx} (see Fig.~\ref{phases}), as in a liquid-gas phase transition where one can go continuously from liquid to gas around the critical point  (in water at 373 C).  If the low temperature part of the first order line is in fact a crossover, there must exist a second, low temperature critical point (as shown in Fig.~\ref{phases}) \cite{Hatsuda2006,Kitazawa:2002bc,Zhang:2008wx}.   It is also possible that the crossover at high temperature and low baryon density is directly connected to the low temperature crossover, without the need for the conventional first order line.\footnote{ The Asakawa-Yazaki critical point is being searched for in experimental programs at the RHIC heavy ion collider at Brookhaven National Laboratory \cite{Aggarwal:2010cw}, the SPS  at CERN \cite{Abgrall:2014xwa}, SIS at GSI in Germany, and will be searched in the future program at FAIR at GSI, NICA at JINR in Dubna \cite{NICA}, and J-PARC in Japan \cite{J-PARC}. Recently possible experimental signatures for the (conventional) critical point were found in analyses for the critical fluctuations \cite{Luo:2015ewa} and the finite volume scaling \cite{Lacey:2014wqa}.  Owing to controversies in the interpretation of those results \cite{Bzdak:2013pha,Bzdak:2016qdc,Kitazawa:2016awu}, further studies are called for.  It should be emphasized that the current state-of-the-art lattice QCD studies based on a Taylor expansion in $\mu_B/T$ around $\mu_B=0$ disfavor a critical point in the region where the expansion is trustworthy, $\mu_B/T \lesssim 2$ \cite{Ding:2015ona}. So far the existence of the first order phase transition has not been 
established.}

   As matter goes from having hadronic to quark degrees of freedom, it may pass through spatially inhomogeneous phases
 \cite{sasaki,buballa3}.  While this is an intriguing possibility, we will not treat it in this review, except to mention one unconventional state,  {\em quarkyonic} matter, which has both aspects of nuclear and quark matter \cite{McLerran:2007qj}.   This state is 
conjectured from studies of dense matter in QCD with a large number of colors $N_c$;  in this limit  screening  of gluons by quarks is suppressed by a numerical factor $1/N_c$, and thus
the gluons remain confined until the quark chemical potential $\mu_q = \mu_B/N_c$ reaches 
 $ \sim N_c^{1/2}\lqcd \gg M_N/N_c$.  In the limit of a large number of colors, the dominant pairing in the quarkyonic matter is,  instead of diquark pairing, the formation of a spatially inhomogeneous chiral condensate of quark particle-hole pairs, called chiral spirals.
 This idea is applied to neutron stars in Ref.~\cite{kenjitoru}.   In the real  world with $N_c=3$,  the 
 extent to which gluons are screened  due to quark excitations remains unclear \cite{RSS,toru-screen}.

\subsection{Unified construction \label{3windowpicture} }

   In the \textit{unified} procedure to construct the equation of state one explicitly restricts the hadronic and quark matter equations of state to their respective domains of validity, avoiding the potentially unphysical implications of the conventional construction. 
 The hadronic equation of state is used only at low densities, $n_B < n_{B L}$ (``L" for lower), where two- and three-body forces dominate and the composite nature of hadrons is not manifest.  A reasonable choice of the maximum density  $n_{B L}$ is $\sim 2 n_0$.  We denote the  corresponding chemical potential as $\mu_{BL}$.   Similarly, the deconfined quark-matter equation of state is used only at relatively high densities, beyond where baryons first percolate and quarks can no longer be thought of as belonging to specific baryons.  For a typical baryon radius of $r_B \sim 0.5$ fm, geometric percolation should occur at a baryon density $\sim 0.08/r_B^3 \sim 4n_0$ \cite{Baym1979}, and thus a reasonable choice of the lowest density at which to use a quark matter equation of state is  $n_{BU}\sim (4-7) n_0$; we label the corresponding chemical potential $\mu_{BU}$ (``U" for upper).  [In the calculations in Sec.~\ref{sec:3-window}  we choose $n_{BU} = 5 n_0$ as a specific illustrative value.]

  In the density range $n_{BL} < n_B < n_{BU}$ neither a purely hadronic nor quark matter picture is applicable.  Given the present intractability of directly calculating the equation of state in this domain, a simple approximate approach is to interpolate $P(\mu_B)$ between the two limiting regimes in a thermodynamically consistent way, requiring that the interpolated pressure matches the hadronic and quark values at $\mu_B^L$ and $\mu_B^U$, while satisfying the thermodynamic constraint $\partial n_B/\partial \mu_B = \partial^2 P/\partial \mu_B^2 >0$, as well as the (reasonable) causality condition that the adiabatic speed of sound at zero frequency, $c_s^2 =\partial P/\partial \varepsilon$ not exceed the speed of light.\footnote{However,  Refs.~\cite{ruderman-bludman,mal,caporaso,Ellis2007} indicate that this requirement of causality on $c_s$ is still suggestive; we are not aware of a rigorous proof that $c_s \le c$ is necessary for causal propagation of signals and information.  In particular, Lorentz invariance itself does not impose such a constraint \cite{ruderman-bludman,Ellis2007}, and it is possible to devise models that exhibit superluminal sound speed $c_s$ (and sometimes even superluminal group velocity as well), yet with causal propagation of signals \cite{ruderman-bludman,mal,caporaso}.   
 The first argument that the sound speed in dense matter can exceed $c/\sqrt3$ is given by Zel'dovich \cite{zeldovich}.  The assumption of $c_s\le c$ was used early on to obtain a maximum possible neutron star mass of 3.2 $M_\odot$ \cite{rhoadesruffini}, compared with a mass of $\sim 5 M_\odot$ in \cite{sabbadini} which did not assume $c_s\le c$;  refined upper bounds were given in \cite{hartle1978,vicky}.}
  These conditions place significant restrictions on the acceptable interpolations of the pressure in the intermediate density regime, and provide  insights into the qualitative properties of this critical domain in neutron star structure.
  
    As noted, the primary distinction between hybrid and unified constructions of the equation of state is that in the latter no direct comparison of the hadronic and quark pressures is made, since the domains of validity of the hadronic and quark descriptions do not overlap.  Accordingly, a number of the stiff quark matter equations of state excluded by the conventional construction (see Fig.~\ref{fig:hybrideg} and related discussion above) are allowed within a unified construction.  Furthermore, the unified construction can encompass hadron-quark continuity. 
    
   A simple but reasonably general function to interpolate the equation of state between the hadronic and the quark matter regimes
is a polynomial which smoothly joins the hadronic and quark pressure curves between $\mu=\mu_{BL}$ and $\mu=\mu_{BL}$,
\beq
\calP (\mu_B) = \sum^N_{m=0} C_m \mu_B^m \hspace{5mm}   \mbox{for}   \hspace{5mm}   \mu_{BL} < \mu_B < \mu_{BU},
\label{Pinterp}    
\eeq
where $\mu_{BL}$ and $\mu_{BU}$ are chosen so that  $n_B(\mu_{BL}) \sim 2n_0$ and $n_B(\mu_{BU}) \sim 5n_0$. The coefficients $C_m$ are chosen to satisfy matching conditions at the boundaries of the interpolating interval.  In general, we require that 
\beq
&&\calP (\mu_{BL}) = P_H (\mu_{BL})\,,~
~ \frac{\partial \calP }{ \partial \mu_B } \bigg|_{\mu_{BL}} = \frac{ \partial P_H }{ \partial \mu_B }\bigg|_{\mu_{BL}} ~,
\cdots
\nonumber\\
&&\calP (\mu_{BU}) = P_Q (\mu_{BU})\,, ~ 
~\frac{\partial \calP }{  \partial \mu_B }\bigg|_{\mu_{BU}} = \frac{ \partial P_Q }{ \partial \mu_B} \bigg|_{\mu_{BU}} \, ,\cdots  .\nonumber\\
\eeq
The number of derivatives to be matched at each boundary is a matter of choice.  Matching up to the second derivative at each boundary  ensures that the pressure, baryon number density, and baryon number compressibility (or susceptibility), $\partial n_B/\partial \mu_B$, are continuous. 
In this case one has six boundary conditions so one needs to include polynomials up to $N=5$. 

\begin{figure}
\includegraphics[width = 0.4\textwidth]{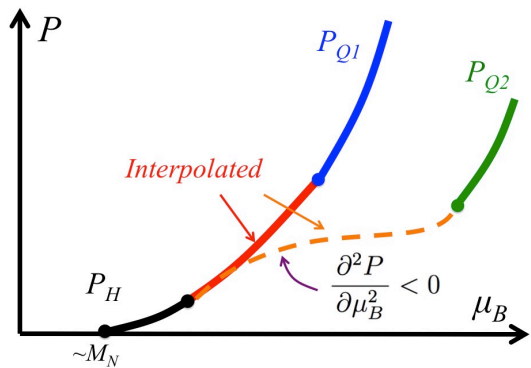}
\vspace{-0.0cm}
\caption{\footnotesize{Schematic interpolation of the hadronic ($P_H$) and quark ($P_{Q1}$, $P_{Q2}$) equations of state.  For $P_{Q1}$ the interpolated pressure is physically acceptable.  However, for $P_{Q2}$ one cannot construct an interpolated pressure without introducing an inflection point.  Such an unphysical feature implies a mechanical instability and $P_{Q2}$ must therefore be discarded from consideration.
}}
\vspace{-0.0cm}
\label{fig:intereg1}
\end{figure}

\begin{figure}
\includegraphics[width = 0.4\textwidth]{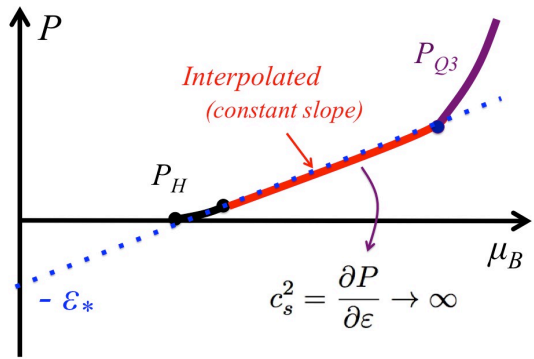}
\vspace{-0.0cm}
\caption{\footnotesize{Schematic interpolation violating the causality conditon.  The linear interpolated pressure implies a constant baryon and energy density (see Fig.~\ref{fig:stiffeg0}); the latter condition leads to the unphysical result $c_s^2 =\partial P/\partial \varepsilon \rightarrow \infty$.
}}
\vspace{-0.0cm}
\label{fig:intereg2}
\end{figure}

   As discussed in Sec.~\ref{3windowpicture}, the interpolated pressure as a function of $\mu_B$ is constrained by the stability condition that $P(\mu_B)$ be without an inflection point, and the requirement that $c_s^2/c^2 = \partial P/\partial \varepsilon = (1/c^2) \partial \ln \mu_B/\partial \ln n_B \le 1$,  so that the thermodynamic sound speed does not exceed the speed of light.   
The stability condition is depicted in Fig,~\ref{fig:intereg1}, where one can interpolate smoothly between the schematic hadronic equation of state $P_H$ and the quark matter equation of state $P_{Q1}$, with $\partial^2 P/\partial \mu_B^2 >0$; however, an interpolation between $P_H$ and the quark matter equation of state $P_{Q2}$ necessarily has an inflection point, violating the stability criterion.   A situation in which the speed of sound exceeds the speed of light is shown in Fig.~\ref{fig:intereg2}, where the slope is constant; in this region $P$ varies but $\varepsilon$ remains constant, so that $c_s^2 = \partial P/\partial \varepsilon \rightarrow \infty$. In general a $P(\mu_B)$ which grows too slowly violates the causality condition. 


\subsection{First order phase transitions  \label{sec:1storder} }

\begin{figure}
\includegraphics[width = 0.4\textwidth]{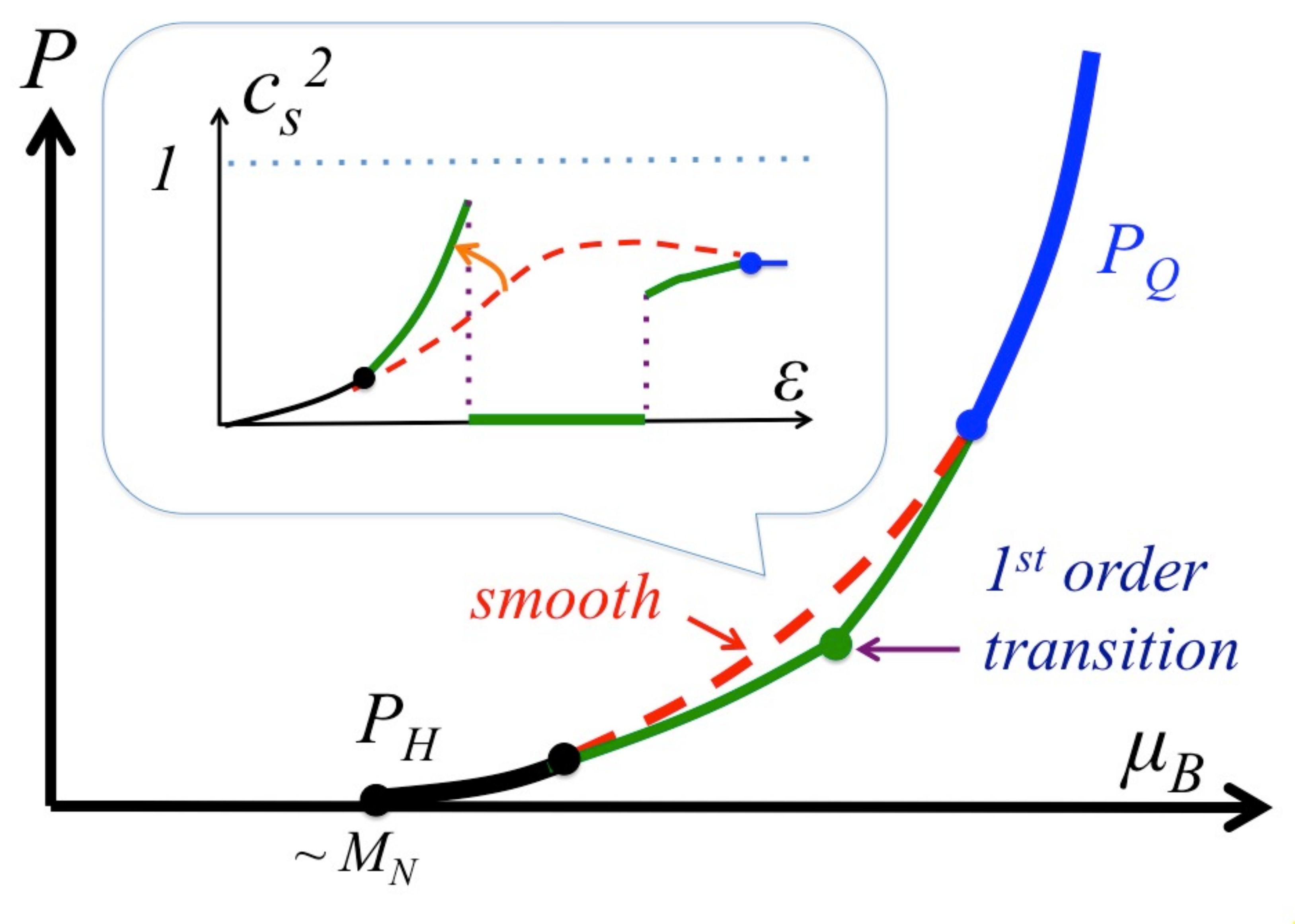}
\vspace{-0.0cm}
\caption{\footnotesize{Interpolated equation of state for both hadron-quark continuity (smooth curve) and a first order phase transition (kinked curve).  In the latter case, for which $P(\mu)$ must lie below the smooth curve, the slope increases discontinuously at the transition.  In the small panel we show the corresponding dependence of the sound speed on the energy or mass density; generally, the kinked curve yields a larger sound speed in the hadronic regime, with the sound speed vanishing in the transition region.
} }
\vspace{-0.0cm}
\label{fig:1storder}
\end{figure}

The possibility of a first order transition is not excluded in the unified construction.  However, such a transition more severely constrains the interpolated pressure than does a continuous hadron-quark evolution.   Figure~\ref{fig:1storder} compares the interpolated pressure curves for a continuous (smooth curve) and a first-order (kinked curve) hadron-quark phase transition.  The boundary matching conditions severely restrict the strength of a possible first order phase transition, as measured by the change in slope of the hybrid pressure curve at the transition.  In particular, except for relatively small changes in slope it is impossible to match the slopes of both the hadronic and quark matter curves without introducing an unphysical inflection point.  In addition, the slope of the hadronic pressure curve is smaller in the case of a first order transition than for a continuous evolution, giving rise to a correspondingly larger sound speed.  Thus, for specified hadronic and quark matter equations of state, an
interpolation with a kink, a first order transition, has a greater chance of violating the causality constraint.  These conditions strongly restrict both the location and strength of a possible first order hadron-quark transition \cite{Alford:2004pf}.  This constraint, a consequence of accounting for massive neutron stars, would indicate that there should not be a strong first order chiral restoration transition in QCD at low temperature (see subsec.~\ref{perc}).

\section{Explicit construction of unified equations of state \label{sec:3-window} }

    The two fundamental ingredients in constructing a unified equation of state that can explain neutron stars of masses $\gtrsim 2 M_\odot$ are first a large vector repulsion $g_V$ and second a large diquark pairing interaction $H$.   
As we will discuss, the characteristic
diquark coupling, $H$, is larger than the conventionally assumed $H/G =3/4 $ expected from the Fierz transform of the one-gluon-exchange interaction.  In fact, 
     en route with decreasing density to the strong two and three quark correlations that eventually become-well defined nucleons, the system may have larger pairing correlations than at higher density, so that
      $H/G > 1 $ (and  $\Delta \sim 100-200\, {\rm MeV}$)   at densities of interest. 
         
\begin{figure}
\begin{minipage}{1.0\hsize}
\includegraphics[width = 0.9\textwidth]{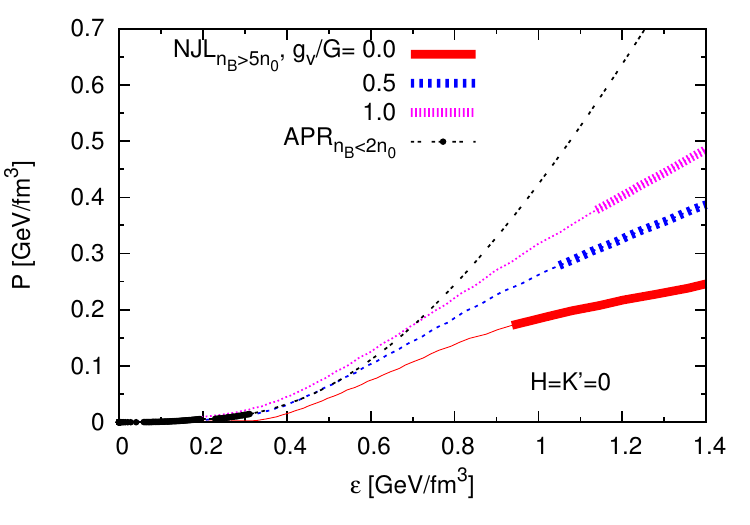}
\end{minipage}\\
\begin{minipage}{1.0\hsize}
\includegraphics[width = 0.9\textwidth]{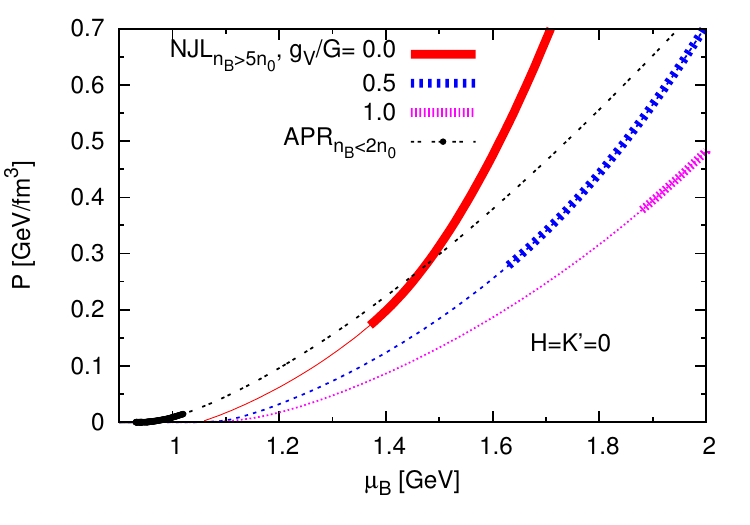}
\end{minipage}
\caption{\footnotesize{ Effects of the vector repulsion on the pressure vs. energy density (top) and pressure vs. quark chemical potential (bottom) for vector couplings $g_V/G=0, 0.5, 1.0$, without pairing, $H=K'=0$.  The NJL curves are shown as bold lines for $n_B>5n_0$, and as thin lines below.   The APR equation of state (solid line for $n_B<2n_0$ and double dotted line above) is also plotted for comparison.  One sees here clearly in the upper panel how increasing $g_V$ stiffens the equation of state.  }}
\label{fig:mu_np_H0}
\end{figure}

\begin{figure}
\includegraphics[width = 0.47\textwidth]{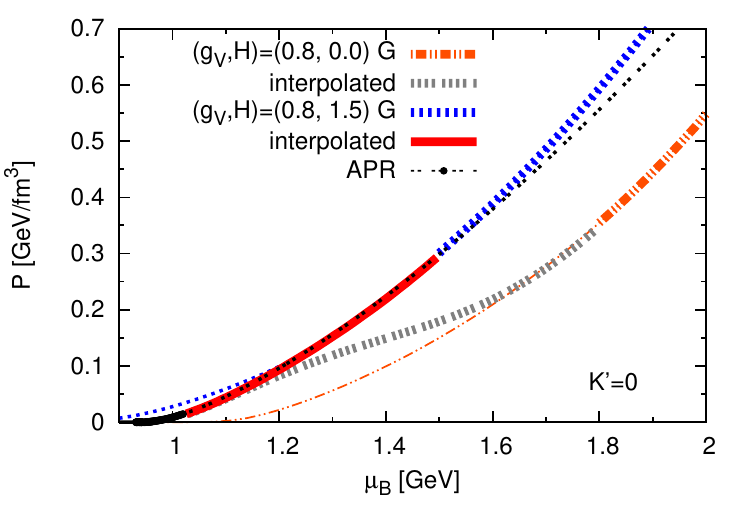}
\caption{\footnotesize{Pressure vs. baryon chemical potential for interpolated equations of state, for parameters  $g_V =0.8 G$, with $H=0$ and 1.5 $G$, and $K'=0$.  At $H=0$ the interpolated equation of state has an unstable region, $ \partial n_B/\partial \mu_B <0$.}}
\label{fig:inter6mu-P}
\end{figure}

\begin{figure}
\includegraphics[width = 0.47\textwidth]{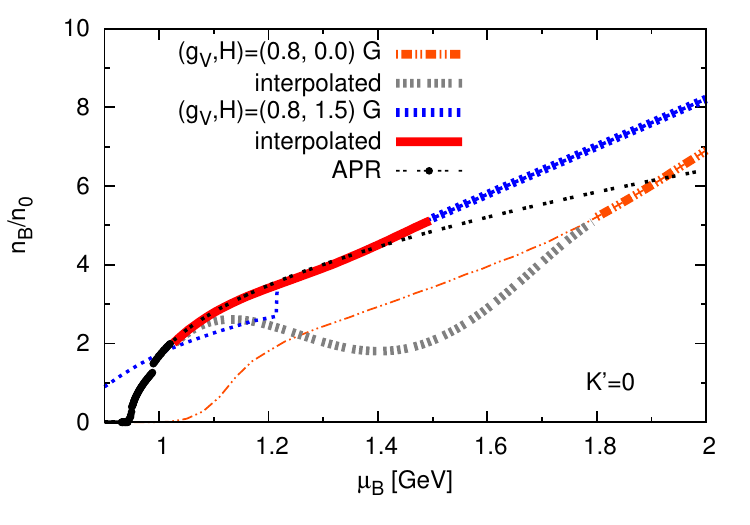}
\caption{\footnotesize{Baryon density vs. chemical potential for interpolated equations of state for the same parameters as in Fig.~\ref{fig:inter6mu-P}. } The discontinuity in the APR curve at $\mu_B \sim 1$ GeV is a result of the onset of neutral pion condensation, and that in the dotted curve at $\mu_B \sim 1.2 $ GeV is the transition from 2SC to CFL color superconductivity.}
\label{fig:inter6mu-n}
\end{figure}

       The vector repulsion (\ref{vec}) plays a major role in stiffening the equation of state calculated in the NJL model, as shown in  Fig.~\ref{fig:mu_np_H0}.    While the equation of state is considerably softer than the nucleon-based APR equation of state for small vector couplings, for sufficiently large $g_V$, it can be as stiff as APR across a wide range of densities,  thus enabling NJL equations of state at sufficiently large $g_V$ to explain massive neutron stars.   However, increasing $g_V$ moves the NJL pressure curve, as a function of $\mu_B$, away from the APR pressure curve in the positive $\mu_B$ direction, making it harder to interpolate between the two phases without introducing an unphysical inflection point, as seen in the $P(\mu_B)$ curve for $H=0$ in Fig.~\ref{fig:inter6mu-P}.  This figure shows the interpolated $P (\mu_B)$ for $H= 0$ and $~1.5\,G$, with $g_V=0.8\,G$ and $K'=0$. For $H=0$, the APR and NJL curves are rather widely separated in $\mu_B$ and it is difficult to construct a sensible interpolated equation of state; Fig.~\ref{fig:inter6mu-n} shows the corresponding plots of $n_B$ vs. $\mu_B$, while Fig.~\ref{fig:mu_np_H0} shows the pressure as a function of energy or mass density.    With increasing $H$ the NJL pressure toward lower chemical potential, enabling one, with large $g_V$ and $H$, to construct a physical interpolation between the hadronic and quark 
regions.\footnote{Were we to include in the quark phase a possible negative residual pressure, $P_g$, originating from a (non-perturbative) condensate of gluons, for $P_g$ as large as $\sim \lqcd^4 \sim 200\, {\rm MeV /fm^3}$,  the total  pressure of the quark matter would become too soft to allow a sensible interpolation between the APR and NJL models.}


\begin{figure}
\includegraphics[width = 0.48\textwidth]{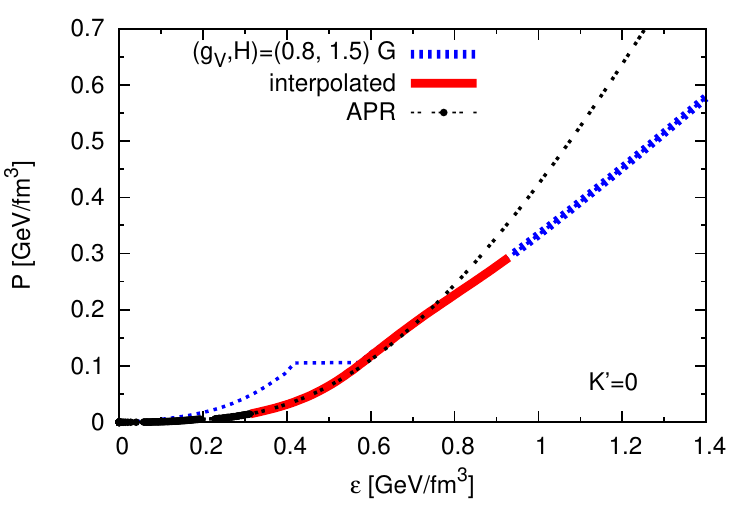}
\caption{\footnotesize{Pressure vs. energy density for interpolated equations of state, with parameters.  $g_V = 0.8G$,,  $H = 1.5G$, $K'=0$.   The flat region in the pressure  reflects the transition from 2SC to CFL color pairing.}} 
\label{fig:inter6e-p}
\end{figure}

    To demonstrate explicitly the construction of a unified equation of state, we assume the APR hadronic equation of state below a baryon density, $n_L \equiv n_B(\mu_{BL}) \simeq 2n_0$, above which the underlying hadronic description begins to break down.  We also assume a quark matter equation of state above a density  $n_U \equiv n_B(\mu_{BU}) \simeq 5 n_0$, to choose a specific representative value, and carry out a polynomial interpolation for $\mu_B$ between $\mu_{BL}$ and $\mu_{BU}$.   Within the range of the NJL model parameters we discuss, the variation of $n_U$ from $4n_0$ to $10n_0$ does not produce significant qualitative changes in the resulting equation of state.  
        
\begin{figure}
\includegraphics[width = 0.45\textwidth]{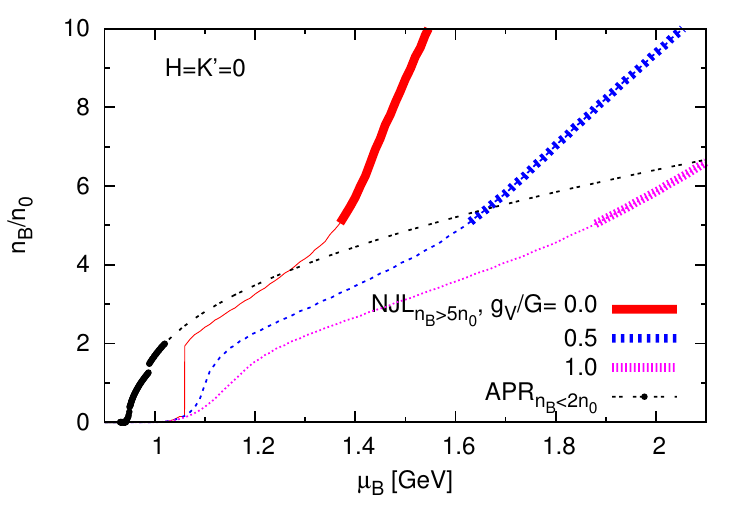}
\caption{
\footnotesize{Effects of the vector repulsion on the baryon density $n_B$, in units of nuclear matter density,  $n_0=0.16\, {\rm fm}^{-3}$, as a function of the baryon chemical potential, for vector couplings $g_V/G$ = 0, 0.5 and 1.0. The baryon density in the APR nucleon equation of state is shown for comparison. The bold line at  $n_B >5n_0$ is the quark pressure, and at $n_B<2n_0$ the APR pressure.  The discontinuity in the solid part of the APR curve is at their onset of pion condensation. The sudden rise of the curve for $g_V=0$ indicates chiral symmetry restoration transition in the NJL model.}
}
\label{fig:gV_n} 
\end{figure}

    The detailed effects of the vector repulsion on $n_B$ as a function of $\mu_B$ are shown in Fig.~\ref{fig:gV_n}; as expected, increasing the repulsion decreases $n_B$ for given $\mu_B$.   To investigate the effects of vector repulsion on the equation of state, we explore a range of couplings, $0.5G \le   g_V \le 1.0 G$.
Increasing the vector repulsion also reduces the tendency of high densities to restore chiral symmetry, an effect discussed in Appendix \ref{masses}.  When $g_V$ exceeds a critical value, dependent on $H$, one cannot interpolate between the hadronic and quark regimes without introducing a mechanical instability, as seen in Fig.~\ref{fig:inter6mu-P}.

   We next discuss how the interpolation depends on the model parameters. Figure~\ref{fig:inter6mu-P} shows the interpolated $P (\mu_B)$ for $H= 0$ and $1.5\,G$, with $g_V=0.8\,G$ and $K'=0$. For $H=0$, the APR and NJL curves are rather widely separated in $\mu_B$ and it is difficult to construct a sensible interpolated equation of state; Fig.~\ref{fig:inter6mu-n} shows the corresponding $n_B$ vs. $\mu_B$.    With increasing $H$ the NJL pressure curve shifts toward lower chemical potential, enabling one to construct a physical interpolation between the hadronic and quark regions by employing large $g_V$ and $H$.

\begin{figure}
\begin{minipage}{1.0\hsize}
\includegraphics[width = 0.92\textwidth]{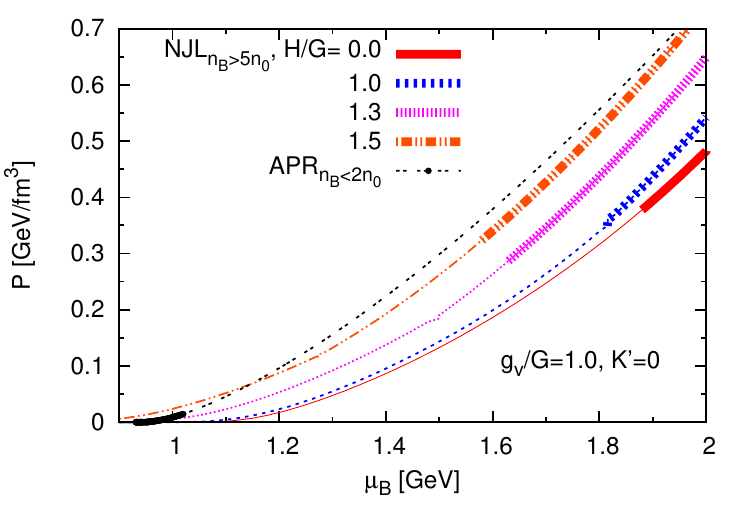}
\end{minipage}\\
\begin{minipage}{1.0\hsize}
\includegraphics[width = 0.9\textwidth]{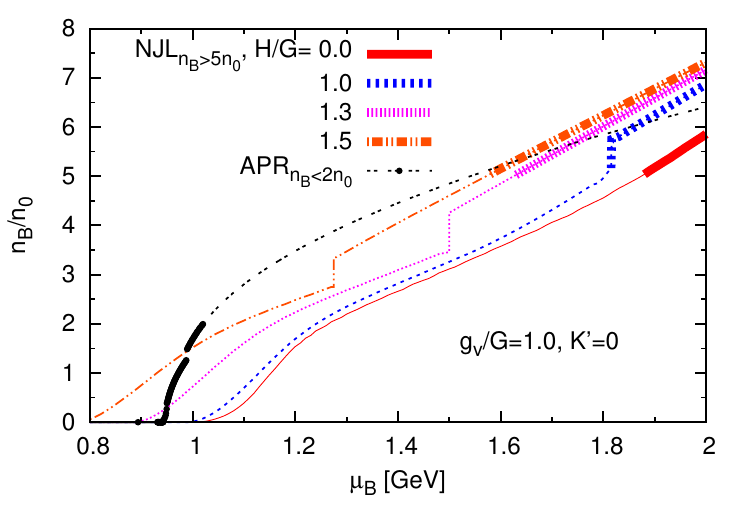}
\end{minipage}
\caption{
\footnotesize{Effects of diquark pairing on the pressure $P$ (top panel) and normalized baryon density $n_B/n_0$ (bottom panel) as functions of the baryon chemical potential, for vector coupling $g_V=1.0\,G$, axial anomaly coupling $K^\prime = 0$, and diquark couplings $H/G$ = 0, 1.0, 1.3, and 1.5.  With increasing $H$, the $P$ and $n_B$ curves shift toward lower chemical potential.  }
}\label{fig:mu_np_H}
\vspace{-0.2cm}
\end{figure}

   The diquark pairing interaction significantly affects the pressure; a larger $H$ leads to larger pairing gaps, which as discussed in subsec.~\ref{stiffequation of state}, leads to a stiffer equation of state  (for $c_s^2 > 1/3$, as is the case in the high density quark regime for $g_V>0$).   This effect is illustrated in Fig.~\ref{fig:mu_np_H}, where with increasing $H$ the pressure as a function of $\mu_B$ is shifted toward lower chemical potential, tending to eliminate the unphysical inflection point.   In addition,  increasing $H$ increases the baryon density at given $\mu_B$.  Both behaviors result from the reduction, with increasing density, of the average single quark energy by the pairing, or color-magnetic, interaction. Such a reduction is expected from constituent quark models \cite{DeRujula:1975qlm}, where the color-magnetic interaction reduces the baryon mass from approximately three times the constituent quark mass ($\sim 3\times336$ MeV) down to the nucleon mass, 938 MeV; in uniform quark matter pairing near the Fermi surface leads to an energy reduction $\delta \varepsilon \sim - p^2_F \Delta^2$.
   
%
\begin{figure}
\includegraphics[width = 0.45\textwidth]{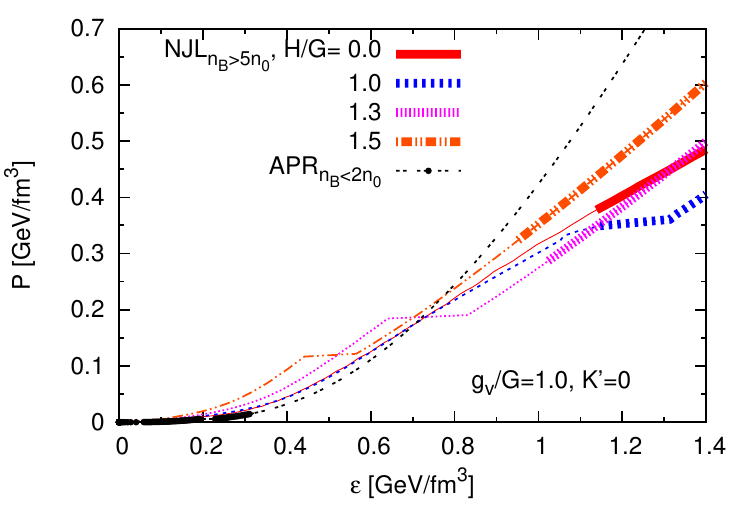}
\caption{\footnotesize{Effects of diquark pairing on the equation of state, $P(\varepsilon)$,  for the same parameter sets as in Fig.~\ref{fig:mu_np_H}. 
The discontinuous change of $\varepsilon$ at fixed $P$ reflects the first order 2SC-CFL phase transition. For $0 < H < 1.5\,G$, the equation of state is softened immediately following the 2SC-CFL phase transition, but as density increases, the equation of state eventually stiffens relative to that at $H = 0$.  For $H=1.5\, G$, the equation of state is stiffer than in unpaired matter for all densities. 
}}
\label{fig:ep_gV_H}
\vspace{-0.2cm}
\end{figure}

   We examine here a range of diquark pairing strengths,  $0.8G \le  H \le 1.5 G$, notably larger as mentioned in subsec.~\ref{params}, than  in previous studies of hybrid equations of state.   Note that the pressure $P(\mu_B)$ of the quark phase for $H=1.5\, G$ with $g_V= G$ (top panel, Fig.~\ref{fig:mu_np_H}) is larger than that of the hadronic APR equation of state.  In the hybrid construction, these pressures must be rejected, thus restricting $H$ to a smaller range  \cite{Buballa2005}.    The direct effects of increasing $H$ on the  stiffness of the equation of state $P(\varepsilon)$ are shown in Fig.~\ref{fig:ep_gV_H}, for $H/G=0, 1.0, 1.3, 1.5$, with $g_V=G$ and $K'=0$.  The flat line, which appears at $n_B \lesssim 5n_0$, reflects the first order phase transition from the 2SC pairing state to CFL pairing, with accompanying softening of the  equation of state.   In typical NJL studies with small $H$, the softening associated with the appearance of condensates leads to a smaller maximum neutron star mass \cite{Buballa2005}.  As seen in Fig.~\ref{fig:ep_gV_H},  the paired phase becomes stiffer than the unpaired phase at high density, a behavior consistent with the schematic discussion in subsec.~\ref{stiffequation of state};  at larger $H$, the stiffening occurs at lower density.  In particular, at $H\ge 1.5\, G$, the paired phase at $n_B \gtrsim 5n_0$ is stiffer than the phase at $H=0$.
   
   As noted in~\cite{Kojo2014}, increasing $K^\prime$ slightly stiffens the quark matter equation of state; however, its impact is much smaller than those of $g_V$ and $H$, as long as we consider a reasonable value of $K' \sim K$ \cite{Powell,Powell2}.  
Thus,  for the sake of simplicity, we restrict the present considerations to $K'=0$, but note that a non-zero $K'$ allows one to choose larger $g_V$.

\begin{figure}
\includegraphics[width = 0.47\textwidth]{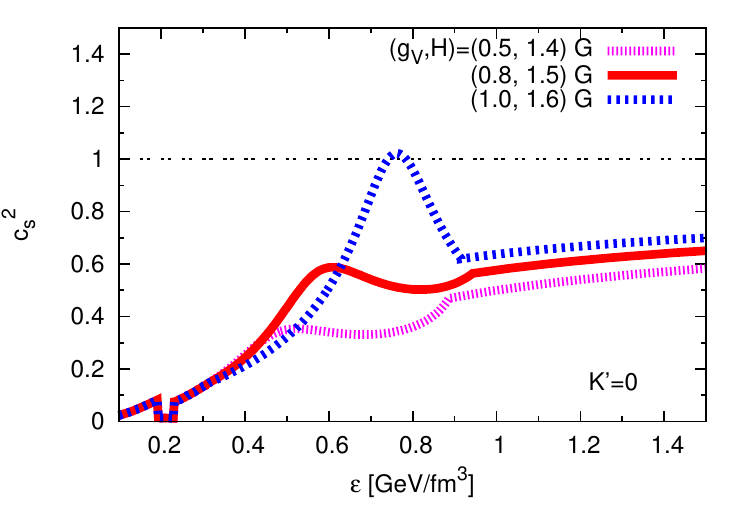}
\caption{\footnotesize{The  squared speed of sound $c_s^2=\partial P/\partial \varepsilon$ as a function of $\varepsilon$.  We compare $(g_V, H)/G$ = (0.5, 1.4), (0.8, 1.5), and (1.0, 1.6).  For each $g_V$, the choice of $H$ is fixed to within $\sim 10\%$ to produce physical interpolations. The dip at $\varepsilon \sim 0.2 $ GeV/fm$^3$ is at the onset of pion condensation in APR. 
}}
\label{fig:inter6cs}
\end{figure}

  Figure~\ref{fig:inter6cs} shows the squared sound speed as a function of $\varepsilon$; in this figure the interpolation region is roughly from $\varepsilon \sim$ 0.3-1 Gev/fm$^3$.  The large sound velocity in the high density quark regime is driven both by the large $g_V$ and $H$.

     In Appendix \ref{masses} we review the effects of $g_V$ and $H$ on the quark effective masses generated by chiral symmetry breaking, on the restoration of chiral symmetry, and on the pairing gaps generated by the diquark condensation.

  As Fig.~\ref{fig:inter6mu-P} shows, for large $g_V$ with large $H$, the equation of state satisfies the stability constraint, and in addition
can support neutron stars with masses above 2.0 $M_\odot$, with a subluminal sound velocity.  
Figure \ref{fig:inter6e-p} shows the interpolated $P (\varepsilon)$ for parameters, $g_V = 0.8\,G$, $H = 1.5\,G$, $K'=0$, with beta equilibrium included; the equation of state in this form directly enters the TOV equation.   For the given interpolation range, we note that the interpolated pressure $P(\varepsilon)$ increases rather rapidly to merge into $P_{ {\rm NJL} } (\varepsilon)$. 
This rapid stiffening is a rather generic feature of hadron-quark interpolations that yield equations of state stiff enough to satisfy the $2M_\odot$ constraint.  However the causality constraint, $\partial P/\partial \varepsilon \lesssim 1$, restricts the rate at which matter can stiffen, so the freedom to choose the model parameters is significantly limited. In fact, model studies show that the physical interpolation almost uniquely fixes the value of $H$ for a given $g_V$. For $g_V/G=0.5, 0.8$, and $1.0$, we are required to choose $H/G \simeq 1.4$, $1.5$, and $1.6$, respectively.  Below, we show the results for these sets of parameters.

  In Appendix~\ref{sec:para_EoS} we give parametrized forms for representative unified equations of state in terms of simple functions.  We call the set of such equations of state QHC18 -- for quark-hadron crossover (2018 version) -- and give them together with instructions for use on the website:  {\em Home Page of Relativistic EOS table for supernovae} (http://user.numazu-ct.ac.jp/$\sim$sumi/eos/index.html).

\section{Neutron stars with unified equations of state \label{sec:constraints}}

    We turn now to the implications of the unified construction of the equation of state (QHC18) on astrophysical properties of neutron stars, and the constraints indicated by current observations, notably the measurements of the two neutron stars with masses $\sim 2 M_\odot$ \cite{Demorest,Antoniadis2013}, early inferences of the mass-radius relation \cite{OzelFreire,Ozel2015,Steiner2015}, and the tidal deformability of neutron stars bounded from above by the binary neutron star merger, GW170817 \cite{GW170817}.
  
   We integrate the TOV equation (\ref{tov}) for a given value of the central baryon density to construct a family of stars whose masses and radii are functions of $n_{Bc}$.
As illustration we use the interpolated equations of state in the liquid interior for three sets of parameters $(g_V,H) = (0.5, 1.4) \,G$, $(0.8, 1.5)\,G$, and $(1.0, 1.6)\,G$, for which we we able to construct sensible interpolated equations of state (see Sec.~\ref{sec:3-window}) with the interpolation window from $n_L= 2.0 n_0$ to $n_U = 5.0 n_0$. 

    At densities $n_B$ from 0.26 $n_0$ to 2.0 $n_0$ in the liquid interior we use the APR equation of state, and
in the crust for $n_B \lesssim 0.26 n_0$, we take the Togashi equation of state \cite{Togashi:2017mjp}, which includes the same detailed physics for the inner crust nuclei as APR includes in the nuclear matter liquid in the interior.  Combining these two equations of state provides a consistent physical description from the inner crust into the liquid interior.\footnote{As discussed in Appendix \ref{sec:para_EoS}, the SLy(4) equation of state for the crust, based on Skyrme effective interactions, does not join in a thermodynamically consistent manner onto the APR equation of state.}

     The unified equations of state and hence the neutron star models constructed from them will be refined over time with improving certainty in our theoretical understanding of the nuclear matter equation of state,  the quark matter parameters -- and indeed the quark model itself -- as well as in the interpolation from nuclear matter to quark matter.

\begin{figure}
\includegraphics[width = 0.47\textwidth]{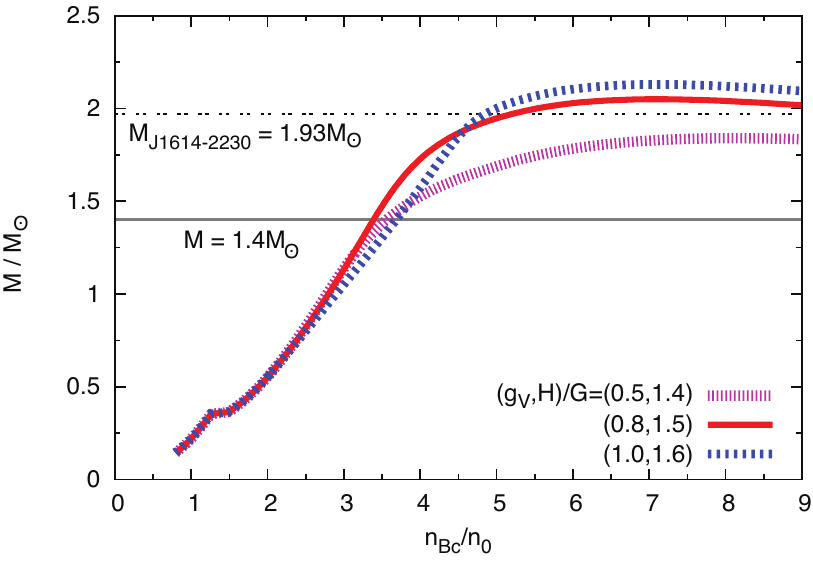}
\caption{\footnotesize{Neutron star mass as a function of central baryon density, $n_{Bc}$, using the unified equation of state (QHC18) for $g_V = 0.8G$, $H = 1.5G$ and  $K' =0$, with $n_U=5 n_0$. }}
\label{fig:M-n}
\vspace{-0.2cm}
\end{figure}

  The neutron star mass as a function of central baryon density is shown in Fig.~\ref{fig:M-n}.  The choice $g_V=0.5\, G$ is not stiff enough to satisfy the $2M_\odot$ constraint, and we are required to take larger values of $g_V$.   At the maximal mass the baryon density is $\gtrsim 5n_0$ or higher, where we expect a quark matter description to be valid.

\begin{figure}
\begin{center}
\includegraphics[width = 0.44\textwidth]{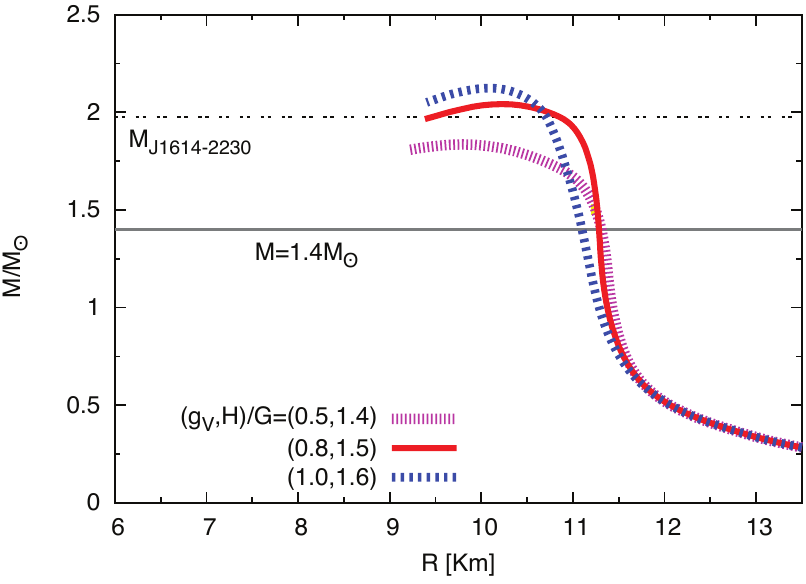}
\caption{\footnotesize{Mass-radius relation for neutron stars with the same equations of state  (QHC18) as in Fig.~\ref{fig:M-n}.}
}
\label{fig:M-R}
\end{center}
\vspace{-0.2cm}
\end{figure}


    Figure \ref{fig:M-R} shows the neutron star mass-radius relation.  Since the overall radii of neutron stars are primarily determined by the equation of state at $n_B\lesssim 2n_0$, neutron stars with masses $\sim 1.4\,M_\odot$  have similar radii, $11.3$-$11.5\,{\rm km}$, 
    for the three parameter sets.   The relative smallness of the radii, compared to those found with typical relativistic mean field equations of state, e.g., \cite{MS}, reflects the relative stiffness of mean-field equations of state at low density.   The mass-radius relations of the unified equations of state reviewed here are reasonably similar to that obtained with a pure nuclear matter equation of state, e.g., APR.  The similarity of radii in hybrid and pure nuclear matter stars is analyzed in \cite{Alford:2004pf}.    
    
       While the masses of certain neutron stars, e.g., those in binary orbits with another neutron star or a white dwarf, can be inferred observationally to good accuracy, observational determination of neutron star radii is much less precise, and does not at this stage permit  a detailed comparison with the neutron star models shown in Fig.~\ref{fig:M-R}.  Data from the NICER experiment \cite{nicer} will make such a comparison more feasible.  A complete and accurate mass vs. radius curve would allow a determination of the equation of state \cite{lindblom}; in this spirit, observations of burst and quiescent low mass X-ray binaries in globular clusters have been used to infer masses and radii of some fourteen neutron stars \cite{OzelFreire,Ozel2015,Steiner2015,Steiner2016} (and references therein); see discussion in subsec.~\ref{MR}.  These measurements, while not sufficiently accurate to determine the equation of state, do indicate constraints on it.   Reference~\cite{inversion} discusses the Bayesian analysis and accuracy of inference of an equation of state from $M$~vs.~$R$ data sets.  
     
\begin{figure}[t]
\includegraphics[width = 0.49\textwidth]{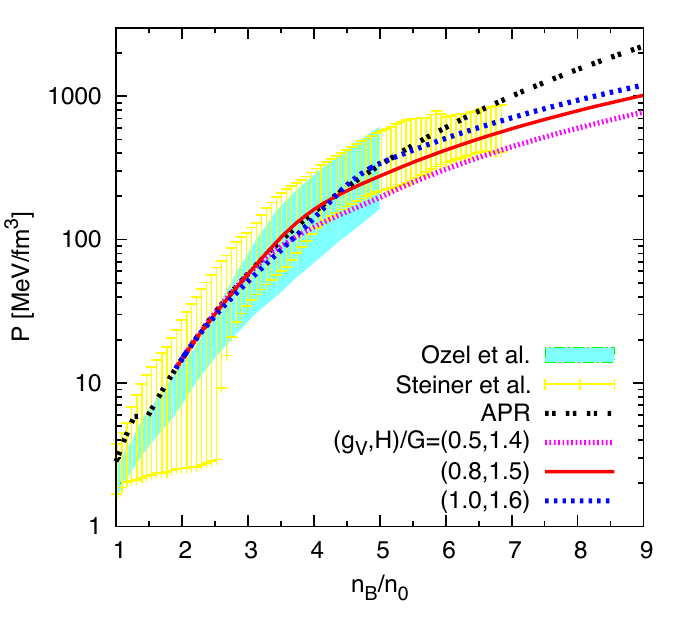} 
\caption{\footnotesize{Pressure vs. baryon density for hadronic APR and three versions of the unified hadron-quark equations of state  (QHC18).  All the equations of state include beta equilibrium.  The shaded regions are those inferred from $M$-$R$ measurements of neutron stars,
at an effective $2\sigma$ level \cite{Ozel2015,Steiner2015}.} }
\label{fig:Lattimer_Ozel_and_Us}
\end{figure}

      Figure~\ref{fig:Lattimer_Ozel_and_Us} shows the pressure vs. baryon density corresponding to the three curves in Fig.~\ref{fig:M-R}.   As implied by the arguments in subsec.~\ref{stiffequation of state}, increasing $g_V$ as well as $H$ tends to increase the pressure at fixed baryon density.   The unified equations of state with quark matter at high density are softer than the APR equation of state;  such asymptotic softening is expected from consistency with perturbative calculations valid at $n_B \gtrsim 100n_0$.       The shaded regions in this figure show a range of equations of state \cite{Ozel2015,OzelFreire,Steiner2015} that are compatible with the available mass-radius data inferred from bursts in low mass X-ray binaries; overall, the unified equations of state are consistent with this range.  While there are possible discrepancies using the equations of state of Ref.~\cite{Ozel2015} in the vicinity of $n_0$, analysis of the differences in the inferences from the data is beyond the scope of this review.   Nonetheless, resolving discrepancies with observation both through better mass-radius determinations, as NICER will provide, as well as through improvements to the theory of nuclear matter in beta equilibrium remains an open challenge.

\begin{figure}[t]
\includegraphics[width = 0.49\textwidth]{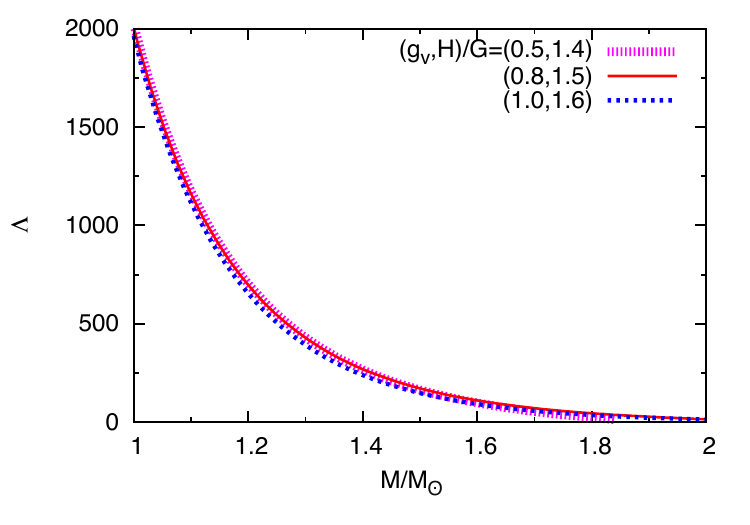} 
\caption{\footnotesize{The dimensionless tidal deformability $\Lambda$ of a neutron star as a function of the neutron star mass $M$, for the same equations of state (QHC18)  as in Fig.~\ref{fig:M-n}. }} 
\label{fig:dimlessLambda}
\end{figure}

   Figure~\ref{fig:dimlessLambda} shows the dimensionless neutron star tidal deformability $\Lambda$, Eq.(\ref{eq:dimlessLambda}),  as a function of the neutron star mass for the three parameter sets used in computing the $M$-$R$ relations; $\Lambda$ was calculated using  the procedure summarized in \cite{Hinderer2008}.   For a $1.4M_\odot$ star, the range of $\Lambda$ is 240-270, similar to what one finds using the APR equation of state throughout the star. 

  The gravitational waveforms as detected in binary neutron star mergers are sensitive to the combination,
\beq
\tilde{\Lambda}\, = \frac{16}{13}\frac{\, (M_1+12M_2)M_1^4\Lambda_1 + (12M_1+M_2)M_2^4\Lambda_2}{ (M_1 + M_2)^5 },  \nonumber\\
\eeq
of the masses and $\Lambda$'s of the individual neutron stars, as derived in a post-Newtonian calculation \cite{Hinderer:2009ca}.

\begin{figure}[t]
\includegraphics[width = 0.45\textwidth]{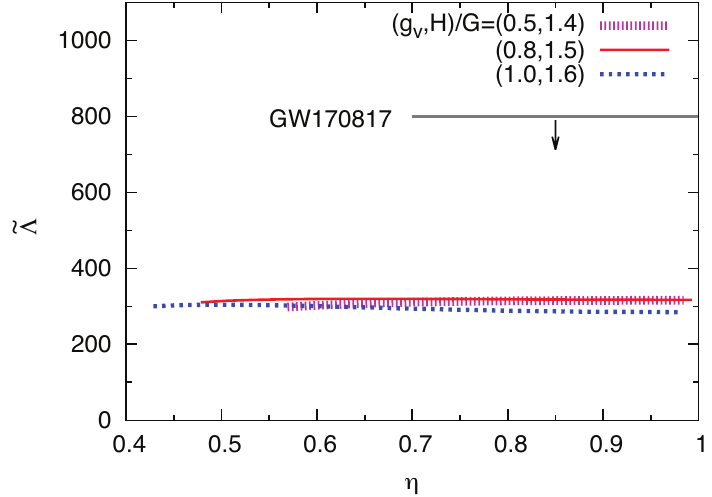} 
\caption{\footnotesize{The combined dimensionless tidal deformability $\tilde{\Lambda}$ of a pair of neutron stars as a function of the mass ratio $\eta = M_1/M_2 (\le 1)$, for the same equations of state (QHC18) as in Fig.~\ref{fig:M-n}, with $\mathcal{M}_{ {\rm chirp} } = 1.188 M_\odot$.  The upper bound from GW170817 is shown.  The curves terminate at low $\eta$ when $M_2$ reaches the  maximum mass for the given equation of state.}} 
\label{fig:Lambda_obs}
\end{figure}

The GW170817 merger measured the chirp mass, $\mathcal{M}_{ {\rm chirp} } = (M_1 M_2)^{3/5} (M_1+M_2)^{-1/5} \simeq 1.188^{+0.004}_{-0.002} M_\odot$. The assumption of small spin ($\lesssim 0.05$) for each neutron star, which is probable from the population analyses, weakly constrains the mass ratio $\eta =M_1/M_2$ (for $M_1 \le M_2$) to 0.7-1.0, but even with this uncertainty, $\eta$ together wtih $\mathcal{M}_{ {\rm chirp} }$ tightly constrains the total mass, $M_1+M_2 \simeq 2.74^{+0.04}_{-0.01} M_\odot$.   
Figure \ref{fig:Lambda_obs} shows the result of our equations of state for $\tilde{\Lambda}$ as a function of $\eta$ with fixed 
$\mathcal{M}_{ {\rm chirp} } = 1.188 M_\odot$.  The resulting $\Lambda_{ {\rm obs} }$ depends weakly on $\eta$ and is $\simeq$ 290-320, consistent with the upper bound $\tilde{\Lambda} \le 800$ (90\% confidence level) found from analysis of GW170817 \cite{GW170817}.

%

\section{Summary \label{sec:massive_stars}}

  As we have reviewed here, the unified construction of equations of state (as exemplified by QHC18) avoids artifacts arising from extrapolating the hadronic and non-confining quark matter equations of state outside their ranges of validity, as in the conventional hybrid construction of the equation of state.  The unified construction allows quark matter to be sufficiently stiff to produce at least $2M_\odot$ neutron stars, and at the same time takes strange quarks into account.  The requirement of stiffness disfavors a strong first order phase transition (although a weak one in the interpolated domain cannot be ruled out), indicating  that hadronic and quark matter are likely not distinctly different, but rather smoothly connected, with the quark interactions as strong as those in the QCD vacuum.

   In the quark matter equations of state discussed here, both the repulsive vector and attractive color-magnetic interactions play important roles in constructing physically acceptable unified equations of state.   The $M>2M_\odot$ constraint requires those interactions to be as strong as the NJL scalar interaction responsible for chiral symmetry breaking. 
   
    As discussed throughout this review, many outstanding questions remain regarding neutron star properties in general, and in the context of hadron-quark continuity in particular. First, an exhaustive analysis of the parameter ranges of interactions in quark matter that is consistent with both hadron-quark continuity and current astrophysical data has yet to be performed.  Second, in order to describe the location and width of the crossover region accurately and  further refine neutron star models, one needs better understanding of the interactions, e.g., the axial anomaly-induced Kobayashi-Maskawa-'t Hooft (KMT) interaction, that drive the hadron-quark crossover.  Third, much work remains to be done in addressing the possibility of additional quark pairing structures and inhomogeneous phases which may exist in the cores of the densest neutron stars (e.g., pion condensation, quarkyonic matter, etc.), and how such structures can span the hadronic to quark matter crossover.  Eventually one would like to go beyond simple NJL models of quark matter.    Finally, as both the quantity and quality of observational neutron star data are improved, constraints on quark models and their parameters will continue to become more stringent.   It is encouraging that current observational inferences are consistent with hadron-quark continuity.   Through precise astrophysical observation, assessment of such continuity can be continually refined.
    
    The effects of the underlying quark picture on the dynamical properties of neutron stars have yet to be fully explored.    The color-magnetic interaction and the resulting quark pairing correlations are expected to play a significant role in the non-equilibrium properties in neutron star interiors,  e.g., in neutrino emission and thermal transport processes involved in cooling of stars.  It is important to delineate possible signatures of large quark matter cores in the cooling.  
        
      A major issue is to determine the effects of finite temperature on the
equation of state of neutron stars,  in order to understand how signals of neutron star structure are produced in the gravitational radiation emerging from binary neutron star and neutron star--black hole mergers \cite{takami,baiotti,sekiguchi,stu1,stu2,bernuzzi}.  
Neutron star temperatures in mergers can be as large as 10$^2$ MeV \cite{janka,sekiguchi, bernuzzi}.   A initial estimate of the effects of finite temperature on the equation of constraint consistent with the present point of view of the hadronic to quark matter crossover is given in Ref. \cite{masuda-T}.

\section{Acknowledgments}
First and foremost we thank Chris Pethick for his invaluable input, insights, and support during the writing of this review, including his substantial contributions to  the discussion of neutron star crusts and the nuclear matter liquid interior.  We thank Feryal \"Ozel, Andrew Steiner, Hajime Togashi for providing us with the numerical tables for their equations of state, Dimitrious Psaltis and  Sanjay Reddy for very helpful input,  and Charalampos Markakis and Michael O'Boyle for their calculations of tidal deformability.
Research reported here was supported in part by NSF Grants PHY0969790 and PHY1305891.  In addition, authors TH and TT were supported by JSPS Grants-in-Aid for Scientific Research (B) No.~24340054 and No.~25287066; 
GB and TH were partially supported by the RIKEN iTHES Project and iTHEMS Program; and TK was supported by NSFC grant 11650110435.       Authors GB, TH, and TK  are grateful to the Aspen Center for Physics, supported in part by NSF Grants PHY1066292 and PHY1607611,  where parts of this review were written.

\appendix

\section{Scaling the TOV equation \label{sec:scaling} }

Here we derive a simple expression for the scale of masses and radii of neutron stars, as well as  quark stars made purely of quarks.  We assume that the equation of state is governed by a
basic energy scale, $\epsilon_0$, which is if the order of the QCD scale parameter $\Lambda_{ {\rm QCD}} \sim$ 200 MeV,
and rescale the mass density and pressure by
\beq
    \rho = \epsilon_0^4 \tilde \rho , \quad    P = \epsilon_0^4 \tilde P,
\eeq 
where $\tilde \rho$ and $\tilde P$ are dimensionless in units with $\hbar=c=1$,  In addition we let $\zeta$  be the scale of the radius,
writing
\beq 
   r = \zeta \tilde r;
\eeq
Then the mass within radius $r$, $m(r) = \int_0^r 4\pi r^2 \rho(r) dr$, scales as $\zeta^3 \epsilon_0^4$.
Clearly by choosing $G \zeta^2 \epsilon_0^4 =1$, or restoring $\hbar$ and $c$,
\beq
   \zeta  = \frac{\hbar^{3/2} c^{7/2}}{\epsilon_0^2 G_N^{1/2}} = \left(\frac{m_pc^2}{\epsilon_0}\right)^2 \frac{\hbar}{m_p c\,\alpha_G^{1/2}},
\eeq
all the dimensional factors in the TOV equation cancel out; here
\beq
    \alpha_{G} = \frac{m_p^2 G_N}{\hbar c} \simeq 0.589 \times 10^{-38} 
\eeq
is the gravitational fine structure constant, with $m_p$ the proton mass.   After rescaling, the TOV equation (\ref{tov}) reduces to the dimensionless form:
\beq
  \frac{\partial \tilde P(r)}{\partial \tilde r} = \frac{1}{\tilde r^2}\frac{(\tilde \rho + \tilde P)(\tilde m(\tilde r) 
    + 4\pi \tilde r^3 \tilde P)}{1-2\tilde m(\tilde r)/\tilde r},
\eeq
where $\tilde m(\tilde r) = \int_0^{\tilde r} 4\pi \tilde r^2 \tilde \rho(\tilde r) d\tilde r$.

    We see then that the total mass scales as 
\beq
  M \propto  \frac{m_p}{\alpha_G^{3/2}}\left(\frac{m_pc^2}{\epsilon_0}\right)^2
  = 1.86\left(\frac{m_pc^2}{\epsilon_0}\right)^2 M_{\odot}.
  \label{mscale}
  \eeq    
(Note that $M_\odot/m_p = 1.189 \times 10^{57}$.)
Similarly the radius scales as 
\beq
   R\propto \zeta =  17.2 \left(\frac{m_pc^2}{\epsilon_0}\right)^2 \, \rm km.  
   \label{rscale}
\eeq   
The actual masses and radii found from integrating the TOV equation are given by the scales set by Eqs.~(\ref{mscale}) and (\ref{rscale}) times numerical factors of order unity;  $\tilde M = \int 4\pi \tilde r^2 \tilde \rho d \tilde r$ for the mass and $\tilde r$ for the radius.  With the choice $\epsilon_0 = m_pc^2$ the 
scale of neutron star masses is close to the expected maximum neutron mass, somewhat above two solar masses, while the scale of the radius is also consistent with expected neutron star radii, $\sim 10-12$ km.
Note also that the compactness $M/R$ scales as $c^2/G$  = 6.9 $M_\odot$/10 km,  where $M_\odot /10 \,\rm{km} = 2.0 \times 10^{27} \rm\, g/cm$.    

  For the particular choice of the free quark equation of state, Eq.~(\ref{freeqeos}), the natural scale is $\epsilon_0 = B^{1/4}$, and we find that the mass and radius scale as $1/B^{1/2}$ as in Eqs.~(\ref{idealgasM}) and (\ref{idealgasR}); the prefactors there result from actual integration of the TOV equation.  
 
 \section{Effects of vector repulsion and the pairing  on the constituent quark masses in quark matter
\label{masses} }

\begin{figure}
\includegraphics[width = 0.45\textwidth]{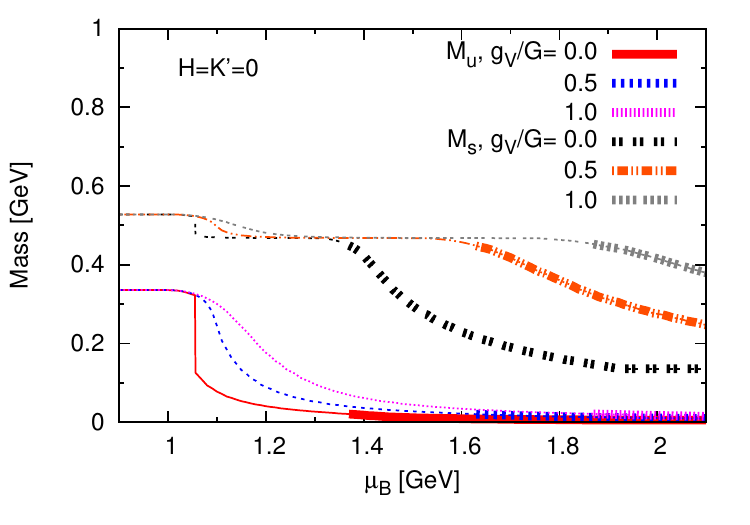}
\caption{
\footnotesize{ The constituent masses of the $u$ and $s$ quarks,  in the pure quark regime in the absence of diquark pairing.   The $d$-quark mass has essentially the same behavior as that of the $u$ quark.  The region $\mu_B$ below $\sim $1.5 GeV is in the intermediate regime, where one must interpolate between the nucleonic and quark pictures; the pure quark behavior shown in this regime is only to illustrate the restoration of chiral symmetry.
}
}
\label{fig:gV_chiral}
\end{figure}

 The phenomenological vector repulsion between quarks in quark matter has significant effects not
only on the pressure but also on the  broken chiral symmetry.   For simplicity, we consider quark matter alone, and neglect the transition to hadronic matter  at low densities.  We also neglect the anomaly-induced coupling 
between the chiral and diquark condensates, $K'$, for the moment.  Then with increasing quark chemical potential $\mu_q = \mu_B/3$, a non-zero quark density begins to develop when $\mu_q$  exceeds the constituent $u$ and $d$ quark masses, $M_u, \,M_d \simeq 336\,{\rm MeV}$.  
  
For $g_V = 0$, the quark density increases rapidly 
    and causes the chiral condensate $\sigma$ and hence the constituent quark mass 
     $M$ to decrease rapidly, as illustrated in Fig.~\ref{fig:gV_chiral} for $H=0$.
  Such a rapid change results in a first order chiral phase transition in which both the quark number and the chiral condensate become discontinuous; see Figs.~\ref{fig:gV_n} and \ref{fig:gV_chiral}.   
  On the other hand, increasing $g_V$ from zero makes it more energetically costly for additional quarks to enter the system; the vector repulsion requires a larger chemical potential 
  in order to achieve a specified baryon density.  As a result, the slope of $n_B(\mu_B)$ decreases, as seen in Fig.~\ref{fig:gV_n},  and the melting of the chiral condensate proceeds more smoothly, as seen in Fig.~\ref{fig:gV_chiral}.
For $g_V$ exceeding a critical value $ \simeq 0.4G$, the first order chiral transition becomes a  smooth crossover \cite{asakawa,Bratovic:2012qs,steinheimer,Kitazawa:2002bc}.

\begin{figure}
\includegraphics[width = 0.45\textwidth]{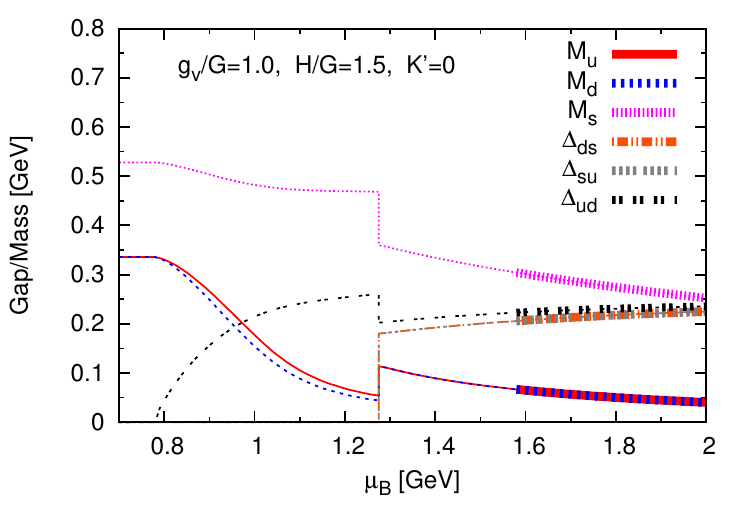}
\caption{\footnotesize{
The constituent  masses and the pairing gaps as a function of $\mu_B$ 
for  $H = 1.5G$,  $g_V= 1.0G$, and $K'=0$,   With increasing density, the system undergoes a first order phase transition from the 2SC phase to the CFL phase.   Just prior to this phase transition, the  strange quarks begin to appear. Thick lines denote the density region $n_B>5n_0$ which enters the construction of the unified 
equation of state.  }}
\label{fig:mu_con_H}
\vspace{-0.2cm}
\end{figure}


The constituent quark masses $M_{u,d,s}$ and the pairing gaps $\Delta_{ud,ds,su}$  for $g_V/G=1$ and $H/G=1.5$
    are shown in Fig.~\ref{fig:mu_con_H}.  The first order  transition from the 2SC phase  to the CFL phase occurs around $\mu_B \sim 1.3 $ GeV in the 
  figure.   If  the anomaly-induced coupling $K'$ is increased from zero, the coexistence of the chiral condensate and diquark condensates becomes energetically favorable.  As a result, the chiral condensate survives to higher chemical potential, while diquark pairing begins to develop at lower chemical potential.  The quantities
  $M_{u,d,s}$ and $\Delta_{ud,ds,su}$ in the region $n_B > 5n_0$ are denoted  by the thick
  lines.  All the  details associated with the onset of the strange quarks and  the 2SC-CFL transition 
   at $n_B < 5n_0$ are not relevant for the unified  equation of state, since they occur in the interpolation region.

\section{Parameterized equations of state \label{sec:para_EoS} }

\begin{figure}
\includegraphics[width = 0.48\textwidth]{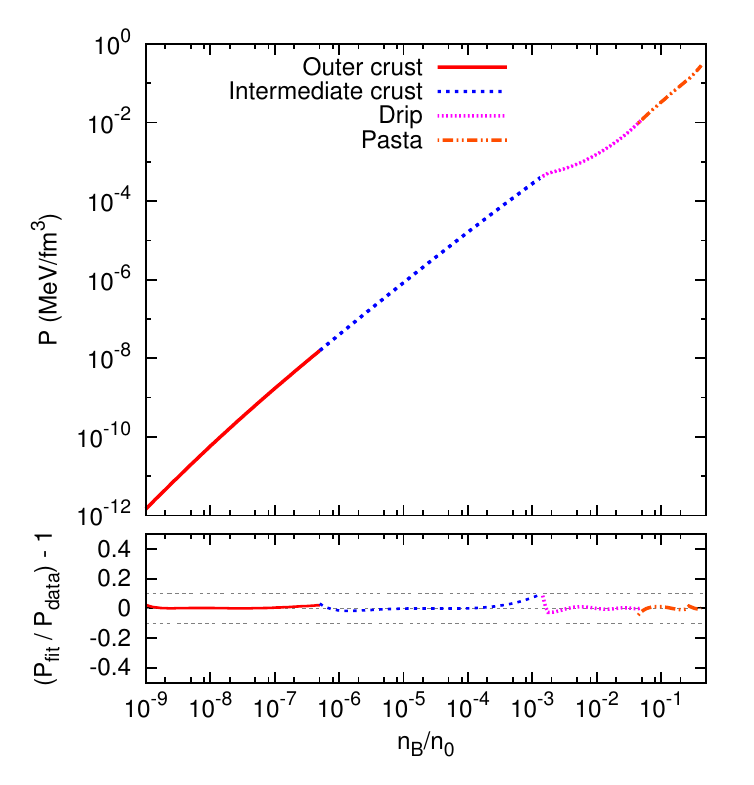}
\caption{\footnotesize{
Pressure $P$ vs. $n_B/n_0$ in the crust from $n_B=10^{-9} n_0$ to $n_B=0.26n_0$. 
The quality of fit, $\delta \equiv P_{ {\rm fit} }/P_{ {\rm data}}-1$,  is shown in the lower panel.  The parameterizations for the four regions, outer crust, intermediate crust, drip regime and nuclear pasta regime, are given in Table \ref{tab:fit_Togashi}. } }
\label{fig:fit_SLyA}
\vspace{-0.2cm}
\end{figure}

In this Appendix we present parameterized forms of equations of state, $\varepsilon(n_B)$, where $\varepsilon = \rho c^2$ is the energy density, and $\rho$ is the mass density, as well as numerical tables, to use in numerical modeling of neutron stars.  Separating the star into domains -- the outer and the intermediate crust regions below neutron drip, the neutron drip region, and the nuclear pasta phase region;  the low and high density regions of the liquid nuclear matter; the crossover; and higher density quark matter -- we use accurate polynomial fits of the numerical equations of state, with the coefficients in the polynomials tuned to each domain.  

     To satisfy thermodynamic relations in numerical computation, and to avoid artifacts, we parametrize only the single thermodynamic function, $\varepsilon(n_B)$, writing
\beq
\varepsilon (\xi) =  a \xi + d_0 + d_1\xi \ln \xi + \sum_{\nu=2}^{ \nu_{ {\rm max} } } d_\nu \xi^{l_\nu} \,,
\label{e_para}
\eeq
where $\xi \equiv n_B/n_0$, with $n_0= 0.16$ fm$^{-3}$, and $a$, $d_n$'s, $ \nu_{ {\rm max} } $, and $l_\nu$'s fitting parameters dependent on the domains.   Then the chemical potential, $\mu_B = \partial \varepsilon/\partial n_B $, and pressure,   $P = n_B^2 (\partial \varepsilon /n_B)/\partial n_B$  are given by
\beq
\!\! \mu_B(\xi) = \frac{1}{n_0} \left[ a +d_1(1+\ln \xi) + \sum_{\nu=2}^{ \nu_{ {\rm max} } }  l_\nu  d_\nu \xi^{l_\nu -1} \right],
\label{mu7}
\eeq
and
\beq
P(\xi) = - d_0 + d_1 \xi + \sum_{\nu=2}^{ \nu_{ {\rm max} } }  (l_\nu - 1) d_\nu \xi^{l_\nu} \,.
\label{P3}
\eeq
Note how the linear term in the pressure ($\sim d_1$) corresponds to a logarithmic term in $\varepsilon$ and in $\mu_B$.\footnote{The reader may be concerned that the parametrization of pressure in the interpolation region Eq.~(\ref{Pinterp}) and that given here are not obviously consistent.
However, we use the present parametrization only to fit numerically the interpolated $P(\mu_B)$.}
In practice, we use the pressure, rather than energy density, to determine the $d_i$ because it is more sensitive to the values of fitting parameters.
\begin{figure}
\includegraphics[width = 0.48\textwidth]{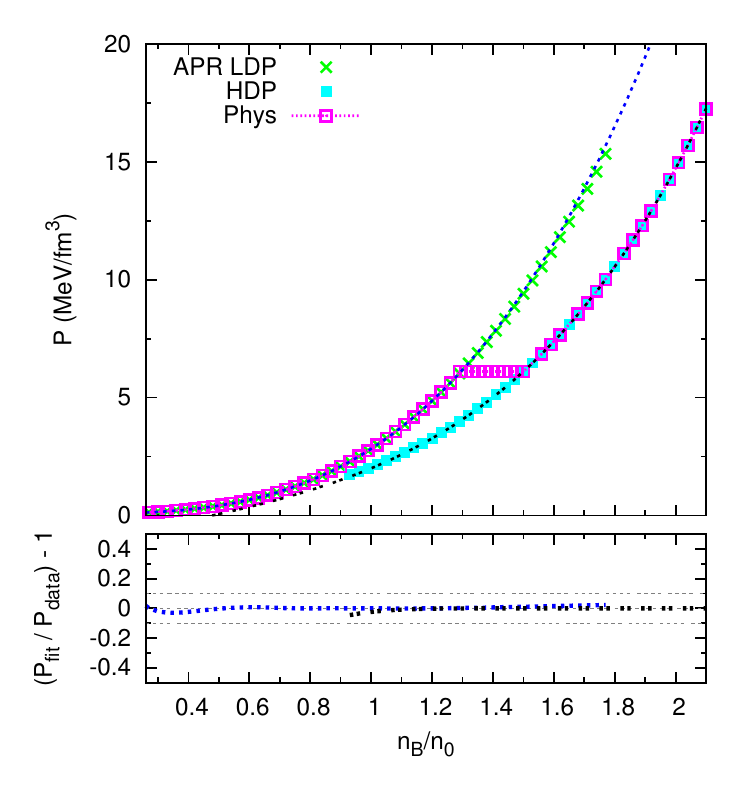}
\caption{\footnotesize{Fitting of $P$ vs $n_B/n_0$ for the APR equation of state in $\beta$-equilibrium in the range $n_B=0.26n_0$ to $2n_0$.     At the phase transition from low to high density}  the APR density changes discontinuously from $n_B\simeq 1.29n_0 \simeq 0.21\,{\rm fm}^{-3}$ to $\simeq 1.50n_0\simeq 0.24\,{\rm fm}^{-3}$. The purple squares show the physical equation of state. The parameterization is given in Table \ref{tab:fit_Togashi}. 
}
\label{fig:fit_APR}
\vspace{-0.2cm}
\end{figure}

\subsection{Crust}

  In the crust we compare the SLy(4) equation of state \cite{sly,HZD} as given in the numerical tables of Haensel and Potekhin \cite{potekhin}, and the Togashi  equation of state \cite{Togashi:2017mjp} determined via variational calculations based on the microscopic AV18 two-body plus the UIX three-body potentials (numerical tables can be found in CompOSE data base, http://compose.obspm.fr/eos/105/). These equations of state are consistent below the nuclear drip regime,  while at higher densities the two equations of state start to deviate because of the differences in interactions and treatments of the nuclei in a neutron gas.  The difference is important for matching the crust equation of state to that in the nuclear liquid.  
     The Togashi equation of state, being based on the same input physics, can be matched with  APR in a thermodynamically consistent manner, while SLy(4) cannot, in the absence of ad-hoc smoothening and reduction of data points.  
     
     Taking APR to describe the nuclear liquid, we adopt the Togashi equation of state for the crust.   We chose the matching point with APR to be $n_B = 0.26n_0$, around which the two equations of state overlap fairly well.
The fit (\ref{P3}) is compared with the tabulated equation of state in Fig. \ref{fig:fit_SLyA}, where only the data points from $n_B = (10^{-9}-10^{-6})n_0$ are used (corresponding to $\sim (3\times10^5- 3\times10^8)$ g/cm$^{3}$).  The region $n_B \lesssim 10^{-9}n_0$ does not affect the overall structure of neutron stars and can be ignored here. The region $10^{-9}n_0 \lesssim n_B \lesssim 10^{-6}n_0$ contributes $\sim$ 20-100 m to the radii and $\sim 10^{-9} M_\odot$ to the masses.  At $n_B \gtrsim 10^{-7}n_0$ ($\sim 3\times 10^{7}$ g/cm$^{3}$) the equations of state are unaffected by magnetic fields as strong as $10^{14}$ G, and by temperatures  $\lesssim 10^9$ K \cite{haensel}.

\begin{center}
\begin{table*}[t]

\caption{\footnotesize{ Fitting parameters for the crust and nuclear liquid regions (see text). The unit of the coefficients are ${\rm MeV/fm^3}$. Each domain starts at the threshold $\xi_{{\rm th}}=(n_B/n_0)_{ {\rm th}}$ shown in the table.}  
}
\begin{tabular}{|c| c|| c|c|c|c| c || c |c|c}
\hline
   &~~~~~$\xi_{{\rm th}}$~~~~~&~~~~~$a$~~~~~&~~~~$d_0$~~~~&~~~~$d_1$~~~~&~~~~$d_2$~~~~&~~~~$d_3$~~~~&~~~~$l_2$~~~~&~~~~$l_3$~~~~\\
\hline 
~~Outer crust ~~&~$10^{-9}$~&~$149.9$~~&~ $-$ ~&~$-7.112\cdot 10^{-2}$~~&~$9.168$~~&~$-1.522$~~&~$1.271$~~&~$0.97508$~\\
%
~~Intermediate crust~~&~$5\cdot 10^{-7}$~&~$148.2$~~&~ $-$ ~&~$-3.391\cdot 10^{-2}$~~&~$6.922$~~&~$-3.591\cdot 10^{-9}$~&~$1.224$~~&~$1.0\cdot10^{-2}$~\\
%
~~Drip     ~~&~$1.5\cdot 10^{-3}$~&~$150.9$~~&~ $-$ ~&~$9.940\cdot 10^{-2}$~~&~$3.941$~~&~ $-6.422\cdot 10^{-4}$~~&~$2.205$~~&~$9.353\cdot 10^{-2}$~~\\
%
~~Pasta  ~~&~$5\cdot 10^{-2}$~&~$150.1$~~&~$-$~&~ $0.9845$ ~&~$1.443 $~~&~$1.861$~~&~$5.059$~~&~$0.7864$~~\\
\hline 
~~APR (LDP) ~~ &~$0.26$~&~$151.5$~~&$-6.690\cdot 10^{-2}$~&~$4.914\cdot 10^{-2}~$~&~$1.320$~~&~$-$~&~$3.056$~~&~$-$~\\
~~ -- 1st order --  ~~&~$1.29$~&~$-$~&~$-$& ~$-$~&~$-$~~&~ $-$~ & ~$-$~&~$-$~\\
~~APR (HDP) ~~ &~$1.50$~&~$151.7$~~&~$1.220$~&~$2.461$~&~$0.2666$~~&~$-$~&~$3.856$~~&~$-$~\\
\hline 
\end{tabular} 
\label{tab:fit_Togashi}

\vspace{1cm}
\caption{\footnotesize{Fitting parameters for the unified equations of state for various sets of $(g_V,H)/G$ with $K'=0$ (see text). The unit of coefficients are ${\rm MeV/fm^3}$. The parameters $l_\nu$'s are fixed to integers, $l_\nu = \nu$. To have a good quality fit in the interpolated domain, it is important to include all the digits shown. }  
}
\begin{tabular}{|c| c|| c|c|c|c| c | c |c|c|}
\hline
   &~~~$\xi_{{\rm th}}$~~~&~~~~~$a$~~~~~&~~~~$d_0$~~~~&~~~~$d_1$~~~~&~~~~$d_2$~~~~&~~~~$d_3$~~~~&~~~~$d_4$~~~~&~~~~$d_5$~~~~&~~~~$d_6$~~~~\\
\hline 
~~(0.5,1.4)  ~~&~~&~$514.390$~~&~ $1038.33  $ ~&~$2088.56$~~&~$-1688.70$~~&~$350.282$~&~$-51.8325$~~&~$4.42276$~&~$-0.162045$~ \\
%
~~(0.8,1.5)  ~~&~$2.0$~&~$-495.283$~~&~ $-715.928$ ~&~$-1761.03$~~&~$1722.13$~~&~$-428.157$~~&~$76.4943 $~~&~$-7.78341 $~&~$0.335614$~\\
%
~~(1.0,1.6)  ~~&~~&~$1575.17$~~&~$3029.17$~&~$6352.80$ ~&~$-5460.10$~~&~$ 1230.94$~&~$-204.048$~~&~$19.9068$~&~$-0.843217$~ \\
\hline
~~(0.5,1.4)  ~~&~~&~$116.0$~~&~ $ 54.43$ ~&~$-1.989$~~&~$10.44$~~&~$-$~~&~$-$~~&~$-$~&~$-$~\\
~~(0.8,1.5)  ~~&~$5.0$~&~$102.0$~~&~$55.85$~&~ $0.6177 $ ~~&~$13.15$~~&~$-$~~&~$-$~~&~$-$~&~$-$~\\
~~(1.0,1.6)  ~~&~~&~$87.28$~~&~$54.58$~&~ $3.639$ ~~&~$15.03$~~&~$-$~~&~$-$~~&~$-$~&~$-$~\\
\hline 
\end{tabular}
\label{tab:fit_quark}

\vspace{1cm}
\caption{\footnotesize{ Fitting parameters for the quark matter equations of state at $\xi =n_B/n_0 \ge 5.0$ for various sets of $(g_V,H)/G$ with $K'=0$ (see text).  The coefficients are in units of ${\rm MeV/fm^3}$, and $l_\nu = \nu$.}   }  
\begin{tabular}{|c| c|| c|c|c|c| c | c |c|c|}
\hline ~$g_V/G$~
   &~$H/G$~~&~~~~~$a$~~~~~&~~~~$d_0$~~~~&~~~~$d_1$~~~~&~~~~$d_2$~~~~\\
\hline 
~         &~$1.4$~&~$116.0$~~&~$54.31$ ~&~$-2.020$~~&~$12.06$~~\\
~ 0.7 ~&~$1.5$~&~$103.6$~~&~$53.07$ ~&~$-0.1710$~~&~$12.39$~~\\
         ~&~$1.6$~&~$88.36$~~&~$52.73$ ~&~ $3.047$ ~  &~$12.65$~~\\
\hline 
         ~&~$1.4$~&~$115.9$~~&~$54.50$ ~&~$-2.000$~~&~$12.87$~~\\
~ 0.8 ~&~$1.5$~&~$102.0$~~&~$55.85$ ~&~ $0.6177 $ ~~&~$13.15$~~\\
         ~&~$1.6$~&~$85.49$~~&~$57.77$ ~&~ $4.537$ ~~&~$13.35$~~\\
\hline 
         ~&~$1.4$~&~$116.8$~~&~$52.79$ ~&~$-2.422$~~&~$13.71$~~\\
~ 0.9 ~&~$1.5$~&~$105.3$~~&~$50.06$ ~&~ $-1.081 $ ~~&~$14.08$~~\\
         ~&~$1.6$~&~$87.03$~~&~$55.01$ ~&~ $3.749$ ~~&~$14.22$~~\\
\hline 
         ~&~$1.4$~&~$114.4$~~&~$57.00$ ~&~$-1.146$~~&~$14.42$~~\\
~ 1.0 ~&~$1.5$~&~$99.61$~~&~$60.04$ ~&~ $1.892 $ ~~&~$14.68$~~\\
         ~&~$1.6$~&~$87.28$~~&~$54.58$ ~&~ $3.640$ ~~&~$15.03$~~\\
\hline 
\end{tabular}
\label{tab:fit_quark_more}

\end{table*}
\end{center}

   Table \ref{tab:fit_Togashi} summarizes the crustal fitting coefficients. To obtain good fitting quality within the present simple parameterization, 
we divide the crust into four regions: the outer crust, from $n_B = (10^{-9}-5\cdot 10^{-7})n_0$, the intermediate crust from $n_B = 5\cdot 10^{-7}n_0$ to the neutron drip regime $n_B= (5\cdot 10^{-7}-1.5 \cdot 10^{-3})n_0$;
the neutron drip regime, $n_B= (1.5\cdot10^{-3}-5\cdot10^{-2})n_0$; and the fourth interval, $n_B=(5\cdot10^{-2}-0.26)n_0$, containing the various nuclear pasta phases.

\begin{figure}[h]
\includegraphics[width = 0.48\textwidth]{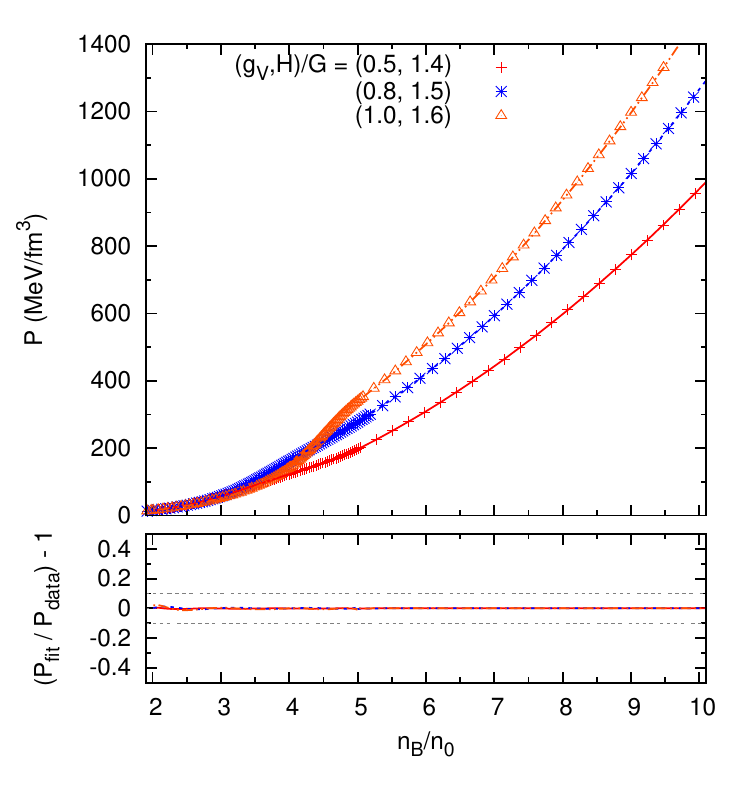}
\caption{\footnotesize{Pressure vs. $n_B/n_0$ for the interpolated and quark matter equations of state. The parameter sets are  $(g_V/G,H/G)=(0.5,1.4), (0.8,1.5), (1.0,1.6)$, all with $K'=0$.  
}}
\label{fig:fit_interCSCA}
\vspace{-0.2cm}
\end{figure}

\subsection{APR in beta equilibrium}

  We detail here the APR equation of state in the range $0.26n_0 < n_B < 2n_0$. 
  The fitting coefficients are summarized in Table \ref{tab:fit_Togashi}. From the parametrizations of the equation of state for pure neutron matter and symmetric nuclear matter in the APR paper  \cite{APR}, one can approximately calculate the equation of state in beta equilibrium by quadratic interpolation, Eq.~(\ref{quadinterp}); in interpolating with the APR parametrized forms  we use the average of the proton and neutron masses, $(m_p+m_n)/2 \simeq 938.92\, {\rm MeV}$.

The APR equation of state has a low and a high density regime, distinguished by a first order phase transition associated with the onset of neutral pion condensation; in matter in beta equilibrium the density changes discontinuously from $n_B\simeq 1.29 n_0\simeq 0.21\,{\rm fm}^{-3}$ to $\simeq 1.50 n_0\simeq 0.24\,{\rm fm}^{-3}$.  The low density parametrization (LDP) is for the range
$0.26n_0 \lesssim  n_B \lesssim   1.29 n_0$, 
and the high density parametrization (HDP), the range $ 1.50 n_0 \lesssim n_B \lesssim  2.0 n_0$.  The error estimator  $\delta \equiv P_{ {\rm fit} }/P_{ {\rm data}}-1$, shown in Fig.~\ref{fig:fit_APR}, does not exceed $\pm 0.01$, for the fits.

\subsection{Unified equations of state (QHC18)}

In the interpolation region $n_L = 2n_0 \le n_B \le  n_U=5 n_0$, we parametrize the equations of state by taking the exponents in Eq.~\ref{e_para} to be integers, 
$
l_\nu = \nu =2, \ldots, \nu_{ {\rm max} }$, with $\nu_{ {\rm max} } = 6$. In the quark matter domain, we fit the data for $5n_0 \le n_B \le 10n_0$ and 
find that taking only one term, $\nu_{ {\rm max} } = 2$, yields adequate fits (Fig.~\ref{fig:fit_interCSCA}). Table~\ref{tab:fit_quark} summarizes the fitting parameters for unified equations of state for $K'=0$ and parameter sets
$(g_V/G,H/G)=$(0.5,1.4), (0.8,1.5), and (1.0,1.6), as used in the calculations shown in Fig.~\ref{fig:M-R}.  The coefficients for the interpolated domain strongly correlate one another. To maintain the good accuracy of fits as in Fig.~\ref{fig:M-R}, we must keep 5-6 digits in the fitting parameters.

   The quark model parameters $(g_V,H)$ chosen here are suitable for interpolating with the APR equation of state. For interpolation with other nuclear equations of state or other possible interpolation schemes, it would be better to use slightly different sets of parameters. For this reason we have generated numerical equations of state for a wider range of $(g_V,H)$ which are fitted by Eq.~\ref{e_para}. Table \ref{tab:fit_quark_more} summarizes the fitting parameters.

The present equation of state QHC18, for several choices of QCD parameters, is given together with instructions for use on the  website:  {\em Home Page of Relativistic EOS table for supernovae} (http://user.numazu-ct.ac.jp/$\sim$sumi/eos/index.html).

\clearpage


\begin{thebibliography}{99}

\bibitem{haensel} See, e.g., P. Haensel,  A.~Y. Potekhin, and D.~G. Yakovlev, 
{\em Neutron stars 1:~Equation of state and structure} (Springer-Verlag, New York ) 2007).

\bibitem{annrev1}  G. Baym and  C.~J. Pethick, ``Neutron Stars,"  Annu.  Rev.  Nucl.  Sci.
25, 27-77 (1975);  ``Physics of neutron stars," Annu.~Rev. Astron.  Astrophys. {\bf 17}, 415-43 (1979).

\bibitem{Hell:2014xva}
  T.~Hell and W.~Weise,
  ``Dense baryonic matter: constraints from recent neutron star observations,''
  Phys.\ Rev.\ C {\bf 90} (2014),  045801 (2014).


\bibitem{Lattimer:2012nd}
  J.~M.~Lattimer,
  ``The nuclear equation of state and neutron star masses,''
  Ann.\ Rev.\ Nucl.\ Part.\ Sci.\  {\bf 62},  485-515  (2012). 
  
\bibitem{OzelFreire}
F. {\"O}zel and P. Freire, ``Masses, radii, and the equation of state of neutron stars,
 Annu. Rev. Astron. Astrophys. {\bf 54}, 401-440 (2016).


\bibitem{fredcole}  C. M. Miller, and F. K. Lamb, ``Observational constraints on neutron star masses and radii," Eur.\ Phys.\ J.\ A {\bf 52}, 69, 63: 1-20 (2016).

\bibitem{Ozel2010}F. {\"O}zel, G. Baym, and T.~G\"{u}ver, ``Astrophysical measurement of the equation of state of neutron star matter,'' Phys. Rev. D \textbf{82}, 101301(R) (2010). 

\bibitem{Steiner2012}A. W. Steiner, J. M. Lattimer, and E. F. Brown, ``The equation of state from observed masses and radii of neutron stars,'' Astrophys. J, {\bf 722},  33-54 (2010).

\bibitem{Steiner2012-2}A. W. Steiner, J. M. Lattimer, and E. F. Brown,  ``The neutron star mass-radius relation and the equation of state of dense matter,'' Astrophys. J. Letters \textbf{765}, L5: 1-5 (2013). 

\bibitem{Lattimer2014} J. M. Lattimer and A, W. Steiner, ``Neutron star masses and radii from quiescent low-mass X-ray binaries, "Astrophys. J, {\bf 784}, 123-137 (2014).

\bibitem{Ozel2015}F. \"{O}zel, D. Psaltis, T. G\"{u}ver, G. Baym, C. Heinke, and S. Guillot, ``The dense matter equation of state from neutron star radius and mass measurements,''   Astrophys. J. \textbf{820}, 28:1-25 (2016).

\bibitem{Steiner2015} 
  A.~W.~Steiner, S.~Gandolfi, F.~J.~Fattoyev and W.~G.~Newton,
  ``Using neutron star observations to determine crust thicknesses, moments of inertia, and tidal deformabilities,''
  Phys.\ Rev.\ C {\bf 91}, 015804:1-7 (2015).
 

\bibitem{Steiner2016} A. W. Steiner, J. M. Lattimer, and E. F. Brown, ``Neutron star radii, universal relations, and the role of prior distributions," Euro. Phys.. J, {\bf A52},  18:1-16  (2016).


\bibitem{Alvarez-Castillo:2016oln}
  D.~Alvarez-Castillo, A.~Ayriyan, S.~Benic, D.~Blaschke, H.~Grigorian and S.~Typel,
  ``New class of hybrid EoS and Bayesian M-R data analysis,''
  Eur.\ Phys.\ J.\ A {\bf 52}, 69: 1-11 (2016).

\bibitem{nicer}  K. C. Gendreau et al.,  ``The Neutron star Interior Composition Explorer (NICER): design and development,"
Proc. SPIE 9905, Space Telescopes and Instrumentation 2016: Ultraviolet to Gamma Ray, 99051H (July 22, 2016).

\bibitem{michi}  M. Baub\"ock, D. Psaltis, and F. \"Ozel,  ``Effects of spot size on neutron-star radius measurements from pulse profiles, 
Astrophys. J. {\bf 811}, 144:1-8 (2015).

\bibitem{Miller:2016kae}
  M.~C.~Miller,  ``The case for PSR J1614-2230 as a NICER target,''
  Astrophys.\ J.\  {\bf 822}, 27:1-7 (2016).

\bibitem{ozel-nicer} F. \"Ozel, D. Psaltis,, Z. Arzoumanian, S. Morsink, and M. Baub\"ock, ``Measuring neutron star radii via pulse profile modeling with NICER,"  Astrophys. J. {\bf 832}, 92:1-8 (2016).

\bibitem{Bogdanov_2008}
S. Bogdanov, J.E. Grindlay, and G.B. Rybicki, ``Thermal X-rays from millisecond pulsars: constraining the fundamental properties of neutron stars," Astrophys. J. {\bf 689}, 407-415 (2008).

\bibitem{timing} A. L. Watts, N. Andersson, D. Chakrabarty, M. Feroci, J. Hebeler, G.
Israel, F. K. Lamb, M. C. Miller, S. Morsink, F. \"Ozel, A.
Patruno, J. Poutanen, D. Psaltis, A. Schwenk, A. W. Steiner, L. Stella,
L. Tolos, and M. van der Klis, ``Measuring the neutron star equation of state using x-ray timing," Rev. Mod. Phys. {\bf 88}, 021001: 1-23 (2016).

\bibitem{kokkotas2001}
K. D. Kokkotas, T. A. Apostolatos, and N. Andersson, ``The inverse problem for pulsating neutron stars: a `fingerprint analysis' for the supranuclear equation of state.  Mon. Not. R. Astron. Soc. 320, 307-315 (2001).


\bibitem{GW170817} B. P. Abbott et al.  (LIGO Scientific Collaboration and Virgo Collaboration), ``GW170817: Observation of gravitational waves from a binary neutron star inspiral," Phys. Rev. Lett. {\bf 119}, 161101:1-18 (2017).

\bibitem{GW170817A} B. P. Abbott et al., ``Multi-messenger observations of a binary neutron star merger," Astrophys. J. Letters {\bf 848}, L12:1-59 (2017), and following papers in Astrophys. J. Letters {\bf 848}.

\bibitem{Hotokezaka:2011dh} 
  K.~Hotokezaka, K.~Kyutoku, H.~Okawa, M.~Shibata, and K.~Kiuchi,
``Binary Neutron Star Mergers: Dependence on the Nuclear Equation of State,''  Phys.\ Rev.\ D {\bf 83}, 124008 (2011);
arXiv:1105.4370 [astro-ph.HE].

\bibitem{takami} K. Takami, L. Rezzolla, and L. Baiotti, ``Constraining the equation of state of neutron stars from binary mergers,"  Phys. Rev. Letters {\bf 113}, 091104 (2014).
 
\bibitem{baiotti} L. Baiotti and L. Rezzolla, ``Binary neutron-star mergers: a review of Einstein's richest laboratory," Repts. Prog. Phys. (in press) (2017), https://doi.org/
arXiv:1607.03540 [gr-qc].

\bibitem{sekiguchi}  
M. Shibata, "Merger of binary neutron stars: Gravitational waves and electromagnetic counterparts," Nucl. Phys. A {\bf 956}, 225-232 (2016). 

\bibitem{stu1} 
V. Paschalidis, M. Ruiz and S. L. Shapiro, ``Relativistic simulations of black hole-neutron star coalescence:
the jet emerges," Astrophys. J.  Letters {\bf 806}, L14:1-5, (2015).

\bibitem{stu2} M. Ruiz, R. Lang, V. Paschalidis and S. L. Shapiro,
Binary neutron star mergers: a jet engine for short gamma-ray bursts,
Astrophys. J. Letters {\bf 824}, L1:1-5 (2016).

\bibitem{bernuzzi}    S.~Bernuzzi, D.~Radice, C.~D.~Ott, L.~F.~Roberts, P.~Moesta and F.~Galeazzi,
  ``How loud are neutron star mergers?,''
  Phys.\ Rev.\ D {\bf 94}, 024023:1-6 (2016).
  

\bibitem{aLIGO} B. P. Abbott et al.,  "GW150914: The advanced LIGO detectors in the era of first discoveries, Phys. Rev. Lett. {\bf 116}, 131103:1-12 (2016).
  
\bibitem{ligo-1}  B. P. Abbott et al.  (LIGO Scientific Collaboration and Virgo Collaboration), ``GW151226: Observation of Gravitational Waves from a 22-Solar-Mass Binary Black Hole Coalescence," Phys. Rev. Letters \textbf{116}, 241103 (2016).

\bibitem{ligo-2}  B. P. Abbott et al.  (LIGO Scientific Collaboration and Virgo Collaboration), ``Astrophysical implications of the binary black hole merger GW150914,"  Astrophys. J.  Letters \textbf{818}, L22:1-15 (2016). 

\bibitem{virgo} F. Acernese et al., ``Advanced Virgo: A second-generation interferometric gravitational wave detector," Class. Quantum Grav. {\bf 32}, 024001:1-55 (2015).

\bibitem{GEO} J. Hough et al. ``Proposal for a Joint German-British Interferometric Gravitational Wave Detector,"  available at eprints.gla.ac.uk/114852/7/114852.pdf.

\bibitem{Kagra}  Y. Aso et al. ``Interferometer design of the KAGRA gravitational wave detector,? Phys Rev D {\bf88}, 043007:1-15 (2013).

\bibitem{LIGO-India}  B. Iyer et al.  ``LIGO-INDIA: Proposal for an interferometric gravitational-wave observatory," LIGO Public Documents, Available at https://dcc.ligo.org/public/0075/M1100296/002/LIGO-India{\_}lw-v2.pdf (2011).

\bibitem{LISA}  See, Proc. 11th Int, LISA Symposium, J. Phys. Conf. Series {\bf 840}, 2017. 

\bibitem{PTA} M. McLaughlin, ``The international pulsar timing array: a galactic scale gravitational wave observatory," Gen Relativ Gravit. {\bf46} 1810:1-22 (2014).

\bibitem{fonseca} E. Fonseca et al.,  ``The NANOGrav nine-year data set: mass and geometric measurements of binary millisecond pulsars,''  Astrophys.\ J.\  {\bf 832}, 167:1-22 (2016).

\bibitem{Demorest} P. Demorest, T. Pennucci, S. M. Ransom, R. M. S. E., and J. W. T. Hessels, ``A two-solar-mass neutron star measured using Shapiro delay,'' Nature (London) \textbf{467}, 1081-3 (2010).

\bibitem{Antoniadis2013}J. Antoniadis, et al., ``A massive pulsar in a compact relativistic binary,'' Science \textbf{340}, 1233232 (2013).

\bibitem{vankerwijk} M. H. van Kerkwijk, R. P. Breton, and S. R. Kulkarni, ``Evidence for a massive neutron star from a radial-velocity study of the companion to the black widow pulsar PSR B1957+20,"  Astrophys. J. {\bf 728}, 95:1-8 (2011).
  
\bibitem{Schroeder-Halperin} Joshua Schroeder and Jules Halpern,  ``Observations and modeling of the companions of short period binary millisecond pulsars: Evidence for high-mass neutron stars,"  Astrophys. J. 793, 78:1-12 (2014).

\bibitem{Romani2012}R. W. Romani, A. V. Filippenko, J. M. Silverman, S. B. Cenko, J. Greiner, A. Rau, J. Elliott, and H. J. Pletsch, ``PSR J1311-3430: A heavyweight neutron star with a flyweight helium companion,''  Astrophys. J. Letters \textbf{760}, L36: 1-6 (2012). 

\bibitem{Romani2015} R. W. Romani, A. V. Filippenko, and S. B. Cenko, "A spectroscopic study of the extreme black widow PSR J1311-3430,"  Astrophys. J. {\textbf 804}, 115R: 1-10  (2015).    

\bibitem{Fukushima2011} K. Fukushima and T. Hatsuda, ``The phase diagram of dense QCD,'' Repts. Prog. Phys. \textbf{74}, 014001 (2011).

\bibitem{asakawa} M. Asakawa and K. Yazaki, ``Chiral restoration at finite density and temperature," Nucl. Phys. A {\bf 504},   668-84 (1989).

\bibitem{Hatsuda2006} T. Hatsuda, M. Tachibana, N. Yamamoto, and G. Baym, ``New critical point Induced by the axial anomaly in dense QCD,'' Phys. Rev. Letters \textbf{97}, 122001 (2006). 

\bibitem{Tolman1939}R. C. Tolman, ``Static solutions of Einstein's field equations for spheres of fluid,'' Phys. Rev. \textbf{55}, 364-73 (1939).
 
\bibitem{Oppenheimer1939}J. Oppenheimer and G. Volkoff, ``On massive neutron cores,'' Phys. Rev. \textbf{55}, 374-81 (1939).

\bibitem{APR}A. Akmal, V. R. Pandharipande, and D.G. Ravenhall, ``Equation of state of nucleon matter and neutron star structure,'' Phys. Rev. C \textbf{58}, 1804-28 (1998).

\bibitem{mpr}J. Morales, Jr., V.R. Pandharipande, and D.G. Ravenhall, ``Improved variational calculations of nucleon matter,'' Phys. Rev. C \textbf{66}, 054308 (2002).


\bibitem{lightreview}  J. Carlson, S. Gandolfi, F. Pederiva, S. C. Pieper, R. Schiavill, K.E. Schmidt, and R. B. Wiringa, "Quantum Monte Carlo methods for nuclear physics," Rev. Mod. Phys.  {\bf 87}, 1067-118 (2015).

\bibitem{gandolfi} S. Gandolfi, J. Carlson, and S. Reddy, ``Maximum mass and radius of neutron stars, and the nuclear symmetry energy,''  Phys. Rev. C {\bf 85}, 032801 (2012).

\bibitem{Gandolfi:2013baa}
  S.~Gandolfi, J.~Carlson, S.~Reddy, A.~W.~Steiner and R.~B.~Wiringa, ``The equation of state of neutron matter, symmetry energy, and neutron star structure,''
  Eur.\ Phys.\ J.\ A {\bf 50} 10:1-11 (2014).

\bibitem{HebelerBogner} See, e.g., K. Hebeler, S. K. Bogner, R. J. Furnstahl, A. Nogga, and A. Schwenk, Phys. Rev. C \textbf{83}, 031301(R) (2011). 

\bibitem{kaiser2} N. Kaiser,  ``Chiral four-body interactions in nuclear matter," Euro. Phys. J. A \textbf{48}, 135-45 (2012).

\bibitem{kaiser1} S. Fritsch, N. Kaiser and W. Weise, ``Chiral approach to nuclear matter: Role of two-pion exchange with virtual delta-isobar excitation," Nucl. Phys. A \textbf{750}, 259-93 (2005).

\bibitem{Bodmer1971} A. Bodmer, ``Collapsed Nuclei,'' Phys. Rev. D \textbf{4}, 1601-6 (1971).

\bibitem{Witten1984} E. Witten, ``Cosmic separation of phases,'' Phys. Rev. D \textbf{30}, 272-85 (1984).

\bibitem{Farhi1984}E. Farhi and R. L. Jae, ``Strange matter,'' Phys. Rev. D \textbf{30}, 2379-90 (1984).


\bibitem{Baldo2000}M. Baldo, G. Burgio, and H. Schulze, ``Hyperon stars in the Brueckner-Bethe-Goldstone theory,'' Phys. Rev. C \textbf{61}, 055801 (2000). 

\bibitem{Nishizaki2001}S. Nishizaki, Y. Yamamoto, and T. Takatsuka, ``Effective YN and YY interactions and hyperon-mixing in neutron star matter  -- Y $\equiv \Lambda$ case,'' Prog. Theor. Phys. \textbf{105}, 607-26 (2001).

\bibitem{Nishizaki2002}S. Nishizaki, Y. Yamamoto, and T. Takatsuka, ``Hyperon-mixed neutron star matter and neutron stars,'' Prog. Theor. Phys. \textbf{108}, 703-18 (2002).

\bibitem{Schulze2011} H. Schulze and T. Rijken, ``Maximum mass of hyperon stars with the Nijmegen ESC08 model,'' Phys. Rev. C \textbf{84}, 035801 (2011). 

\bibitem{BECnstars} C. J. Pethick, T. Sch\"afer, and A. Schwenk, ``Bose-Einstein condensates in neutron stars,"  in 
 in {\em Universal Themes of Bose-Einstein Condensation}, D. Snoke, N. Proukakis, and P. Littlewood, eds, Cambridge Univ.
Press,  Cambridge 2017). pp.~573-592;  arXiv:1507.05839.

\bibitem{migdalpi}  A.~B. Migdal, ``Stability of vacuum and limiting fields," Zh. Eksp. Teor. Fiz. 61, 2209-24 (1971) [Sov. Phys. JETP 34, 1184-91 (1972)]; 
``Pion Fields in Nuclear Matter," Rev. Mod. Phys. 50, 107-72 (1978).

\bibitem{barshay} S. Barshay, G. Vagradov, and G. E. Brown, ``Possibility of a phase transition to a pion condensate in neutron stars,'' Phys. Letters B \textbf{43} 5, 359 {2013}61 (1973).  

\bibitem{GBpi}   G. Baym, ``Pion condensation in nuclear and neutron star matter,''  Phys.
Rev.  Letters 30, 1340-2 (1973).

\bibitem{sawyer} R~.F. Sawyer and D. J. Scalapino, ``Pion condensation in superdense nuclear matter,'' Phys. Rev. D \textbf{7} 4, 953-64 (1973). 

\bibitem{AM}A. Mukherjee, ``Variational theory of hot nucleon matter. II. Spin-isospin correlations and equation of state of nuclear and neutron matter,'' Phys. Rev. C \textbf{79}, 045811 (2009).

\bibitem{kaplan} D. B. Kaplan and A. E. Nelson, ``Strange goings on in dense nucleonic matter,'' Phys. Letters B \textbf{175} 1, 57-63 (1986). 

\bibitem{PPT}V. R. Pandharipande, C. J. Pethick, and V. Thorsson, ``Kaon energies in dense matter,'' Phys. Rev. Letters \textbf{75}, 4567-70 (1995).

\bibitem{BB}H.A. Bethe and G.E. Brown, ``Evolution of binary compact objects that merge,'' Astrophys. J. \textbf{506}, 780-9 (1998), and references therein.

\bibitem{GS}N.K. Glendenning and J. Schaffner-Bielich, ``First order kaon condensate,'' Phys. Rev. C \textbf{60}, 025803 (1999). 


\bibitem{wambach-hyp}  H. Dapo, B.-J. Schaefer, and J. Wambach,
``Appearance of hyperons in neutron stars," Phys. Rev. C {\bf 81}, 035803 (2010).


\bibitem{Weissenborn2011}S. Weissenborn, D. Chatterjee and J. Schaffner-Bielich, ``Hyperons and massive neutron stars: Vector repulsion and SU(3) symmetry,'' Phys. Rev. C \textbf{85}, 065802 (2012). 

\bibitem{Weissenborn2012}S. Weissenborn, D. Chatterjee and J. Schaffner-Bielich, ``Hyperons and massive neutron stars: the role of hyperon potentials,'' Nucl. Phys. A \textbf{881}, 62-77 (2012).

\bibitem{lonardoni} D. Lonardoni, A. Lovato, S. Gandolfi, and F. Pederiva, ``Hyperon puzzle: hints from quantum Monte Carlo Calculations,"  Phys. Rev. Letters \textbf{114}, 092301 (2015). 

\bibitem{Ukawa:2015eka} 
 A.~Ukawa,  ``Kenneth Wilson and lattice QCD,''
  J.\ Statist.\ Phys.\  {\bf 160}, 1081-1124 (2015).

\bibitem{halqcd}   
T. Inoue (HAL QCD Collaboration), ``Hyperon single-particle potentials from QCD on lattice," Proc. 26th Int. Nucl. Phys. Conf. (INPC2016), Adelaide, Australia,  Sept. 2016, arXiv:1612.08399. 


\bibitem{dirk} S.A. Chin and J.D. Walecka, `` An equation of state for nuclear and higher-density matter based on a relativistic mean-field theory," Phys. Letters {\bf B52} 24-28 (1974).

\bibitem{MS}H. M\"uller and B.D. Serot, ``Relativistic mean-field theory and the high-density nuclear equation of state,'' Nucl. Phys. A \textbf{606}, 508-37 (1996).

\bibitem{radius} J. M. Lattimer and M. Prakash, ``Neutron star observations: Prognosis for equation of state constraints,'' Phys. Rep. \textbf{442}, 109-65 (2007).

\bibitem{Baym1979}G. Baym, ``Confinement of quarks in nuclear matter,'' Physica \textbf{96A}, 131-135 (1979).

\bibitem{Satz1} T. Celik, F. Karsch, and H. Satz, ``A percolation approach to strongly interacting matter,'' Phys. Letters B {\bf 97}, 128-30 (1980).

\bibitem{Satz2}  H. Satz, ``Deconfinement and percolation,"  Nucl. Phys. A {\bf 642}, c130-c142 (1998);  Repts. Prog. Phys. {\bf  63} ,1511 (2000);
{\em Extreme States of Matter in Strong Interaction Physics} (Springer 2012). 

\bibitem{Nambu:1961tp}
  Y.~Nambu and G.~Jona-Lasinio, ``Dynamical model of elementary particles based on an analogy with superconductivity. 1,'' Phys.\ Rev.\  {\bf 122}, 345-58 (1961).
  
\bibitem{Holt:2014hma}
  J.~W.~Holt, M.~Rho and W.~Weise,
  ``Chiral symmetry and effective field theories for hadronic, nuclear and stellar matter,''
  Phys.\ Rept.\  {\bf 621}, 2-75 (2016).
  
\bibitem{friman} T. Kr{\"u}ger, I. Tews, B. Friman, K. Hebeler, and A. Schwenk, ``The chiral condensate in neutron matter, Phys. Letters B726,  412-416 (2013).
 
 \bibitem{dtsonReversal}D.T. Son and M. A. Stephanov, ``Inverse meson
mass ordering in the color-flavor-locking phase of high-density QCD,''
Phys. Rev. D {\bf 61}, 074012 (2000). 

\bibitem{fukushimaConstruct}K. Fukushima, ``Quark description of
the Nambu-Goldstone Bosons in the color-flavor-locked phase,'' Phys.
Rev. D {\bf 70}, 094014 (2004).
   
\bibitem{Yamamoto} N. Yamamoto, M. Tachibana, T. Hatsuda, and G. Baym, ``Phase structure, collective modes, and the axial anomaly in dense QCD,'' Phys. Rev. D \textbf{76}, 074001 (2007).

\bibitem{song} Y. Song and G. Baym, ``Generalized Nambu-Goldstone pion in dense matter: a schematic NJL model,'' Phys. Rev. C (submtted 2017); arXiv:1703.08236 [nucl-th].

\bibitem{Alford:2007xm}   M.~G.~Alford, A.~Schmitt, K.~Rajagopal and T.~Sch$\ddot{ {\rm a} }$fer, ``Color superconductivity in dense quark matter,'' Rev.\ Mod.\ Phys.\  {\bf 80},  1455-515 (2008).  

\bibitem{Shifman:1978bx}
  M.~A.~Shifman, A.~I.~Vainshtein and V.~I.~Zakharov, ``QCD and resonance physics. Theoretical foundations,'' Nucl.\ Phys.\ B {\bf 147} 385-447 (1979).
  
\bibitem{ivanenko} D. Ivanenko and D. F. Kurdgelaidze, ``Remarks on quark stars," Lett. Nuovo Cim. {\bf 2}. 13-16 (1969), and references therein.
  
\bibitem{itoh}  N. Itoh, `Hydrostatic equilibrium of hypothetical quark stars,"  Prog. Theor. Phys. {\bf 44}, 291-292 (1970).
    
\bibitem{chin} G. Baym and S. A. Chin, ``Can a neutron star be a giant MIT bag?,''  Phys. Letters B \textbf{62}, 241-4 (1976).

\bibitem{Buballa2005} M. Buballa, ``NJL-model analysis of dense quark matter,'' Phys. Rep. \textbf{407}, 205-376 (2005).


\bibitem{Benic:2014jia}
  S.~Benic, D.~Blaschke, D.~E.~Alvarez-Castillo, T.~Fischer and S.~Typel,
  ``A new quark-hadron hybrid equation of state for astrophysics - I. High-mass twin compact stars,''
  Astron.\ Astrophys.\  {\bf 577}, A40: 1-8 (2015).


\bibitem{Alvarez-Castillo:2016wqj}
  D.~Alvarez-Castillo, S.~Benic, D.~Blaschke, S.~Han and S.~Typel,
  ``Neutron star mass limit at $2M_\odot$ supports the existence of a CEP,''  Eur. Phys. J. A 52, 232: 1-8 (2016).
  arXiv:1608.02425 [nucl-th].
  
\bibitem{Ranea-Sandoval:2015ldr}
  I.~F.~Ranea-Sandoval, S.~Han, M.~G.~Orsaria, G.~A.~Contrera, F.~Weber and M.~G.~Alford,
  ``Constant-sound-speed parametrization for Nambu-Jona-Lasinio models of quark matter in hybrid stars,''
  Phys.\ Rev.\ C {\bf 93},  045812  (2016).

\bibitem{Schafer:1998ef}  T.~Sch$\ddot{ {\rm a} }$fer and F.~Wilczek, ``Continuity of quark and hadron matter,'' Phys.\ Rev.\ Letters\  {\bf 82}, 3956-9 (1999).

\bibitem{Bratovic2012} N. Bratovic, T. Hatsuda, and W. Weise, ``Role of vector interaction and axial anomaly in the PNJL modeling of the QCD phase diagram,'' Phys. Letters B \textbf{719}, 131-5 (2012).

\bibitem{Lourenco2012} O. Lourenco, M. Dutra, T. Frederico, A. Delfino, and M. Malheiro, ``Vector interaction strength in Polyakov-Nambu-Jona-Lasinio models from hadron-quark phase diagrams,'' Phys. Rev. D \textbf{85}, 097504 (2012). 

\bibitem{Masuda2013-0} K. Masuda, T. Hatsuda, and T. Takatsuka, ``Hadron-quark crossover and massive hybrid stars with strangeness,'' Astrophys. J. \textbf{764},12: 1-5 (2013). 

\bibitem{Masuda2013}K. Masuda, T. Hatsuda, and T. Takatsuka, ``Hadron-quark crossover and massive hybrid stars,'' Prog. Theor. and Exp. Phys. \textbf{7}, 073D01 (2013). 

\bibitem{Klahn2013}T. Kl\"{a}hn, R. {\L}astowiecki, and D. Blaschke, ``Implications of the measurement of pulsars with two solar masses for quark matter in compact stars and heavy-ion collisions: A Nambu-Jona-Lasinio model case study,''  Phys. Rev. D \textbf{88}, 085001 (2013). 

\bibitem{Masuda:2015kha}
  K.~Masuda, T.~Hatsuda and T.~Takatsuka,
 ``Hyperon puzzle, hadron-quark crossover and massive neutron stars,''
  Eur.\ Phys.\ J.\ A {\bf 52}, 65-79 (2016). 
  

\bibitem{potekhin}  P.  Haensel and A. Y. Potekhin, ``Analytical representations of unified equations of state of neutron-star matter,"    Astron.\  Astrophys. 428 ,191-197 (2004); tabulations given at http://www.ioffe.ru/astro/NSG/NSEOS.  

\bibitem{bbp} G. Baym, H. A. Bethe, and C. J. Pethick, ``Neutron star matter,'' Nucl. Phys. A {\bf 175}, 225-71 (1971).

\bibitem{sly} F. Douchin and P. Haensel,  ``A unified equation of state of dense matter and neutron star structure,'' Astron.\ Astrophys.\  {\bf 380}, 151-67 (2001).
 
\bibitem{Togashi:2017mjp}
  H.~Togashi, K.~Nakazato, Y.~Takehara, S.~Yamamuro, H.~Suzuki, and M.~Takano, ``Nuclear equation of state for core-collapse supernova simulations with realistic nuclear forces,''
  Nucl.\ Phys.\ A {\bf 961}, 78-105 (2017).  (\protect{http://www.np.phys.waseda.ac.jp/EOS/})
 
\bibitem{Fraga:2013qra}E. S. Fraga, A. Kurkela and A. Vuorinen, ``Interacting quark matter equation of state for neutron stars,'' Astrophys. J. Letters, \textbf{781}, L25: 1-5 (2014).

\bibitem{Kurkela:2014vha}A. Kurkela, E. S. Fraga, J. Schaffner-Bielich and A. Vuorinen, ``Constraining neutron star matter with Quantum Chromodynamics,'' Astrophys.\ J.\  {\bf 789}, 127-33 (2014).
  
\bibitem{Kojo2014}T. Kojo, P. D. Powell, Y. Song, and G. Baym, ``Phenomenological QCD equation of state for massive neutron stars," Phys. Rev. D \textbf{91}, 045003 (2015). 

\bibitem{Kojo:2015fua} T.~Kojo, ``Phenomenological neutron star equations of state: 3-window modeling of QCD matter,'' Eur.\ Phys.\ J.\ A {\bf 52},  51-69 (2016). 

\bibitem{qm2015-tk} T. Kojo, P. D. Powell, Y. Song, and G. Baym, ``Phenomenological QCD equations of state for neutron stars," Proc. QM2015 Kobe, Nucl. Phys. A {\bf 956}, 821-5 (2016).
  
 
\bibitem{Whittenbury:2013wma}
  D.~L.~Whittenbury, J.~D.~Carroll, A.~W.~Thomas, K.~Tsushima and J.~R.~Stone,
  ``Quark-Meson Coupling Model, Nuclear Matter Constraints and Neutron Star Properties,''
  Phys.\ Rev.\ C {\bf 89}, 065801 (2014). 
   
\bibitem{Whittenbury:2015ziz}
  D.~L.~Whittenbury, H.~H.~Matevosyan and A.~W.~Thomas,
  ``Hybrid stars using the quark-meson coupling and proper-time Nambu-Jona-Lasinio models,''
  Phys.\ Rev.\ C {\bf 93},  035807 (2016).
   
\bibitem{Polyakov} A. M. Polyakov, ``Thermal properties of gauge fields and quark liberation,'' Phys. Letters {\bf B} 72, 477-80 (1978).

\bibitem{Susskind:1979up} L.~Susskind, ``Lattice models of quark confinement at high temperature,'' Phys.\ Rev.\ D {\bf 20},  2610-8 (1979).

\bibitem{Banks:1979fi} T.~Banks and E.~Rabinovici, ``Finite temperature behavior of the lattice Abelian Higgs model,'' Nucl.\ Phys.\ B {\bf 160}, 349-79 (1979).

\bibitem{Svetitsky} B. Svetitsky, ``Symmetry aspects of finite temperature confinement transitions,'' Phys. Rep. \textbf{132}, 1-53 (1986).

\bibitem{Aoki:2006we} Y.~Aoki, G.~Endrodi, Z.~Fodor, S.~D.~Katz and K.~K.~Szabo, ``The order of the quantum chromodynamics transition predicted by the standard model of particle physics,''  Nature {\bf 443} 675-678 (2006).

\bibitem{bazarov-2012} A. Bazavov et al. 
(HotQCD Collaboration),  ``Chiral and deconfinement aspects of the QCD transition," Phys. Rev. D {\bf 85} 054503  (2012). 

\bibitem{hotQCD} A. Bazavov (HotQCD collaboration), ``Equation of state in ( 2+1 )-flavor QCD,'' Phys. Rev. D \textbf{90}, 094503 (2014). 

\bibitem{WB} S. Borsanyi, Z. Fodor, C, Hoelbling, S. D. Katz, S. Krieg, and K. K. Szabo, ``Full result for the QCD equation of state with 2+1 flavors'', Phys. Letters B \textbf{370}, 99-104 (2014).


\bibitem{DeTar:1985kx}
  C.~E.~Detar,
  ``A conjecture concerning the modes of excitation of the quark-gluon plasma,''
  Phys.\ Rev.\ D {\bf 32}, 276-83  (1985).
 
\bibitem{Hatsuda1985} 
T.~Hatsuda and T.~Kunihiro, 
``Fluctuation effects in hot quark matter: precursors of chiral transition at finite temperature,'' 
Phys.\ Rev.\ Letters\  {\bf 55} 158-61 (1985). 

\bibitem{Pisarski:2000eq} R.~D.~Pisarski, ``Quark gluon plasma as a condensate of SU(3) Wilson lines,'' Phys.\ Rev.\ D {\bf 62}, 111501  (2000).

\bibitem{Dumitru:2010mj} A.~Dumitru, Y.~Guo, Y.~Hidaka, C.~P.~K.~Altes and R.~D.~Pisarski, ``How wide is the transition to deconfinement?,'' Phys.\ Rev.\ D {\bf 83}, 034022  (2011). 

\bibitem{Pisarski:2016ixt}  R.~D.~Pisarski and V.~V.~Skokov, ``A chiral matrix model of the semi-quark gluon plasma in QCD,'' Phys. Rev. D {\bf 94}, 034015 (2016). 

\bibitem{Blaizot:2000fc}
  J.~P.~Blaizot, E.~Iancu and A.~Rebhan, ``Approximately selfconsistent resummations for the thermodynamics of the quark gluon plasma. 1. Entropy and density,'' Phys.\ Rev.\ D {\bf 63} 065003  (2001). 
  
\bibitem{Andersen:2011sf}
  J.~O.~Andersen, L.~E.~Leganger, M.~Strickland and N.~Su, ``Three-loop HTL QCD thermodynamics,'' JHEP {\bf 2011}, 53-77  (2011).  

\bibitem{kenjitoru}  K. Fukushima and T. Kojo,  ``The quarkyonic star," Astrophys.\ J.\  {\bf 817},  180-8  (2016). 

\bibitem{Vogl:1991qt}  U.~Vogl and W.~Weise, ``The Nambu and Jona Lasinio model: Its implications for hadrons and nuclei,''Prog.\ Part.\ Nucl.\ Phys.\  {\bf 27}, 195-272  (1991).

\bibitem{Lutz1992}M. Lutz, S. Klimt, and W. Weise, ``Meson properties at finite temperature and baryon density,'' Nucl. Phys. A  \textbf{542}, 521-58 (1992).

\bibitem{Klevansky:1992qe}  
 S.~P.~Klevansky,``The Nambu-Jona-Lasinio model of quantum chromodynamics,''
  Rev.\ Mod.\ Phys.\  {\bf 64}, 649-708 (1992).
 
\bibitem{Hatsuda1994}  T. Hatsuda and T. Kunihiro, ``QCD phenomenology based on a chiral effective Lagrangian,'' Phys. Rep. \textbf{247}, 221-367 (1994).   

\bibitem{Rehberg} P.~Rehberg, S.~P.~Klevansky and J.~Hufner, ``Hadronization in the SU(3) Nambu-Jona-Lasinio model,'' Phys.\ Rev.\ C {\bf 53}, 410-29  (1996).

\bibitem{Powell} P. D. Powell and G. Baym, ``Axial anomaly and the three-flavor Nambu--Jona-Lasinio model with confinement: Constructing the QCD phase diagram,'' Phys. Rev. D \textbf{85}, 074003 (2012).

\bibitem{Powell2}P. D. Powell and G. Baym, ``Asymmetric pairing of realistic mass quarks and color neutrality in the Polyakov--Nambu--Jona-Lasinio model of QCD,'' Phys. Rev. D \textbf{88}, 014012 (2013).

\bibitem{Abuki} H. Abuki, G. Baym, T. Hatsuda, and N. Yamamoto, ``Nambu--Jona-Lasinio model of dense three-flavor matter with axial anomaly: The low temperature critical point and BEC-BCS diquark crossover,'' Phys. Rev. D \textbf{81}, 125010 (2010).

\bibitem{Takatsuka} T. Takatsuka, T. Hatsuda, and K. Masuda, ``Massive hybrid stars with strangeness,'' [arXiv:1402.4677 [nucl-th]] (2013). 
7th Intl. Symp. on Chiral Symmetry in Hadrons and Nuclei (CHIRAL 13) Oct. 2013, Beijing (World
Scientific Publ. Co., 2015).


\bibitem{Alford:2015gna}
  M.~G.~Alford and S.~Han,
  ``Characteristics of hybrid compact stars with a sharp hadron-quark interface,''
  Eur.\ Phys.\ J.\ A {\bf 52},  62-76 (2016). 
   

\bibitem{pod2000} D. Psaltis, F. \"Ozel, and S. DeDeo, "Photon propagation around compact objects and the inferred properties of thermally emitting neutron stars," Astrophys. J. {\bf 544} 390-6 (2000).
  
\bibitem{fozel}  F. {\"O}zel, ``Surface emission from neutron stars and implications for the physics of their interiors," Repts. Prog. Phys. {\bf 76}, 016901 (2013).

\bibitem{lai} D. Lai, E.~E. Salpeter, and S.~L. Shapiro, ``Hydrogen molecules and chains in a superstrong magnetic field, "Phys. Rev. A {\bf 45} 4832-4847 (1992).

\bibitem{feryalmag} F. {\"O}zel, ``Surface emission properties of strongly magnetic neuton stars,"    Astrophys.~J. {\bf 563}, 276-288 (2001). 

\bibitem{smitha} W. DeGottardi, Q. Meng, S. Vishveshwara, and G. Baym, ``Surface radiation from magnetized neutron stars,'' to be submitted.

\bibitem{shapiro17A}  S.~L. Shapiro, ``Black holes, disks, and jets following binary mergers and stellar collapse:
the narrow range of electromagnetic luminosities and accretion rates," Phys. Rev. D {\bf 95}, 
101303(R):1-6 (2017).

\bibitem{m17}   B.~D. Metzger, ```Welcome to the multi-messenger era!
Lessons from a neutron star merger and the landscape ahead,"  arXiv:1710.05931.

\bibitem{mm17} B. Margalit and B.~D. Metzger, ``Constraining the maximum mass of neutron stars from multi-messenger observations of GW170817,"  arXiv:1710.05938.

\bibitem{stu17B} M. Ruiz, S.~L. Shapiro, and A. Tsokaros, ``GW170817, General relativistic magnetohydrodynamic simulations, and the neutron star maximum mass," arXiv:1711.00473.

\bibitem{rezzolla17} L. Rezzolla, E.~R. Most, and L.~R. Weih, ``Using gravitational-wave observations and quasi-universal relations to constrain the maximum mass of neutron stars, arXiv:1711.00314.

\bibitem{lawrence2015}
S. Lawrence, J.~G. Tervala, P.~F. Bedaque, and M.~C. Miller, ``An upper bound neutron star masses from models of short gamma-ray bursts," Astrophys. J. {\bf 808}, 186:1-7 (2015). 

\bibitem{shibata17} M. Shibata, S. Fujibayashi, K. Hotokezaka, K. Kiuchi, K. Kyutoku, Y. Sekiguchi, and M. Tanaka, ``GW170817: Modeling based on numerical relativity and its implications," arXiv:1710.07579.

\bibitem{janka17} A. Bauswein, O. Just, H.-T. Janka, and N. Stergioulas, ``Neutron-star radius constraints from GW170817 and future detections," arXiv1710.06843.

\bibitem{radice17} D. Radice, A. Perego, F. Zappa, and S. Bernuzzi, ``GW170817: Joint contstraint on the neutron star equation of state from multimessenger observations," arXiv:1711.03647.

\bibitem{poisson} E. Poisson and C. M. Will, {\em Gravity: newtonian, post-newtonian, relativistic},
  Cambridge Univ. Press, Cambridge, 2014, Eq. (2.249).

\bibitem{kip} K.~S. Thorne, ``Tidal stabilization of rigidly rotating, fully relativistic neutron stars,"	Phys. Rev. D {\bf 58}, 124031(1998).

\bibitem{monk} W.~H. Munk and G.~J.~F. MacDonald, {\em The rotation of the earth: a geophysical discussion}, Cambridge Univ. Press, Cambridge, 1960, Eq. (5.7.1).

\bibitem{Hinderer2008} T. Hinderer, ``Tidal Love numbers of neutron stars,"
Astrophys. J.  {\bf 677} 1216-20 (2008); erratum Astrophys. J. {\bf 697}, 964 (2009).

\bibitem{Hinderer2008a} {\'E}.~{\'E}. Flanagan and T. Hinderer, ``Constraining neutron-star tidal Love numbers with gravitational-wave detectors," Phys. Rev. D{\bf 77}, 021502(R)
(2008).

\bibitem{Hinderer:2009ca}
T.~Hinderer, B.~D.~Lackey, R.~N.~Lang and J.~S.~Read, ``Tidal deformability of neutron stars with realistic equations of state and their gravitational wave signatures in binary inspiral,''  Phys. Rev. D{\bf 81}, 123016 (2010). 
 
\bibitem{read2013} J~S.Read, L. Baiotti, J.~D.~E. Creighton, J~.L. Friedman, B. Giacomazzo, K. Kyutoku, C Markakis, L. Rezzolla, M, Shibata, and K. Taniguchi, ``Matter effects on binary neutron star waveforms," Phys. Rev. D{\bf 88}, 044042 (2013). 

 

\bibitem{tsuruta} 
S.Tsuruta, ``Thermal properties and detectability of neutron stars, II. 
Thermal evolution of rotation-powered neutron stars,"  Phys. Rep. {\bf 292}, 1 (1998). 

\bibitem{yakovlev} 
D.~G. Yakovlev and C.~J. Pethick, ``Neutron star cooling,'' Annu.~Rev. Astron.  Astrophys. {\bf 42} 169-210  (2004). 

\bibitem{Potekhin:2015qsa} 
  A.~Y.~Potekhin, J.~A.~Pons and D.~Page,
  ``Neutron stars - cooling and transport,''
  Space Sci.\ Rev.\  {\bf 191}, 239 (2015).
 
\bibitem{Pethick92} C.~J. Pethick, ``Cooling of neutron stars," Rev. Mod. Phys. {\bf 64}, 1133-40 (1992).
 
\bibitem{yls} D.G. Yakovlev, K.P. Levenfish, Yu.A. Shibanov, ``Cooling neutron stars and superfluidity in their interiors,"   Uspekhi Fizicheskikh Nauk {\bf 169} 825-68 (1999); Physics Uspekhi {\bf 42}, 737-778 (1999).   
 
\bibitem{Page:2013hxa} 
  D.~Page, J.~M.~Lattimer, M.~Prakash and A.~W.~Steiner,
  ``Stellar superfluids''  in {\em Novel superfluids}, Vol. 2, Ed. by K-H. Bennemann and J. B. Ketterson
 (Int. Series Monographs on Phys., Oxford Univ. Press).

\bibitem{Wijnands:2017jsc} 
  R.~Wijnands, N.~Degenaar and D.~Page,
  ``Cooling of accretion-heated neutron stars,''
  J.\ Astrophys.\ Astron.\  {\bf 38}, 49 (2017).
 
\bibitem{cumming} A. Cumming, E. F. Brown, F, J. Fattoyev, C. J. Horowitz, D. Page, and S, Reddy, ``Lower limit on the heat capacity of the neutron star core," Phys. Rev. C {\bf 95}, 025806 (2017).
 
\bibitem{Psaltis2014}
D. Psaltis and F.  \"Ozel, ``Pulse profiles from spinning neutron stars in the Hartle-Thorne approximation," Astrophys. J, {\bf 792}, 87-93 (2014).

   
 \bibitem{RLS} L. Rezzolla, F. K. Lamb, S. L Shapiro, 
``R-mode oscillations in rotating magnetic neutron stars," Astrophys. J. Letters {\bf 531}, L139-L142 (2000).

\bibitem{rmodes} M. Alford, ``Analysis of how pulsar timing data can constrain the composition of a neutron star via r-mode oscillations,"   Phys. Rev. Letters {\bf 113} 251102:1-5 (2014).

\bibitem{watts} A.L. Watts, and T. E. Strohmayer (2006), ``Detection with RHESSI of high-hrequency X-ray oscillations in the tail of the 2004 hyperflare from SGR 1806-20," Astrophys. J. {\bf 637}, L117-L120 (2006).

\bibitem{qpo} M. C. Miller, ``QPO constraints on neutron stars, New Astronomy Reviews {\bf 54}, 128-134 (2010).  


\bibitem{bps} G. Baym, C. J. Pethick and P. Sutherland, ``The ground state of matter at high densities: Equation of state and stellar models,'' Astrophys.\ J.\  {\bf 170}, 299-317 (1971).

\bibitem{HZD} P. Haensel, J. L. Zdunik, and F.  Douchin, ``Equation of state of dense matter and the minimum mass of cold neutron stars,"  Astron.\ Astrophys. {\bf 385}, 301-307 (2002).  

\bibitem{dgr} D. G. Ravenhall, C. J. Pethick, and J. R. Wilson, ``Structure of matter below nuclear saturation density,'' Phys. Rev. Letters \textbf{50}, 2066-9 (1983).

\bibitem{hsy} M. Hashimoto, H. Seki, and M. Yamada, ``Shape of nuclei in the crust of neutron star," Prog. Theor. Phys. \textbf{71}, 320-6 (1984). 

\bibitem{ohy} K. Oyamatsu, M. Hashimoto, and M. Yamada, ``Further study of the nuclear shape in high-density matter," Prog. Theor. Phys. \textbf{72}, 373-5 (1984). 

\bibitem{wk} R. D. Williams and S. E. Koonin, ``Sub-saturation phases of nuclear matter,'' Nucl. Phys. A \textbf{435}, 844-58 (1985).

\bibitem{wis} G. Watanabe, K. Iida, and K. Sato, ``Thermodynamic properties of nuclear 'pasta' in neutron star crusts,''   Nucl. Phys. A \textbf{676}, 455-73 (2000);  {\it ibid.} A \textbf{687}, 512-31 (2001); Erratum-ibid. A \textbf{726}, 357-65 (2003) .

\bibitem{Horowitz:2004yf} C.~J.~Horowitz, M.~A.~Perez-Garcia and J.~Piekarewicz,
  ``Neutrino-pasta scattering: The opacity of nonuniform neutron - rich matter,''
  Phys.\ Rev.\ C {\bf 69} , 045804 (2004).  
  
\bibitem{Watanabe:2004tr}
  G.~Watanabe, T.~Maruyama, K.~Sato, K.~Yasuoka and T.~Ebisuzaki,
  ``Simulation of transitions between `Pasta' phases in dense matter,''
  Phys.\ Rev.\ Letters\  {\bf 94}, 031101 (2005).
  
  \bibitem{Maruyama:2005vb}
  T.~Maruyama, T.~Tatsumi, D.~.N.~Voskresensky, T.~Tanigawa and S.~Chiba,
  ``Nuclear pasta structures and the charge screening effect,''
  Phys.\ Rev.\ C {\bf 72}, 015802 (2005).
  
\bibitem{Maruyama:2012bi}
 T.~Maruyama, G.~Watanabe and S.~Chiba,  ``Molecular dynamics for dense matter,'' Prog. Theor. Exp. Phys. {\bf 2012} 01A201.

\bibitem{Iida:2013fra}
  K.~Iida and K.~Oyamatsu, ``Symmetry energy, unstable nuclei, and neutron star crusts,''
  Eur.\ Phys.\ J.\ A {\bf 50} 42-57  (2014). 
  
\bibitem{urban} N. Martin and M. Urban, ``Liquid-gas coexistence versus energy minimization with respect to the density profile 
in the inhomogeneous inner crust of neutron stars,"  Phys. Rev. C {\bf 92}, 015803 (2015).  
  
\bibitem{pichon} P. Haensel and B. Pichon, ``Experimental nuclear masses and the ground state of cold dense matter,'' Astron.\ Astrophys.\ {\bf 283}, 313-8 (1994).

\bibitem{pudliner} B. S. Pudliner, A. Smerzi, J. Carlson, V. R. Pandharipande, S. C. Pieper, and D. G. Ravenhall, ``Neutron drops and Skyrme energy density functionals,'' Phys. Rev. Letters {\bf 76}, 2416-19 (1996).

\bibitem{sakurai} H. Sakurai, Repts. Prog. Phys. to be published.  

\bibitem{douchin}  F. Douchin , P. Haensel, and J. Meyer, ``Nuclear surface and curvature properties for SLy Skyrme forces and nuclei in the inner neutron-star crust,'' Nucl. Phys. A {\bf 665}, 419-46  (2000).

\bibitem{pearson} J. M. Pearson, N. Chamel, S. Goriely, and C. Ducoin, ``Inner crust of neutron stars with mass-fitted Skyrme functionals,'' Phys. Rev. C {\bf 85}, 065803 (2012).

\bibitem{bennett}   D. G. Ravenhall, C. D. Bennett, and C. J. Pethick, ``Nuclear surface energy and neutron-star matter,''  Phys. Rev. Letters {\bf 28}, 978-81 (1972).

\bibitem{avogadro} P.  Avogadro, F.  Barranco, R. A. Broglia, and E.  Vigezzi, ``Vortex-nucleus interaction in the inner crust of neutron stars,'' Nucl. Phys. A {\bf 811}, 378-412 (2008).

\bibitem{avogadro1}  P. Avogadro, F. Barranco, A. Idini, and E. Vigezzi, ``Medium polarization effects on the superfluidity of finite nuclei and of the Inner crust of neutron stars,"  in  {\em Fifty Years of Nuclear BCS: Pairing in Finite Systems},
R. A. Broglia and V. Zelevinsky, eds. (World Scientific Publ. Co., 2013).

\bibitem{chamel} N. Chamel, J.M. Pearson, and S. Goriely, ``Pairing: from atomic nuclei to neutron-star crusts,"  in  {\em Fifty years of nuclear BCS: Pairing in finite systems},
R. A. Broglia and V. Zelevinsky, eds. (World Scientific Publ. Co., 2013).

\bibitem{apw} T. Ainsworth, D. Pines, and J. Wambach, ``Effective interactions and superfluid energy gaps for low density neutron matter," 
Phys. Letters B {\bf 222}, 173-8 (1989).

\bibitem{gps} A. Gezerlis, C. J. Pethick, and A. Schwenk,
``Pairing and superfluidity of nucleons in neutron stars,"
in {\em Novel Superfluids}, v.~2,
 K. H. Bennemann and J. B. Ketterson, eds. (Oxford University Press, Oxford, 2014) p. 580;
arXiv:1406.6109.

\bibitem{lorenz} C. J. Pethick, D. G. Ravenhall, and C. Lorenz, ``The inner boundary of a neutron star crust,'' Nucl. Phys. A {\bf 584}, 675-703 (1995).

\bibitem{kai} K. Hebeler, J. M. Lattimer, C. J. Pethick, and A. Schwenk, ``Equation of state and neutron star properties constrained by nuclear physics and observation,'' Astrophys.\ J.\   {\bf 773}, 11-24 (2013); arXiv:1303.4662. 

\bibitem{hartle1978}  J. Hartle, ``Bounds on the mass and moment of inertia of non-rotating neutron stars,'' Phys. Reports {\bf 46}, 201-47 (1978). 

\bibitem{zdunik} J. L. Zdunik, M. Fortin and P. Haensel, `Neutron star properties and the equation of state for the core,''
Astron. and Astrophysm {\bf 599} A119:1-8 (2017).

\bibitem{sotani} H. Sotani, K.Iida, and K. Oyamatsu, ``Probing crustal structures from neutron star compactness,"  arXiv 1706.04736.

\bibitem{kobyakov}  D. Kobyakov and C. J. Pethick, ``Towards a metallurgy of neutron star crusts,'' Phys. Rev. Letters {\bf 112}, 112504 (2014).

\bibitem{kobyakov2} D. Kobyakov and C. J. Pethick,  ``Nucleus-nucleus interactions in the inner crust of neutron stars," Phys. Rev. C {\bf 94}, 055806:1-10 (2016).

\bibitem{chiral} E. Epelbaum, H.-W. Hammer, and U.-G. Meissner, ``Modern theory of nuclear forces,'' Rev. Mod. Phys. {\bf 81}, 1773-1825  (2009).

\bibitem{tews}  I. Tews,  S. Gandolfi, A. Gezerlis, and A. Schwenk,  ``Quantum Monte Carlo calculations of neutron matter with chiral three-body forces," Phys. Rev. C {\bf 93}, 024305 (2016).

\bibitem{drischler} C. Drischler,  K. Hebeler, and A. Schwenk, ``Asymmetric nuclear matter based on chiral two- and three-nucleon interactions," Phys. Rev. C {\bf 93}, 054314 (2016).

\bibitem{ggc2015} S. Gandolfi, A. Gezerlis, and J. Carlson, ``Neutron matter from low to high density,"  Ann. Rev. Nucl. Part. Sci. {\bf 65}, 303-328  (2015).

\bibitem{glcs}  S. Gandolfi, A. Lovato, J. Carlson, and K. E. Schmidt,  ``From the lightest nuclei to the equation of state of asymmetric nuclear matter with realistic nuclear interactions,"  Phys. Rev. C {\bf 90}, 061306 (2014).

\bibitem{dyugaev}
  A.~M.~Dyugaev,  ``Character of phase transition in case of $\pi$ condensation,''
  Pisma Zh.\ Eksp.\ Teor.\ Fiz.\  {\bf 22}, 181-5 (1975).  JETP Lett., {\bf 22},  83-5  (1975).

\bibitem{migdal-rpts}
  A.~B.~Migdal, E.~E.~Saperstein, M.~A.~Troitsky and D.~N.~Voskresensky,  ``Pion degrees of freedom in nuclear matter,''
  Phys.\ Rept.\  {\bf 192}, 179-437 (1990).

\bibitem{Book}K. Yagi, T. Hatsuda, and Y. Miake, {\em Quark-Gluon Plasma: From Big Bang to Little Bang}, Cambridge Univ. Press, Cambridge 2005). 

\bibitem{Freedman:1976ub} B. A. Freedman and L. D. McLerran, ``Fermions and gauge vector mesons at finite temperature and density. 3. The ground state energy of a relativistic quark gas,'' Phys.\ Rev.\ D {\bf 16}, 1169-85 (1977).

\bibitem{Freedman:1977gz} B. Freedman and L. D. McLerran, ``Quark star phenomenology,'' Phys.\ Rev.\ D {\bf 17}, 1109-22 (1978).

\bibitem{Fraga:2001id}E. S. Fraga, R. D. Pisarski and J. Schaffner-Bielich, ``Small, dense quark stars from perturbative QCD,'' Phys.\ Rev.\ D {\bf 63}, 121702 (2001).

\bibitem{Kurkela:2009gj}  A. Kurkela, P. Romatschke and A. Vuorinen, ``Cold quark matter,'' Phys.\ Rev.\ D {\bf 81},105021 (2010).

\bibitem{novikov} V. A. Novikov, M. A. Shifman, A. I. Vainshtein and V. I. Zakharov, ``Are all hadrons alike?,"  Nucl. Phys. B {\bf 191}, 301-69 (1981). 

\bibitem{McLerran:2007qj} L. McLerran and R. D. Pisarski, ``Phases of cold, dense quarks at large N(c),''  Nucl.\ Phys.\ A {\bf 796}, 83-100 (2007).

\bibitem{ARW} M. Alford, K. Rajagopal, and F. Wilczek ``Color-flavor locking and chiral symmetry breaking in high density QCD," Nucl. Phys. B{\bf 537}, 443-458 (1999).

\bibitem{pisarskirischke} R. D. Pisarski and D. H. Rischke, ``Why color-flavor Locking is just like chiral symmetry breaking, Proc. Judah Eisenberg Memorial Symposium, `Nuclear matter, hot and cold', Tel Aviv, April 14 - 16, 1999;
arXiv:nucl-th/9907094.


\bibitem{Kunihiro:1991qu}  T.~Kunihiro, ``Quark number susceptibility and fluctuations in the vector channel at high temperatures,'' Phys.\ Letters\ B {\bf 271}, 395-402 (1991).

\bibitem{Alford:2004pf} M.~Alford, M.~Braby, M.~W.~Paris and S.~Reddy, ``Hybrid stars that masquerade as neutron stars,''
  Astrophys.\ J.\  {\bf 629}, 969-78 (2005).
  
\bibitem{damson} D. T. Son, ``Superconductivity by long-range color magnetic interaction in high-density quark matter," Phys. Rev. D {\bf 59}, 094019 (1999).

\bibitem{DeRujula:1975qlm}  A.~De Rujula, H.~Georgi and S.~L.~Glashow, ``Hadron masses in a gauge theory," Phys.\ Rev.\ D {\bf 12} 147-62 (1975).

\bibitem{Anselmino:1992vg} M.~Anselmino, E.~Predazzi, S.~Ekelin, S.~Fredriksson and D.~B.~Lichtenberg,
``Diquarks,'' Rev.\ Mod.\ Phys.\  {\bf 65},  1199-1233 (1993).
  
\bibitem{Selem:2006nd}  A.~Selem and F.~Wilczek, ``Hadron systematics and emergent diquarks,''  hep-ph/0602128.

\bibitem{Jaffe:2004ph} R.~L.~Jaffe, ``Exotica,';  Phys.\ Rept.\  {\bf 409} 1-45 (2005). 

\bibitem{Kobayashi} M. Kobayashi and T. Maskawa, ``Chiral symmetry and eta-X mixing,''  Prog. Theor. Phys. \textbf{44}, 1422-4 (1970). 

\bibitem{'tHooft:1986nc}  G.~'t Hooft, ``How instantons solve the U(1) problem,''  Phys.\ Rept.\  {\bf 142}, 357-387 (1986).  
    

\bibitem{Gerhold} A. Gerhold and A. Rebhan, ``Gauge dependence identities for color superconducting QCD,'' Phys. Rev. D \textbf{68}, 011502 (2003).

\bibitem{Dietrich} D. D. Dietrich and D. H. Rischke, ``Gluons, tadpoles, and color neutrality in a two flavor color superconductor,'' Prog. Part. Nucl. Phys. \textbf{53}, 305-316 (2004).


\bibitem{Buballa2} M. Buballa and I. A. Shovkovy, ``A Note on color neutrality in NJL-type models,'' Phys. Rev. D \textbf{72}, 097501 (2005).  

\bibitem{Iida} K. Iida and G. Baym, ``The superfluid phases of quark matter: Ginzburg-Landau theory and color neutrality,'' Phys. Rev. D \textbf{63}, 074018 (2001).

\bibitem{Steiner} A. W. Steiner, S. Reddy, and M. Prakash, ``Color neutral superconducting quark matter,'' Phys. Rev. D \textbf{66} 094007 (2002).

\bibitem{Abuki:2005ms}H. Abuki, M. Kitazawa, and T. Kunihiro, ``How do chiral condensates affect color superconducting quark matter under charge neutrality constraints?,'' Phys. Letters B \textbf{615}, 102-110 (2005).

\bibitem{OV} J. R. Oppenheimer and G. M. Volkoff, ``On massive neutron cores, 
Phys. Rev. {\bf 55}, 374-381 (1939).

\bibitem{lo} P. M. Lo, B. Friman, and K. Redlich, ``Polyakov loop fluctuations
 and deconfinement in the limit of heavy quarks,"
Phys. Rev. D{\bf 90}, 074035 (2014). 

\bibitem{Fukushima2004}  K.~Fukushima, ``Chiral effective model with the Polyakov loop,'' Phys. Letters B {\bf 591}, 277-284 (2004).

\bibitem{Rossner2007} S. R\"{o}ssner, C. Ratti, and W. Weise, ``Polyakov loop, diquarks and the two-flavour phase diagram,'' Phys. Rev. D {\bf 75}, 034007 (2007).

\bibitem{Bratovic:2012qs}N. M. Bratovic, T. Hatsuda and W. Weise, ``Role of Vector Interaction and Axial Anomaly in the PNJL Modeling of the QCD Phase Diagram,'' Phys.\ Letters\ B {\bf 719}, 131-135 (2013).


\bibitem{steinheimer}  J.~Steinheimer and S.~Schramm, ``The problem of repulsive quark interactions - Lattice versus mean field models,'' Phys. Letters B {\bf 696},  257-261 (2011).

\bibitem{toru-screen} T. Kojo and G. Baym, ``Color screening in cold quark matter,
Phys. Rev. D {\bf 89}, 125008 (2014).

\bibitem{Alcock}  C. Alcock, E. Farhi, and A. Olinto, ``Strange stars,"  Astrophys. J. {\bf 310}, 261-272 (1986).

\bibitem{Alpar} M. A. Alpar,  ``Comment on strange stars,"  Phys. Rev. Letters {\bf 58}, 2152 (1987).

\bibitem{AHP}  M. G. Alford, S. Han, and M. Prakash, ``Generic conditions for stable hybrid stars, Phys. Rev. D 88, 083013 (2013).

\bibitem{Cheng2010} M. Cheng, et al., ``Equation of state for physical quark masses,'' Phys. Rev. D \textbf{81} 054504 (2010).

\bibitem{Borsanyi2010} S. Bors\'{a}nyi, G. Endrodi, Z. Fodor, A. Jakovac, S. D. Katz, S. Krieg, C. Ratti and K. K. Szabo, ``The QCD equation of state with dynamical quarks,'' JHEP 77: 1-32 (2010). 

\bibitem{HRG} V. Vovchenko, D. V. Anchishkin, and M. I. Gorenstein, ``Hadron resonance gas equation of state from lattice QCD," Phys.. Rev. C {\bf 91}, 024905 (2015).

\bibitem{Asakawa1995} M.~Asakawa and T.~Hatsuda, ``What thermodynamics tells about QCD plasma near phase transition,''  Phys.\ Rev.\ D {\bf 55}, 4488-91 (1997).

\bibitem{qm2015-gb}  G. Baym, ``Ultrarelativistic heavy ion collisions: the first billion seconds," Proc. QM2015 Kobe,  Nucl. Phys. A  {\bf 956}, 1-10, (2016).

\bibitem{lottini} S. Lottini and G. Torrieri, ``Quarkyonic percolation and deconfinement at finite density and number of colors,'' Phys. Rev, C {\bf 88} 024912 (2013).

\bibitem{soukoulis} C. M. Soukoulis, Q. Li, and G. S. Grest, 
``Quantum percolation in three-dimensional systems,"  Phys. Rev. B {\bf 45}, 7724-7729 (1992);  A. Kaneko, D. Uema, and T. Ohtsuki, ``Quantum percolation and the Anderson transition,"  J. Phys. Soc. Jpn. {\bf 72}, 141-142 (2003); K. Fukushima, private communication.

\bibitem{iida3} K. Iida and G. Baym, ``Superfluid phases of quark matter, III.
Supercurrents and vortices," Phys.  Rev.  D66, 014015, 1-15 (2002).

\bibitem{balachandran}  
  A.~P.~Balachandran, S.~Digal and T.~Matsuura, ``Semi-superfluid strings in high density QCD,''
  Phys.\ Rev.\ D {\bf 73}, 074009:1-0 (2006).

\bibitem{taeko} E. Nakano, M. Nitta, and T. Matsuura, ``Interactions of non-Abelian global strings, Phys. Lett. 
B {\bf 672}, 61-64 (2009); ``Non-Abelian strings in high density QCD: zero modes and interactions, 
Phys. Rev. D {\bf 78}, 045002 (2008).

\bibitem{alford-vort}  M. G. Alford, S. K. Mallavarapu,  T, Vachaspati, and A, Windisch,  ``Stability of superfluid vortices in dense quark matter,"   Phys. Rev. C 93, 045801 (2016).  

\bibitem{boojum} M. Cipriani, W. Vinci, and M. Nitta, ``Colorful boojums at the interface of a color superconductor, Phys. Rev. D {\bf 86}, 121704(R) (2012).

\bibitem{fluxtube}  G. Baym, K. Fukushima, T. Hatsuda, M. Alford, and M. Tachibana, ``Vortex continuity from the hadronic to the color-flavor locked phase," in preparation.

\bibitem{Stephanov:2004wx}  M.~A.~Stephanov, ``QCD phase diagram and the critical point,'' Prog.\ Theor.\ Phys.\ Suppl.\  {\bf 153}, 139-156 (2004) [Int.\ J.\ Mod.\ Phys.\ A {\bf 20}, 4387-92 (2005)].

\bibitem{Kitazawa:2002bc} M.~Kitazawa, T.~Koide, T.~Kunihiro and Y.~Nemoto, ``Chiral and color superconducting phase transitions with vector interaction in a simple model,'' Prog.\ Theor.\ Phys.\  {\bf 108}, 929-951 (2002).
  
\bibitem{Zhang:2008wx}  Z.~Zhang, K.~Fukushima and T.~Kunihiro, ``Number of the QCD critical points with neutral color superconductivity,'' Phys.\ Rev.\ D {\bf 79}, 014004 (2009).


\bibitem{Aggarwal:2010cw}  M.~M.~Aggarwal  et al. [STAR Collaboration], ``An experimental exploration of the QCD phase diagram: the search for the critical point and the onset of de-confinement,'' arXiv:1007.2613 [nucl-ex].

\bibitem{Abgrall:2014xwa}
  N.~Abgrall et al. [NA61 Collaboration], ``NA61/SHINE facility at the CERN SPS: beams and detector system,'' JINST {\bf 9} P06005 (2014).

\bibitem{NICA} NICA white paper, \\ http://theor0.jinr.ru/twiki-cgi/view/NICA/WebHome.

\bibitem{J-PARC} J-PARC white paper, \\ http://asrc.jaea.go.jp/soshiki/gr/hadron/jparc-hi/.

\bibitem{Luo:2015ewa} X.~Luo [STAR Collaboration], ``Energy dependence of moments of net-proton and net-charge multiplicity distributions at STAR,'' PoS CPOD {\bf 2014} 019 (2015).

\bibitem{Lacey:2014wqa} R.~A.~Lacey, ``Indications for a critical end point in the phase diagram for hot and dense nuclear matter,'' Phys.\ Rev.\ Letters\  {\bf 114}, 142301 (2015).

\bibitem{Bzdak:2013pha} A.~Bzdak and V.~Koch, ``Local efficiency corrections to higher order cumulants,'' Phys.\ Rev.\ C {\bf 91},  027901 (2015).

\bibitem{Bzdak:2016qdc} A.~Bzdak, R.~Holzmann and V.~Koch, ``Multiplicity dependent and non-binomial efficiency corrections for particle number cumulants,'' Phys. Rev. C {\bf 94}, 064907 (2016). 

\bibitem{Kitazawa:2016awu} M.~Kitazawa, ``Efficient formulas for efficiency correction of cumulants,'' Phys.\ Rev.\ C {\bf 93},  044911 (2016).

\bibitem{Ding:2015ona}  H.~T.~Ding, F.~Karsch and S.~Mukherjee, ``Thermodynamics of strong-interaction matter from Lattice QCD,'' Int.\ J.\ Mod.\ Phys.\ E {\bf 24}, 1530007 (2015).


\bibitem{sasaki} K. Fukushima and C. Sasaki, ``The phase diagram of nuclear and quark matter at high baryon density,"  Prog. Part. Nucl. Phys. {\bf 72}, 99-154 (2013). 

\bibitem{buballa3}  M. Buballa and S. Carignano, ``Inhomogeneous chiral condensates," Prog. Part. Nucl. Phys. {\bf 81}, 39-96 (2015).

\bibitem{RSS} D. H. Rischke, D. T. Son, and M. A. Stephanov, ``Asymptotic deconfinement in high-density QCD " Phys. Rev. Letters 87, 062001 (2001).

\bibitem{ruderman-bludman} S. A. Bludman and M. A. Ruderman, ``Possibility of the speed of sound exceeding the speed of light in ultradense matter,'' Phys. Rev. {\bf 170}, 1176-1184 (1968); ``Noncausality and instability in ultradense matter," Phys. Rev. D {\bf 1}, 3243-3246 (1970).

\bibitem{mal}  M. A. Ruderman, ``Causes of sound faster than light in classical models of ultradense matter,"  Phys. Rev. {\bf 172}, 1286-90 (1968).

\bibitem{caporaso} G. Caporaso and K. Brecher, ``Must ultrabaric matter be superluminal," Phys. Rev. D {\bf 20}, 1823-31 (1979).

\bibitem{Ellis2007} G. F. R. Ellis, R. Maartens, and M. A. H. MacCallum, ``Causality and the speed of sound," Gen. Relativ. Gravit. \textbf{39}, 1651-60 (2007).

\bibitem{zeldovich} Ya. B. Zeld'ovich, ``The equation of state at ultrahigh densities and its relativistic limitations," ZhETF (USSR) {\bf 41} 1609-15 (1961) [JETP (Sov. Phys.) {\bf 14}, 1143-7 (1962)].

\bibitem{rhoadesruffini} C. E. Rhoades and R. Ruffini, ``Maximum mass of a neutron star," Phys. Rev. Letters {\bf 32}, 324-7 (1974).

\bibitem{sabbadini}  A. G. Sabbadini and J. B. Hartle, ``Upper bound on the mass of non-rotating neutron stars,"
 Astrophys. Space Sci. {\bf 25}, 117-31  (1973).

\bibitem{vicky}  V. Kalogera and G. Baym, ``The maximum mass of a neutron star,"  Astrophys.
J. Letters {\bf 469},  L61-L64 (1996).

\bibitem{lindblom} L. Lindblom, ``Determining the nuclear equation of state from neutron-star masses and radii,"
Astrophys. J. {\bf 398},  569-573 (1992).

\bibitem{inversion} C. A. Raithel,  F. \"Ozel, and D. Psaltis,
``From neutron star observables to the equation of state: II, Bayesian inference of equation of state pressures, arXiv:1704.00737 [astro-ph] (2017).

\bibitem{janka} A. Bauswein, H.-T. Janka, and R. Oechslin, ``Testing approximations of thermal effects in neutron star merger simulations,"  Phys. Rev. D {\bf 82}, 084043:1-10 (2010).


\bibitem{masuda-T}  K. Masuda, T. Hatsuda, and T. Takatsuka, ``Hadron-quark crossover and hot neutron stars at birth,"  Prog. Theor. Exp. {\bf 2016} Phys. 021D01.


\end{thebibliography}
\end{document}